\documentclass[12pt]{article}

\usepackage{amsthm}
\usepackage{mathtools}
\usepackage{amsmath}
\usepackage{amssymb}
\DeclareMathAlphabet{\mathbbold}{U}{bbold}{m}{n}

\allowdisplaybreaks

\usepackage[T1]{fontenc}
\usepackage[utf8]{inputenc}
\usepackage{newtxtext, newtxmath}
\usepackage[protrusion=true, expansion=true]{microtype}
\usepackage{setspace}

\usepackage[english]{babel}
\usepackage{csquotes}

\usepackage[a4paper, margin=1in]{geometry}

\usepackage{booktabs}
\usepackage[hang, small, labelfont=bf, textfont=it, up]{caption}
\usepackage{enumerate}
\usepackage{graphicx}
\usepackage{longtable}
\usepackage{ragged2e}
\usepackage{rotating}
\usepackage{threeparttable}

\newtheorem{algorithm}{Algorithm}
\newtheorem{assumption}{Assumption}

\newtheorem{remark}{Remark}

\usepackage{abstract}

\usepackage[dvipsnames]{xcolor}

\usepackage[colorlinks=true, linkcolor=blue, urlcolor=blue, citecolor=blue]{hyperref}

\usepackage[authordate, backend=biber]{biblatex-chicago}
\usepackage{titling}

\title{
{\Large\bfseries
Inference in Unbalanced Panel Data Models with Interactive Fixed Effects
}\thanks{
First version from April 7, 2020 (see \url{https://arxiv.org/abs/2004.03414}). We are grateful to the editor, associate editor, and the referee for their helpful comments and suggestions. This work was supported by the Deutsche Forschungsgemeinschaft (DFG) under Grant 462020252. The authors report there are no competing interests to declare. The data that support the findings of this study are openly available at \url{https://github.com/dczarnowske/ife-unbalanced}.
}
}
\author{
\normalsize
Daniel Czarnowske
\thanks{
Heinrich-Heine-Universität Düsseldorf, Universitätsstr. 1, 40225 Düsseldorf, Germany
}
\and
\normalsize
Amrei Stammann
\thanks{
Universität Bayreuth, Universitätsstr. 30, 95447  Bayreuth, Germany; e-mail: \texttt{\href{mailto:amrei.stammann@uni-bayreuth.de}{amrei.stammann@uni-bayreuth.de}}
}
}
\date{\small\today}

\DeclareMathOperator{\argmin}{\arg\,\min\;}

\DeclareMathOperator{\diag}{\text{diag}}
\DeclareMathOperator{\eye}{\mathbb{I}}
\DeclareMathOperator{\iid}{\text{iid.}\;}
\DeclareMathOperator{\ind}{\mathbf{1}}

\DeclareMathOperator{\plim}{\text{plim}\;}

\DeclareMathOperator{\EX}{\mathbb{E}}
\DeclareMathOperator{\MX}{\mathbb{M}}

\DeclareMathOperator{\N}{\mathcal{N}}

\newcommand{\norm}[1]{\lVert #1 \rVert}

\begin{document}

\maketitle

\renewcommand{\abstractname}{\vspace{-5em}}
\begin{abstract}
\noindent We derive the asymptotic theory of \textcite{b2009}'s interactive fixed effects estimator for unbalanced panels in which the source of attrition is conditionally random. For inference, we propose a method of alternating projections algorithm based on straightforward scalar expressions to compute the residualized variables required for bias correction and covariance matrix estimation. Simulation experiments confirm that our asymptotic results provide reliable finite-sample approximations. We also reassess \textcite{anrr2019}. Allowing for a more general form of unobserved heterogeneity, we confirm significant effects of democratization on economic growth.
\vfill
\end{abstract}

\clearpage

\onehalfspacing

\section{Introduction}
\label{sec:introduction}

Economists are often concerned that unobserved heterogeneity is correlated with some regressors, leading to inconsistent estimates of the parameters of interest. When panel data are available, fixed effects models are frequently used to address this issue. A critical assumption of these models is that unobserved heterogeneity enters additively. If this fails, for example, because an unobserved financial crisis shock affects each country's output differently, fixed effects models are no longer appropriate (see \textcite{b2009} for additional motivating examples). This concern motivates interactive fixed effects (IFE) estimators, which model unobserved heterogeneity as a low-rank factor structure $\boldsymbol{\lambda}_{i}^{\prime} \mathbf{f}_{t}^{\phantom{\prime}}$, where $\boldsymbol{\lambda}_{i}$ and $\mathbf{f}_{t}$ are unit- and time-specific effects, respectively (see, among others, \cite{hnr1988}, \cite{p2006}, and \cite{b2009}).\footnote{\textcite{bm2015} suggest a related but different approach. Instead of imposing rank restrictions on the time-varying unobserved heterogeneity, they use a clustering approach to assign each cross-sectional unit to a specific group, where the corresponding group-specific heterogeneity is allowed to vary over time.} Throughout this article, we refer to $\boldsymbol{\lambda}_{i}$ as factor loadings and $\mathbf{f}_{t}$ as common factors.

Inspired by \textcite{ah1982}, \textcite{hnr1988} propose a quasi-differencing approach for panels with large $N$ but small $T$. They first remove factor loadings from the estimation equation and then estimate the remaining common factors and parameters using lagged regressors as instruments. While this estimator is consistent under asymptotic sequences in which $T$ is fixed, it is well known that for large $T$, the number of instruments and parameters causes bias (see \cite{ns2004}). More recent work has considered estimators that require both $N$ and $T$ to be large. \textcite{p2006} proposes a common correlated effects (CCE) estimator in the spirit of \textcites{m1978}{c1982}{c1984}, which uses cross-sectional averages of the dependent variable and the regressors to proxy for the unobserved common factors. \textcite{p2006}'s estimator is at least $\sqrt{N}$-consistent without requiring knowledge of the true rank of the factor structure or strong factor assumptions as in \textcites{b2009}{mw2015}{mw2017}. However, it requires additional parametric assumptions on the joint distribution of the dependent variable and the regressors in order to use cross-sectional averages as valid proxy variables. \textcite{b2009} proposes a different estimator that treats the common factors and factor loadings as additional parameters.\footnote{For a detailed discussion of the different interactive fixed effects estimators, we refer the reader to \textcites{b2009}{mw2015}{mw2017}.} This estimator is closely related to \textcite{b2003}'s principal components estimator for pure factor models and has the advantage of not requiring distributional assumptions about the unobserved heterogeneity. Under the assumption that the true number of factors is known, \textcite{b2009} establishes $\sqrt{NT}$-consistency irrespective of cross-sectional and/or time-serial dependence in the idiosyncratic error term. Such dependence does, however, induce an asymptotic bias in the limiting distribution, which can be corrected (see {\citereset\cite{b2009}}). \textcite{mw2017} derive an additional correction for the feedback bias (essentially a \textcite{n1981}-type bias) that arises from the inclusion of predetermined regressors such as lagged dependent variables. Because the true number of factors is generally unknown, \textcite{mw2015} show that as long as the number of factors used to estimate $\boldsymbol{\beta}$ exceeds the true number, the estimator remains at least $\sqrt{\min(N, T)}$-consistent, at the potential cost of some efficiency loss from including irrelevant factors. Given a consistent estimator of $\boldsymbol{\beta}$, the number of factors can then be estimated using estimators for pure factor models (see, among others, \cite{be1992}, \cite{bn2002}, \cite{hl2007}, \cite{abc2010}, \cite{o2010}, \cite{ah2013}, and \cite{do2019}). A recent comparison of popular estimators for pure factor models is given in \textcite{cj2019}.

In applied work, observations are often missing. A frequent cause is attrition: individuals may drop out of a panel because they move or leave the participating household, and in some cases they are replaced by new survey participants. In macroeconomic panels, countries are sometimes divided into several independent states. Non-response can also lead to the replacement of survey participants. These cases give rise to very different missing data patterns that, in the absence of sample selection, generally do not affect the properties of estimators (see \cite{fw2018}). In the presence of missing data, the principal component estimator of \textcite{b2009} requires an additional imputation step based on the EM algorithm of \textcites{sw1998}{sw2002} (see the appendix of \cite{b2009} and \cite{bly2015}). \textcite{bly2015} demonstrate consistency of the EM-type principal component estimator through simulation studies, but provide no guidance on inference. The asymptotic properties of the EM algorithm for factor models were recently studied by \textcite{jms2021}.

We make the following contributions. First, we extend \textcite{mw2017} using the insights of \textcite{fw2018} to derive the asymptotic distribution of the IFE estimator in unbalanced panels under the assumption that attrition is conditionally random. Second, we propose a novel method of alternating projections algorithm to compute the residuals required for inference. The algorithm relies on straightforward scalar expressions and is particularly suited to settings with missing data, though it can also be applied to balanced panels. We also propose an alternative estimation procedure to those of \textcites{b2009}{bly2015}. Specifically, we combine the profile-objective-function reformulation of \textcites{mw2015}{mw2017} with matrix completion methods such as the EM algorithm. This procedure eliminates the need to optimize explicitly over the high-dimensional nuisance parameters $(\boldsymbol{\lambda}_{1}^{\prime}, \ldots, \boldsymbol{\lambda}_{N}^{\prime})^{\prime}$ and $(\mathbf{f}_{1}^{\prime}, \ldots, \mathbf{f}_{T}^{\prime})^{\prime}$, and typically converges in few iterations. We also present a regularization-based matrix completion approach as an alternative to the EM algorithm, which is particularly advantageous for larger-scale panels, as noted by \textcite{fll2021}. An \texttt{R} package implementing all proposed methods is available at \url{https://github.com/dczarnowske/InteractiveEffects}. Third, we analyze the finite-sample properties of the IFE estimator for a dynamic model through simulation experiments that explore different shares of missing data, confirming that our asymptotic results provide a reliable approximation to finite-sample behavior. Additional Monte Carlo results for static models, covering various error term configurations and missing data patterns, are reported in Section \ref{os:additional_simulation} of the Online Supplement. Fourth, given that our results assume the true number of factors is known, we also examine the performance of various estimators for this quantity. For sufficiently long panels, all estimators perform similarly regardless of the share of missing observations. In configurations with high persistence, only a few estimators achieve reliable predictions; without high persistence, all estimators predict the correct number of factors almost perfectly. Fifth, we reassess the baseline analysis of \textcite{anrr2019} using the IFE estimator. Our findings qualitatively confirm their main results. However, in their preferred specification, the estimated short-run and long-run effects are roughly halved relative to those they report. Sixth, our findings and algorithms extend to several related estimators, including the minimum distance estimator of \textcites{mw2017}{msw2018} for endogenous regressors, the nuclear norm regularized estimators of \textcite{mw2026}, and the estimator for nonlinear factor models of \textcite{cfw2021}.

Related work to ours is  \textcite{sww2026}.\footnote{Some of the ideas in \textcite{sww2026} build on our earlier work. The bias correction formulas and algorithms presented in our paper first appeared in an \textit{arXiv} preprint circulated in April 2020 (see \cite{cs2020}).} Their approach extends the asymptotic expansion of \textcite{fw2016}, while ours builds on the expansion of \textcite{mw2017}. Both expansions eliminate the effects of high-dimensional nuisance parameters through projections, resulting in related inference procedures.

The paper is organized as follows. Section \ref{sec:model_estimation} introduces the model and presents estimation and inference procedures. Section \ref{sec:nof} briefly reviews estimators for the number of factors. Section \ref{sec:simulation} presents simulation results. Section \ref{sec:empirical_example} reassesses \textcite{anrr2019} using the IFE estimator. Section \ref{sec:other_estimators} presents algorithms for related estimators. Section \ref{sec:conclusion} concludes.

Throughout this article, we follow standard notation: scalars are in roman type, vectors and matrices in boldface, and all vectors are column vectors. Let $\mathbf{A}$ be an $M \times N$ matrix. We write $[\mathbf{A}]_{ij}$ for the $(i,j)$-th element of $\mathbf{A}$, where $i$ is a row index and $j$ is a column index. $\eye_{M}$ denotes the $M \times M$ identity matrix.

\section{Estimation and Inference}
\label{sec:model_estimation}

\subsection{Model, Estimator, and Asymptotic Distribution}
\label{sec:model_consistency}

We consider the following unobserved effects model:
\begin{equation}
\label{eq:ifemodel}
    y_{it} = \mathbf{x}_{it}^{\prime} \boldsymbol{\beta} + \boldsymbol{\lambda}_{i}^{\prime} \mathbf{f}_{t}^{\phantom{\prime}} + e_{it}  \, ,
\end{equation}
where $i$ and $t$ index individuals and time periods, respectively, $\mathbf{x}_{it} \coloneqq (x_{it, 1}, \ldots, x_{it, K})^{\prime}$ is a vector of $K$ regressors, $\boldsymbol{\beta}$ is the corresponding parameter vector, and $e_{it}$ is the idiosyncratic error term. Let $N$ and $T$ denote the number of individuals and time periods. To accommodate unbalanced panels, let $\mathcal{D} \subseteq \{1, \ldots, N\} \times \{1, \ldots, T\}$ denote the set of index pairs for which a complete observation is available; that is, $(i, t) \in \mathcal{D}$ if and only if $y_{it}$ and every component of $\mathbf{x}_{it}$ are observed. The panel is unbalanced whenever $y_{it}$, any element of $\mathbf{x}_{it}$, or both are missing for some $(i, t)$, and $n \coloneqq \lvert\mathcal{D}\rvert$ denotes the total sample size. The unobserved effects in \eqref{eq:ifemodel} follow a factor structure, where $\boldsymbol{\lambda}_{i} \coloneqq (\lambda_{i1}, \ldots, \lambda_{iR})^{\prime}$ is a vector of factor loadings and $\mathbf{f}_{t} \coloneqq (f_{t1}, \ldots, f_{tR})^{\prime}$ is a vector of common factors. We assume the factor structure is of low rank, with $R \ll \min(N, T)$.

Given the number of factors $R$, the estimator of the common parameters is defined as:
\begin{equation}
    \label{eq:ife_estimator}
    \hat{\boldsymbol{\beta}} \coloneqq \underset{\boldsymbol{\beta} \in \mathbb{R}^{K}}{\argmin} Q\left(\boldsymbol{\beta}\right) \, ,
\end{equation}
where
\begin{equation}
    \label{eq:ife_objfunction}
    Q\left(\boldsymbol{\beta}\right) \coloneqq \underset{\boldsymbol{\Lambda}, \mathbf{F}}{\min} \; \frac{1}{NT} \sum_{(i, t) \in \mathcal{D}} \left(y_{it} - \mathbf{x}_{it}^{\prime} \boldsymbol{\beta} - \boldsymbol{\lambda}_{i}^{\prime} \mathbf{f}_{t}^{\phantom{\prime}}\right)^{2}
\end{equation}
is the profile objective function. Here, $\boldsymbol{\Lambda} \coloneqq (\boldsymbol{\lambda}_{1}, \ldots, \boldsymbol{\lambda}_{N})^{\prime}$ is an $N \times R$ matrix of factor loadings and $\mathbf{F} \coloneqq (\mathbf{f}_{1}, \ldots, \mathbf{f}_{T})^{\prime}$ is a $T \times R$ matrix of common factors.

Let $\mathcal{C}$ be a conditioning set containing the sigma-algebra generated by the true factor loadings and common factors, and let $\mathcal{Z}_{i}^{t} \coloneqq \sigma(\{(\mathbf{x}_{is}, e_{i(s - 1)}) \colon s \leq t\})$ hold for all $i, t, N, T$. For balanced panels, \textcite{mw2017} derived the asymptotic distribution of the interactive fixed effects estimator \eqref{eq:ife_estimator} under an asymptotic framework in which $N, T \rightarrow \infty$ at the constant rate $N / T \rightarrow \kappa^{2}$ with $0 < \kappa < \infty$. Their assumptions further require that the true number of factors is known, that $\{(\mathbf{x}_{it}, e_{it}) \colon t = 1, \ldots, T\}$ is independent across $i$ (conditional on $\mathcal{C}$), that $\EX[e_{it} \mid \mathcal{C} \vee \mathcal{Z}_{i}^{t}] = 0$ holds for all $i, t, N, T$, and that the regressors are not fully absorbed by the factor structure (a non-collinearity condition).

As argued by \textcite{fw2018} in Section 4.1, missing observations do not pose major theoretical challenges if the attrition process is deterministic or conditionally random. Let $\mathcal{I}_{t} \coloneqq \{i \colon (i, t) \in \mathcal{D}\}$, $\mathcal{T}_{i} \coloneqq \{t \colon (i, t) \in \mathcal{D}\}$, and $\delta_{it}$ be an attrition indicator for all $i, t, N, T$. To derive the asymptotic distribution of \eqref{eq:ife_estimator} for unbalanced panels, we augment the assumptions of \textcite{mw2017} with one of the following assumptions.

\begin{assumption}[Stochastic Attrition Process]
    \label{ass:missing_data_stochastic}
    i) $\{(\mathbf{x}_{it}, e_{it}, \delta_{it}) \colon t = 1, \ldots, T\}$ is independent across $i$ (conditional on $\mathcal{C}$). ii) $\delta_{it}$ is independent of $(\mathbf{x}_{it}, e_{it})$ conditional on $\mathcal{C}$. iii) $\sum_{t^{\prime} = 1}^{T} \EX[\delta_{it^{\prime}} \delta_{it} \mid \mathcal{C}] - \EX[\delta_{it^{\prime}} \mid \mathcal{C}] \EX[\delta_{it} \mid \mathcal{C}] \leq c_{\max} < \infty$ a.\,s. uniformly over $i, t, N, T$. iv) $\EX[\delta_{it} \mid \mathcal{C}] \geq c_{\min} > 0$ a.\,s. uniformly over $i, t, N, T$.
\end{assumption}

\begin{assumption}[Deterministic Attrition Process]
    \label{ass:missing_data_deterministic}
    i) $\lvert \mathcal{T}_{i} \rvert / T \rightarrow c_{i} > 0$ as $T \rightarrow \infty$ for all $i$. ii) $\lvert \mathcal{I}_{t} \rvert / N \rightarrow c_{t} > 0$ as $N \rightarrow \infty$ for all $t$.
\end{assumption}

\begin{remark}[Additional Assumptions]~ 
\begin{itemize}
    \item Assumption \ref{ass:missing_data_stochastic} is a conditional missing-at-random assumption. It excludes endogenous sample selection, where missingness is associated with the contemporaneous idiosyncratic error term. Assumption \ref{ass:missing_data_stochastic} i) strengthens the conditional cross-sectional independence assumption of \textcite{mw2017} (Assumption 5 (iii)). Both assumptions are standard in the panel data econometrics literature. Assumption \ref{ass:missing_data_stochastic} ii) restricts the attrition process by requiring that, conditional on $\mathcal{C}$, observations are missing at random and independently of $(\mathbf{x}_{it}, e_{it})$. This assumption could be relaxed to a mean independence condition (see, for example, \textcite{w2010} Section 19). Assumption \ref{ass:missing_data_stochastic} iii) is a summability condition that restricts temporal dependence in the attrition process (conditional on $\mathcal{C}$). It provides a flexible characterization of weakly dependent processes and could be replaced by a strong mixing condition. Assumption \ref{ass:missing_data_stochastic} iv) ensures that every $(i, t)$ pair is observed with positive probability, which guarantees that certain matrices are positive definite, including the non-collinearity condition.
    \item Assumption \ref{ass:missing_data_deterministic} imposes regularity conditions on deterministic attrition processes, such as network settings where at least one observation per unit is missing by design, as studied by \textcite{cfw2021}. Parts i) and ii) ensure that the number of observations associated with each common factor and its loading grows with the sample size.
\end{itemize}
\end{remark}

Let $\bar{p}_{itt^{\prime}}^{f} \coloneqq \mathbf{f}_{t}^{\prime} \overline{\boldsymbol{\Phi}}_{i}^{- 1} \mathbf{f}_{t^{\prime}}$, $\bar{\xi}_{it}^{\dagger} \coloneqq \boldsymbol{\lambda}_{i}^{\prime} \overline{\boldsymbol{\Psi}}_{t}^{- 1} \overline{\boldsymbol{\Phi}}_{i}^{- 1} \mathbf{f}_{t}^{\phantom{\prime}}$, $\overline{\boldsymbol{\Phi}}_{i} \coloneqq \sum_{t = 1}^{T} \EX[\delta_{it} \mid \mathcal{C}] \, \mathbf{f}_{t}^{\phantom{\prime}} \mathbf{f}_{t}^{\prime}$, and $\overline{\boldsymbol{\Psi}}_{t} \coloneqq \sum_{i = 1}^{N} \EX[\delta_{it} \mid \mathcal{C}] \, \boldsymbol{\lambda}_{i}^{\phantom{\prime}} \boldsymbol{\lambda}_{i}^{\prime}$. Let $\mathbf{A}^{\cdot} \coloneqq (\mathbf{a}_{1}^{\cdot}, \ldots, \mathbf{a}_{T}^{\cdot})^{\prime}$ and $\mathbf{C}^{\cdot} \coloneqq (\mathbf{c}_{1}^{\cdot}, \ldots, \mathbf{c}_{N}^{\cdot})^{\prime}$, where the dot in the exponent is a placeholder. Under the assumptions of \textcite{mw2017} augmented by Assumption \ref{ass:missing_data_stochastic}, the estimator in \eqref{eq:ife_estimator} has the following asymptotic distribution when data are conditionally missing at random:
\begin{equation}
    \label{eq:asympotic_distribution}
    \sqrt{NT} (\hat{\boldsymbol{\beta}} - \boldsymbol{\beta}) + \kappa \, \mathbf{W}^{- 1} \mathbf{B}_{1} + \kappa^{- 1} \, \mathbf{W}^{- 1} \mathbf{B}_{2} + \kappa \, \mathbf{W}^{- 1} \mathbf{B}_{3} \overset{d}{\rightarrow} \N(0, \mathbf{W}^{- 1} \boldsymbol{\Omega} \mathbf{W}^{- 1})
\end{equation}
where
\begin{align*} 
    \mathbf{W} \coloneqq& \, \underset{N, T \rightarrow \infty}{\plim} \, \frac{1}{NT} \sum_{i = 1}^{N} \sum_{t = 1}^{T} \EX[\delta_{it} \mid \mathcal{C}] \, \EX\left[\ddot{\mathbf{x}}_{it}^{\lambda f} (\ddot{\mathbf{x}}_{it}^{\lambda f})^{\prime} \mid \mathcal{C}\right] \, , \\
    \boldsymbol{\Omega} \coloneqq& \, \underset{N, T \rightarrow \infty}{\plim} \, \frac{1}{NT} \sum_{i = 1}^{N} \sum_{t = 1}^{T} \EX[\delta_{it} \mid \mathcal{C}] \, \EX\left[e_{it}^{2} \, \ddot{\mathbf{x}}_{it}^{\lambda f} (\ddot{\mathbf{x}}_{it}^{\lambda f})^{\prime} \mid \mathcal{C}\right] \, ,\\
    \mathbf{B}_{1} \coloneqq& \, \underset{N, T \rightarrow \infty}{\plim} \, \frac{1}{N} \sum_{i = 1}^{N} \sum_{t = 1}^{T - 1} \sum_{t^{\prime} = t + 1}^{T} \EX[\delta_{it^{\prime}} \delta_{it} \mid \mathcal{C}] \, \bar{p}_{itt^{\prime}}^{f} \EX[\ddot{\mathbf{x}}_{it^{\prime}}^{f} e_{it} \mid \mathcal{C}] \, , \\
    \mathbf{B}_{2} \coloneqq& \, \underset{N, T \rightarrow \infty}{\plim} \, \frac{1}{T} \sum_{i = 1}^{N} \bigg(\sum_{t = 1}^{T} \EX[\delta_{it} \mid \mathcal{C}] \, \EX[e_{it}^{2} \mid \mathcal{C}] \bigg) \bigg(\sum_{t = 1}^{T} \EX[\delta_{it} \mid \mathcal{C}] \, \EX[\ddot{\mathbf{x}}_{it}^{\lambda} \mid \mathcal{C}] \,  \bar{\xi}_{it}^{\dagger} \bigg) \, , \\
    \mathbf{B}_{3} \coloneqq& \, \underset{N, T \rightarrow \infty}{\plim} \, \frac{1}{N} \sum_{t = 1}^{T} \bigg(\sum_{i = 1}^{N} \EX[\delta_{it} \mid \mathcal{C}] \, \EX[e_{it}^{2} \mid \mathcal{C}] \bigg) \bigg(\sum_{i = 1}^{N} \EX[\delta_{it} \mid \mathcal{C}] \, \EX[\ddot{\mathbf{x}}_{it}^{f} \mid \mathcal{C}] \,  \bar{\xi}_{it}^{\dagger} \bigg) \, , 
\end{align*}
with 
\begin{align*}
    &\ddot{x}_{it, k}^{\lambda f} \coloneqq \, x_{it, k} - \boldsymbol{\lambda}_{i}^{\prime} \mathbf{a}_{t}^{\ast\ast} - \mathbf{f}_{t}^{\prime} \mathbf{c}_{i}^{\ast\ast} \, , \\
    &(\mathbf{A}^{\ast\ast}, \mathbf{C}^{\ast\ast}) \in \, \underset{\mathbf{A} \in \mathbb{R}^{T \times R}, \, \mathbf{C} \in \mathbb{R}^{N \times R}}{\argmin} \sum_{i = 1}^{N} \sum_{t = 1}^{T} \EX[\delta_{it} \mid \mathcal{C}] (\EX[x_{it, k} \mid \mathcal{C}] - \boldsymbol{\lambda}_{i}^{\prime} \mathbf{a}_{t}^{\phantom{\prime}} - \mathbf{f}_{t}^{\prime} \mathbf{c}_{i}^{\phantom{\prime}})^{2} \, ,  \\
    &\ddot{x}_{it, k}^{\lambda} \coloneqq \, x_{it, k} - \boldsymbol{\lambda}_{i}^{\prime} \mathbf{a}_{t}^{\ast} \, , \quad \mathbf{A}^{\ast} \in \, \underset{\mathbf{A} \in \mathbb{R}^{T \times R}}{\argmin} \sum_{i = 1}^{N} \sum_{t = 1}^{T} \EX[\delta_{it} \mid \mathcal{C}] (\EX[x_{it, k} \mid \mathcal{C}] - \boldsymbol{\lambda}_{i}^{\prime} \mathbf{a}_{t}^{\phantom{\prime}})^{2} \, ,  \\
    &\ddot{x}_{it, k}^{f} \coloneqq \, x_{it, k} - \mathbf{f}_{t}^{\prime} \mathbf{c}_{i}^{\ast} \, , \quad \mathbf{C}^{\ast} \in \, \underset{\mathbf{C} \in \mathbb{R}^{N \times R}}{\argmin} \sum_{i = 1}^{N} \sum_{t = 1}^{T} \EX[\delta_{it} \mid \mathcal{C}] (\EX[x_{it, k} \mid \mathcal{C}] - \mathbf{f}_{t}^{\prime} \mathbf{c}_{i}^{\phantom{\prime}})^{2} \, , 
\end{align*}
denoting residuals from population projections. When the attrition process is deterministic, we augment the assumptions of \textcite{mw2017} by Assumption \ref{ass:missing_data_deterministic}, and the asymptotic distribution follows immediately by replacing $\EX[\delta_{it} \mid \mathcal{C}]$ with $\delta_{it}$ in \eqref{eq:asympotic_distribution}. The derivation is provided in Appendix \ref{app:derivation_distribution}.

The bias term $\mathbf{B}_{1}$ represents feedback bias (a generalization of the \textcite{n1981}-bias) arising from potential feedback from past outcomes to future realizations of the regressors. Specifically, $\mathbf{x}_{it}$ may depend on $(e_{i(t - 1)}, e_{i(t - 2)}, \ldots)$, $\boldsymbol{\lambda}_{i}$, and $\mathbf{f}_{t}$ in an arbitrary nonlinear manner. Our framework thus naturally accommodates dynamic specifications such as $\mathbf{x}_{it} = y_{i(t - 1)}$. Feedback bias is ruled out by assumption in \textcite{b2009} and was first introduced by \textcite{mw2017}.

The remaining bias terms, $\mathbf{B}_{2}$ and $\mathbf{B}_{3}$, are also present in \textcite{b2009} and \textcite{mw2017} for balanced panels. They arise when the idiosyncratic error term is heteroskedastic across individuals or over time, respectively. The reason is that $\sum_{i = 1}^{N} \sum_{t = 1}^{T} \EX[\ddot{\mathbf{x}}_{it}^{\lambda} \mid \mathcal{C}] \,  \bar{\xi}_{it}^{\dagger} = 0$ and $\sum_{i = 1}^{N} \sum_{t = 1}^{T} \EX[\ddot{\mathbf{x}}_{it}^{f} \mid \mathcal{C}] \,  \bar{\xi}_{it}^{\dagger} = 0$ follow from the definition of the population residuals. In unbalanced panels, the attrition process can induce a form of heteroskedasticity. Consequently, even when the idiosyncratic error term is homoskedastic, $\mathbf{B}_{2}$ and $\mathbf{B}_{3}$ are generally non-zero, since missing probabilities may also be heterogeneous. This finding is consistent with \textcite{sww2026}.

The covariance matrix $\mathbf{W}^{- 1} \boldsymbol{\Omega} \mathbf{W}^{- 1}$ allows for arbitrary heteroskedasticity. For balanced panels, \textcite{b2009} and \textcite{mw2017} discuss simplifications that arise under homoskedasticity in the cross-section and/or time dimension. However, as noted in the discussion of $\mathbf{B}_{2}$ and $\mathbf{B}_{3}$, the attrition process can induce heteroskedasticity, rendering such simplifications generally invalid.

The bias terms $\mathbf{B}_{1}$ and $\mathbf{B}_{3}$ are of order $\overline{T}^{- 1}$, while $\mathbf{B}_{2}$ is of order $\overline{N}^{- 1}$, where $\overline{T} \coloneqq N^{- 1} \sum_{i = 1}^{N} \sum_{t = 1}^{T} \EX[\delta_{it} \mid \mathcal{C}]$ and $\overline{N} \coloneqq T^{- 1} \sum_{t = 1}^{T} \sum_{i = 1}^{N} \EX[\delta_{it} \mid \mathcal{C}]$. The bias terms are therefore larger, to a degree that depends on the extent of missing data. This is consistent with the results of \textcite{fw2018} on the asymptotic distribution of traditional fixed effects estimators in unbalanced panels.

For balanced panels, the asymptotic distribution reduces to that derived by \textcite{mw2017}, rendering Assumptions \ref{ass:missing_data_stochastic} and \ref{ass:missing_data_deterministic} redundant.

\begin{remark}[Strict exogeneity]
If the regressors are strictly exogenous rather than weakly exogenous, as in \textcite{b2009}, i.e., $\EX[e_{it} \mid \mathcal{C} \vee \mathcal{X}_{i}] = 0$, where $\mathcal{X}_{i} \coloneqq \sigma(\{\mathbf{x}_{is} \colon s \in \{1, \ldots, T\}\})$ holds for all $i, t, N, T$, then there is no feedback bias, i.e., $\mathbf{B}_{1} = \mathbf{0}_{K}$. Hence, the asymptotic distribution in \eqref{eq:asympotic_distribution} simplifies by dropping the first bias term. However, as noted in Remark 6 of \textcite{b2009}, the idiosyncratic errors may still exhibit weak serial correlation. In such settings, the bias term $\mathbf{B}_{3}$ and the covariance matrix $\boldsymbol{\Omega}$ become
\begin{align*}
    \mathbf{B}_{3} =& \, \underset{N, T \rightarrow \infty}{\plim} \, \frac{1}{N} \sum_{t = 1}^{T} \sum_{t^{\prime} = 1}^{T} \bigg(\sum_{i = 1}^{N} \EX[\delta_{it^{\prime}} \delta_{it} \mid \mathcal{C}] \, \EX[e_{it^{\prime}} e_{it} \mid \mathcal{C}] \bigg) \bigg(\sum_{i = 1}^{N} \EX[\delta_{it^{\prime}} \delta_{it} \mid \mathcal{C}] \, \EX[\ddot{\mathbf{x}}_{it^{\prime}}^{f} \mid \mathcal{C}] \,  \bar{\xi}_{it}^{\dagger} \bigg) \, , \\
    \boldsymbol{\Omega} \coloneqq& \, \underset{N, T \rightarrow \infty}{\plim} \, \frac{1}{NT} \sum_{i = 1}^{N} \sum_{t = 1}^{T} \sum_{t^{\prime} = 1}^{T} \EX[\delta_{it^{\prime}} \delta_{it} \mid \mathcal{C}] \, \EX\left[e_{it^{\prime}} e_{it} \, \ddot{\mathbf{x}}_{it^{\prime}}^{\lambda f} (\ddot{\mathbf{x}}_{it}^{\lambda f})^{\prime} \mid \mathcal{C}\right] \, .
\end{align*}
We maintain the assumption that $e_{it}$ is independent across $i$ (conditional on $\mathcal{C}$), thus ruling out cross-sectional correlation, as discussed in Remark 7 of \textcite{b2009}.
\end{remark}

\subsection{Estimation Algorithm}
\label{sec:estimation}

For balanced panels, \textcites{mw2015}{mw2017} showed that the profile objective function \eqref{eq:ife_objfunction} can be reformulated as
\begin{equation*}
Q\left(\boldsymbol{\beta}\right) = \underset{\boldsymbol{\Lambda}, \mathbf{F}}{\min} \; \frac{1}{NT} \sum_{i = 1}^{N} \sum_{t = 1}^{T} \left(y_{it} - \mathbf{x}_{it}^{\prime} \boldsymbol{\beta} - \boldsymbol{\lambda}_{i}^{\prime} \mathbf{f}_{t}^{\phantom{\prime}}\right)^{2} = \frac{1}{NT} \sum_{r = R + 1}^{T} \mu_{r} \big(\boldsymbol{\Gamma}(\boldsymbol{\beta})^{\prime} \boldsymbol{\Gamma}(\boldsymbol{\beta})\big) \, ,
\end{equation*}
where $\boldsymbol{\Gamma}(\boldsymbol{\beta})$ is an $N \times T$ matrix with $[\boldsymbol{\Gamma}(\boldsymbol{\beta})]_{it} = y_{it} - \mathbf{x}_{it}^{\prime} \boldsymbol{\beta}$, and $\mu_{r}(\cdot)$ denotes the $r$-th largest eigenvalue. This reformulation is advantageous because it eliminates the need to optimize explicitly over the high-dimensional nuisance parameters $\boldsymbol{\Lambda}$ and $\mathbf{F}$. Moreover, since modern algorithms for symmetric eigenvalue problems are highly optimized, computing $\hat{\boldsymbol{\beta}}$ remains efficient even for large $T$. Estimates of $\boldsymbol{\Lambda}$ and $\mathbf{F}$ are subsequently recovered by decomposing $\widehat{\boldsymbol{\Gamma}} \coloneqq \boldsymbol{\Gamma}(\hat{\boldsymbol{\beta}})$. Specifically, under the normalizing restrictions $\mathbf{F}^{\prime} \mathbf{F}^{\phantom{\prime}} / T = \eye_{R}$ and $\boldsymbol{\Lambda}^{\prime}\boldsymbol{\Lambda}^{\phantom{\prime}}$ diagonal, $\widehat{\mathbf{F}}$ equals the first $R$ eigenvectors of $\widehat{\boldsymbol{\Gamma}}^{\prime} \widehat{\boldsymbol{\Gamma}}$ multiplied by $\sqrt{T}$, and $\widehat{\boldsymbol{\Lambda}} = \widehat{\boldsymbol{\Gamma}} \widehat{\mathbf{F}} / T$.\footnote{Other valid normalizing restrictions are discussed in \textcite{bn2013}. Moreover, if $T > N$, it is computationally more efficient to minimize $(NT)^{- 1} \sum_{r = R + 1}^{N} \mu_{r} \big(\boldsymbol{\Gamma}(\boldsymbol{\beta}) \boldsymbol{\Gamma}(\boldsymbol{\beta})^{\prime}\big)$ and estimate $\widehat{\boldsymbol{\Lambda}}$ as the first $R$ eigenvectors of $\widehat{\boldsymbol{\Gamma}} \widehat{\boldsymbol{\Gamma}}^{\prime}$ multiplied by $\sqrt{N}$ and $\widehat{\mathbf{F}} = \widehat{\boldsymbol{\Gamma}}^{\prime} \widehat{\boldsymbol{\Lambda}} / N$, imposing $\boldsymbol{\Lambda}^{\prime} \boldsymbol{\Lambda}^{\phantom{\prime}} / N = \eye_{R}$, where $\mathbf{F}^{\prime}\mathbf{F}^{\phantom{\prime}}$ is diagonal.}

The estimation procedure for unbalanced panels is motivated by the following decomposition of \eqref{eq:ifemodel}:
\begin{equation}
    \label{eq:ifemodel_decomposition}
    y_{it} - \mathbf{x}_{it}^{\prime} \boldsymbol{\beta} = [\boldsymbol{\Gamma}(\boldsymbol{\beta})]_{it} = \boldsymbol{\lambda}_{i}^{\prime} \mathbf{f}_{t}^{\phantom{\prime}} + e_{it} \, ,
\end{equation}
where $\boldsymbol{\Gamma}(\boldsymbol{\beta})$ has missing entries corresponding to unobserved index pairs, i.e., entries are missing whenever $(i, t) \notin \mathcal{D}$. The key idea is that, for a given $\boldsymbol{\beta}$, the observed entries of $\boldsymbol{\Gamma}(\boldsymbol{\beta})$ can be used to estimate $\boldsymbol{\Lambda}$ and $\mathbf{F}$, which in turn permit imputation of the missing entries via $\boldsymbol{\lambda}_{i}^{\prime} \mathbf{f}_{t}^{\phantom{\prime}}$.

We introduce two matrix completion algorithms to accomplish this. Algorithm 1 is the classical Expectation-Maximization (EM) algorithm, originally developed by \textcites{sw1998}{sw2002} for pure factor models. Algorithm 2 follows \textcite{fll2021}, combining nuclear norm regularization with a debiasing step to mitigate regularization bias. The second approach is particularly advantageous for large-scale panel data, as it typically offers substantial computational speed gains over the EM algorithm. Numerical comparisons of both algorithms are provided in Appendix \ref{app:num_comp_mc}. We then demonstrate how these matrix completion algorithms integrate into the reformulation of the profile objective function proposed by \textcites{mw2015}{mw2017}.

Before introducing the matrix completion algorithms, we adopt the notation of \textcite{ccs2010} for handling observed and missing data. Let
\begin{equation*}
    [\mathcal{P}_{\mathcal{D}}^{\phantom{\perp}}(\mathbf{M})]_{it} \coloneqq \begin{cases}
    [\mathbf{M}]_{it} & \text{if} \;\; (i, t) \in \mathcal{D}\\
    0 & \text{otherwise}
    \end{cases}
\end{equation*}
denote the projection operator onto the subspace of matrices whose support is contained in $\mathcal{D}$, and let $\mathcal{P}_{\mathcal{D}}^{\perp}$ denote its orthogonal complement, defined analogously with $\mathcal{D}$ replaced by its complement. By construction, $\mathcal{P}_{\mathcal{D}}^{\phantom{\perp}}(\mathbf{M}) + \mathcal{P}_{\mathcal{D}}^{\perp}(\mathbf{M}) = \mathbf{M}$ for any $N \times T$ matrix $\mathbf{M}$.

We now present the first matrix completion algorithm.
\begin{algorithm}\label{alg:mc_em}
	EM algorithm
	
	\noindent Given $\boldsymbol{\Gamma}(\boldsymbol{\beta})$, $\mathcal{D}$, and $R$. Initialize $\mathbf{M} = \mathbf{0}_{N \times T}$ and repeat the following steps until convergence.
	\begin{description}
		\item[Step 1.] Set $\boldsymbol{\Gamma}^{\ast}(\boldsymbol{\beta}) = \mathcal{P}_{\mathcal{D}}^{\phantom{\perp}}(\boldsymbol{\Gamma}(\boldsymbol{\beta})) + \mathcal{P}_{\mathcal{D}}^{\perp}(\mathbf{M})$.
		\item[Step 2.] Update $\mathbf{M} = \boldsymbol{\Gamma}^{\ast}(\boldsymbol{\beta}) \mathbf{F}^{\ast} (\mathbf{F}^{\ast})^{\prime} / T$, where $\mathbf{F}^{\ast}$ are the first $R$ eigenvectors of $\boldsymbol{\Gamma}^{\ast}(\boldsymbol{\beta})^{\prime} \boldsymbol{\Gamma}^{\ast}(\boldsymbol{\beta})$ multiplied by $\sqrt{T}$.
	\end{description}
    Return $\boldsymbol{\Gamma}^{\ast}(\boldsymbol{\beta}) = \mathcal{P}_{\mathcal{D}}^{\phantom{\perp}}(\boldsymbol{\Gamma}(\boldsymbol{\beta})) + \mathcal{P}_{\mathcal{D}}^{\perp}(\mathbf{M})$ after convergence.
\end{algorithm}

Algorithm \ref{alg:mc_em} is the classical approach to missing data for pure factor models. Heuristically, Step 1 is the E-step, where missing entries are imputed using current parameter estimates, and Step 2 is the M-step, which applies eigenvalue decomposition to the completed data, motivated by the decomposition in \eqref{eq:ifemodel_decomposition}. Despite its longstanding use in empirical work, the formal asymptotic properties of this algorithm were established only recently by \textcite{jms2021}. As noted by \textcite{fll2021}, the EM algorithm can be computationally burdensome for large-scale panel data relative to modern regularization-based alternatives.

Hence, we next present the regularized matrix completion algorithm proposed as Algorithm 5 in \textcite{fll2021}.
\begin{algorithm} \label{alg:mc_redebias}
	Regularized matrix completion algorithm with debiasing
	
	\noindent Given $\boldsymbol{\Gamma}(\boldsymbol{\beta})$, $\mathcal{D}$, $R$, and $\nu > 0$. 
	\begin{description}
        \item[Step 1.] Initialize $\mathbf{M} = \mathbf{0}_{N \times T}$ and repeat the following steps until convergence.
        \item[\quad Step 1.1.] Set $\boldsymbol{\Gamma}^{\ast}(\boldsymbol{\beta}) = \mathcal{P}_{\mathcal{D}}^{\phantom{\perp}}(\boldsymbol{\Gamma}(\boldsymbol{\beta})) + \mathcal{P}_{\mathcal{D}}^{\perp}(\mathbf{M})$.
        \item[\quad Step 1.2.] Update $\mathbf{M} = \mathcal{S}_{\nu}(\boldsymbol{\Gamma}^{\ast}(\boldsymbol{\beta}))$, where $\mathcal{S}_{\nu}(\boldsymbol{\Gamma}^{\ast}(\boldsymbol{\beta})) = \mathbf{U} \boldsymbol{\Sigma}_{\nu} \mathbf{V}^{\prime}$, $\boldsymbol{\Gamma}^{\ast}(\boldsymbol{\beta}) = \mathbf{U} \boldsymbol{\Sigma} \mathbf{V}^{\prime}$ is the singular value decomposition of the rank-$r$ matrix $\boldsymbol{\Gamma}^{\ast}(\boldsymbol{\beta})$ with $\boldsymbol{\Sigma} = \diag(\sigma_{1}, \ldots, \sigma_{r})$ and $r \leq \min(N, T)$, $\boldsymbol{\Sigma}_{\nu} = \diag((\sigma_{1} - \nu)_{+}, \ldots, (\sigma_{r} - \nu)_{+})$, and $(b)_{+}$ equals $b$ if $b > 0$ and zero otherwise.
		\item[Step 2.] Set $\boldsymbol{\Gamma}^{\ast}(\boldsymbol{\beta}) = \mathcal{P}_{\mathcal{D}}^{\phantom{\perp}}(\boldsymbol{\Gamma}(\boldsymbol{\beta})) + \mathcal{P}_{\mathcal{D}}^{\perp}(\mathbf{M})$ and compute $\boldsymbol{\Lambda} = (\boldsymbol{\lambda}_{1}, \ldots, \boldsymbol{\lambda}_{N})^{\prime}$ as the first $R$ eigenvectors of $\boldsymbol{\Gamma}^{\ast}(\boldsymbol{\beta}) \boldsymbol{\Gamma}^{\ast}(\boldsymbol{\beta})^{\prime}$ multiplied by $\sqrt{N}$.
        \item[Step 3.] Compute $\widetilde{\mathbf{F}} = (\tilde{\mathbf{f}}_{1}, \ldots, \tilde{\mathbf{f}}_{T})^{\prime}$, where $\tilde{\mathbf{f}}_{t} = (\sum_{i \in \mathcal{I}_{t}} \boldsymbol{\lambda}_{i} \boldsymbol{\lambda}_{i}^{\prime})^{- 1} \sum_{i \in \mathcal{I}_{t}} \boldsymbol{\lambda}_{i} [\boldsymbol{\Gamma}(\boldsymbol{\beta})]_{it}$ for each $t \in \{1, \ldots, T\}$.
        \item[Step 4.] Compute $\widetilde{\boldsymbol{\Lambda}} = (\tilde{\boldsymbol{\lambda}}_{1}, \ldots, \tilde{\boldsymbol{\lambda}}_{N})^{\prime}$, where $\tilde{\boldsymbol{\lambda}}_{i} = (\sum_{t \in \mathcal{T}_{i}} \tilde{\mathbf{f}}_{t} \tilde{\mathbf{f}}_{t}^{\prime})^{- 1} \sum_{t \in \mathcal{T}_{i}} \tilde{\mathbf{f}}_{t} [\boldsymbol{\Gamma}(\boldsymbol{\beta})]_{it}$ for each $i \in \{1, \ldots, N\}$.
		\item[Step 5.] Set $\widetilde{\mathbf{M}} = \widetilde{\boldsymbol{\Lambda}} \widetilde{\mathbf{F}}^{\prime}$.
	\end{description}
    Return $\boldsymbol{\Gamma}^{\ast}(\boldsymbol{\beta}) = \mathcal{P}_{\mathcal{D}}^{\phantom{\perp}}(\boldsymbol{\Gamma}(\boldsymbol{\beta})) + \mathcal{P}_{\mathcal{D}}^{\perp}(\widetilde{\mathbf{M}})$.
\end{algorithm}

Algorithm \ref{alg:mc_redebias} is a modern matrix completion approach combining nuclear norm regularization with a post-estimation debiasing step. Step 1 implements the \textit{SOFT-IMPUTE} algorithm of \textcite{mht2010} to solve the nuclear norm penalized optimization problem. Unlike Algorithm \ref{alg:mc_em}, which imposes a ``hard'' rank constraint by retaining only the first $R$ singular values, Algorithm \ref{alg:mc_redebias} restricts the rank implicitly via the tuning parameter $\nu$. Since the nuclear norm is the convex envelope of the rank operator, Algorithm \ref{alg:mc_redebias} is expected to outperform Algorithm \ref{alg:mc_em} in terms of computational speed in many settings, particularly high-dimensional ones (see \textcite{mht2010} for details). Steps 2 through 4 implement the two-step least squares debiasing procedure of \textcites{chlz2019}{chlz2023} to mitigate regularization bias.

\begin{remark}[Selection of the tuning parameter]
    Following \textcites{chlz2019}{chlz2023}, the tuning parameter $\nu$ must satisfy $\nu > c_{\nu} \max(\sqrt{N}, \sqrt{T})$ for some constant $c_{\nu} > 0$. Analogously to the Lasso literature, $\nu$ must be large enough to dominate the ``score'' $\norm{\mathbf{E}^{\ast}}_{2}$ with high probability, where $\mathbf{E}^{\ast}$ is the $N \times T$ matrix of observed errors with entries $[\mathbf{E}^{\ast}]_{it} = \delta_{it} e_{it}$. Under the assumption that $e_{it}$ is independent across $i$ and $t$ (conditional on $\mathcal{C}$), results from \textcite{l2005} imply $\norm{\mathbf{E}^{\ast}}_{2} \leq c_{e} \max(\sqrt{N}, \sqrt{T})$ for some constant $c_{e} > 0$, provided the fourth moments of the idiosyncratic error are uniformly bounded. \textcite{mw2017} extend this bound to settings with weak temporal and cross-sectional dependence via high-level summability conditions detailed in their supplementary material. For a comprehensive theoretical treatment of spectral norm bounds under different dependence structures, see \textcite{v2012}. In practice, $\nu$ can be selected by cross-validation, as described in \textcite{abdik2021}, or via a plug-in approach proposed by \textcites{chlz2019}{chlz2023}.
\end{remark}

For unbalanced panels, we adapt the profile objective function to accommodate missing observations by using the completed matrix:
\begin{equation}
    \label{eq:ife_objfunction2}
    Q\left(\boldsymbol{\beta}\right) = \frac{1}{NT} \sum_{r = R + 1}^{T} \mu_{r} \big(\boldsymbol{\Gamma}^{\ast}(\boldsymbol{\beta})^{\prime} \boldsymbol{\Gamma}^{\ast}(\boldsymbol{\beta})\big) \, ,
\end{equation}
where $\boldsymbol{\Gamma}^{\ast}(\boldsymbol{\beta})$ is the completed matrix obtained after convergence of Algorithm \ref{alg:mc_em} or \ref{alg:mc_redebias}. After obtaining $\hat{\boldsymbol{\beta}}$ by minimizing \eqref{eq:ife_objfunction2}, $\widehat{\boldsymbol{\Lambda}}$ and $\widehat{\mathbf{F}}$ are recovered by decomposing $\widehat{\boldsymbol{\Gamma}}^{\ast} \coloneqq \boldsymbol{\Gamma}^{\ast}(\hat{\boldsymbol{\beta}})$.

To solve this minimization problem efficiently, we recommend a Quasi-Newton method (e.g., BFGS) with an analytical gradient. Let $\check{\boldsymbol{\beta}}$ denote a trial value, and let $\check{\boldsymbol{\Lambda}}$ and $\check{\mathbf{F}}$ be the estimates of $\boldsymbol{\Lambda}$ and $\mathbf{F}$ obtained from the decomposition of $\check{\boldsymbol{\Gamma}}^{\ast} \coloneqq \boldsymbol{\Gamma}^{\ast}(\check{\boldsymbol{\beta}})$. The analytical gradient is
\begin{equation*}
    \left. \frac{\partial Q\left(\boldsymbol{\beta}\right)}{\partial \boldsymbol{\beta}} \right|_{\boldsymbol{\beta} = \check{\boldsymbol{\beta}}} = - \frac{2}{NT} \sum_{(i, t) \in \mathcal{D}} \big([\check{\boldsymbol{\Gamma}}^{\ast}]_{it} - [\check{\boldsymbol{\Lambda}} \check{\mathbf{F}}^{\prime}]_{it} \big) \, \mathbf{x}_{it} \, .
\end{equation*}
In our simulations and empirical applications, BFGS typically converges in few iterations and substantially reduces the computational overhead introduced by the iterative matrix completion task.

Our estimation procedure is based on a reformulation of the objective function in \textcite{b2009} for unbalanced panels. Convergence of his alternating estimation algorithm was recently established by \textcite{sww2026}, and we expect their results to carry over to our setting.

Because the rank constraint renders the optimization problem in \eqref{eq:ife_objfunction2} non-convex (see \textcite{mw2026}), the choice of starting values is critical. Following the intuition in \textcite{sw2016b}, one could construct initial estimates from balanced sub-panels. However, this approach requires sufficiently large sub-panels and still necessitates testing multiple starting guesses.

To overcome these limitations, we recommend initializing the optimization with the nuclear norm minimizing estimator of \textcite{mw2026}:
\begin{equation}
    \label{eq:nnm_estimator}
	\hat{\boldsymbol{\beta}}^{\star} \coloneqq \underset{\boldsymbol{\beta} \in \mathbb{R}^{K}}{\argmin} Q^{\star}(\beta) \, , \quad Q^{\star}(\beta) \coloneqq \frac{1}{NT} \sum_{j = 1}^{\min(N, T)} \sigma_{j} \big(\mathcal{P}_{\mathcal{D}}^{\phantom{\perp}}(\boldsymbol{\Gamma}(\boldsymbol{\beta}))\big) \, ,
\end{equation}
where $\sigma_{j}(\cdot)$ denotes the $j$-th largest singular value. The key advantage of \eqref{eq:nnm_estimator} is that its objective function is convex. Although \textcite{mw2026} show that this estimator is consistent only at rate $\sqrt{\min(N, T)}$ (rather than the rate $\sqrt{NT}$ of \eqref{eq:ife_estimator}), it provides a reliable and computationally efficient starting guess for minimizing \eqref{eq:ife_objfunction2}. We recommend solving \eqref{eq:nnm_estimator} using a Quasi-Newton method with analytical gradient
\begin{equation*}
    \left. \frac{\partial Q^{\star}\left(\boldsymbol{\beta}\right)}{\partial \boldsymbol{\beta}} \right|_{\boldsymbol{\beta} = \check{\boldsymbol{\beta}}} = - \frac{1}{NT} \sum_{(i, t) \in \mathcal{D}} [\check{\mathbf{U}} \check{\mathbf{V}}^{\prime}]_{it} \, \mathbf{x}_{it} \, ,
\end{equation*}
where $\mathcal{P}_{\mathcal{D}}^{\phantom{\perp}}(\boldsymbol{\Gamma}(\check{\boldsymbol{\beta}})) = \check{\mathbf{U}} \check{\boldsymbol{\Sigma}} \check{\mathbf{V}}^{\prime}$ is the singular value decomposition of $\mathcal{P}_{\mathcal{D}}^{\phantom{\perp}}(\boldsymbol{\Gamma}(\check{\boldsymbol{\beta}}))$ with $\check{\boldsymbol{\Sigma}} = \diag(\check{\sigma}_{1}, \ldots, \check{\sigma}_{\min(N, T)})$.

\begin{remark}[Alternative estimation procedures]
    The supplementary material of \textcite{b2009} introduces an alternative estimation procedure for unbalanced panels. This approach alternates between updating $\check{\boldsymbol{\beta}}$ given $(\check{\boldsymbol{\Lambda}}, \check{\mathbf{F}})$ and updating $(\check{\boldsymbol{\Lambda}}, \check{\mathbf{F}})$ given $\check{\boldsymbol{\beta}}$ until convergence, where $\check{\boldsymbol{\Lambda}}$ and $\check{\mathbf{F}}$ are recovered by decomposing the completed matrix $\check{\boldsymbol{\Gamma}}^{\ast}$. Although \textcite{b2009} uses Algorithm \ref{alg:mc_em} for the matrix completion step, it can be replaced by Algorithm \ref{alg:mc_redebias}. For balanced panels, further estimation procedures are detailed in \textcite{b2009} and \textcite{mw2015}. We expect these methods can be adapted to unbalanced settings by incorporating the algorithms discussed in this paper.
\end{remark}

\subsection{Bias Correction}
\label{sec:bc_inference}

We obtain estimators for $\mathbf{W}$, $\boldsymbol{\Omega}$, $\mathbf{B}_{1}$, $\mathbf{B}_{2}$, and $\mathbf{B}_{3}$ by forming sample analogues, i.e., by dropping expectations and substituting the corresponding estimators for $\boldsymbol{\beta}$, $\boldsymbol{\Lambda}$, and $\mathbf{F}$. Let $L$ denote a bandwidth parameter for the truncation kernel of \textcite{nw1987}, depending on the sample size. Then,
\begin{align*} 
    \widehat{\mathbf{W}} \coloneqq& \, \frac{1}{n} \sum_{(i, t) \in \mathcal{D}} \hat{\mathbf{x}}_{it}^{\lambda f} (\hat{\mathbf{x}}_{it}^{\lambda f})^{\prime} \, , \\
    \widehat{\boldsymbol{\Omega}} \coloneqq& \, \frac{1}{n} \sum_{(i, t) \in \mathcal{D}} \hat{e}_{it}^{2} \, \hat{\mathbf{x}}_{it}^{\lambda f} (\hat{\mathbf{x}}_{it}^{\lambda f})^{\prime} \, ,\\
    \widehat{\mathbf{B}}_{1} \coloneqq& \, \frac{1}{N} \sum_{j = 1}^{L} \sum_{t = j + 1}^{T} \sum_{i \in \mathcal{I}_{t} \cap \mathcal{I}_{t - j}} \bigg(\frac{\lvert\mathcal{T}_{i}\rvert}{\lvert\mathcal{T}_{i}\rvert - j}\bigg) \, \hat{p}_{i(t - j)t}^{f} \hat{\mathbf{x}}_{it}^{f} \hat{e}_{i(t - j)} \, , \\
    \widehat{\mathbf{B}}_{2} \coloneqq& \, \frac{1}{T} \sum_{i = 1}^{N} \bigg(\sum_{t \in \mathcal{T}_{i}} \hat{e}_{it}^{2} \bigg) \bigg(\sum_{t \in \mathcal{T}_{i}} \hat{\mathbf{x}}_{it}^{\lambda} \hat{\xi}_{it}^{\dagger} \bigg) \, , \\
    \widehat{\mathbf{B}}_{3} \coloneqq& \, \frac{1}{N} \sum_{t = 1}^{T} \bigg(\sum_{i \in \mathcal{I}_{t}} \hat{e}_{it}^{2} \bigg) \bigg(\sum_{i \in \mathcal{I}_{t}} \hat{\mathbf{x}}_{it}^{f} \hat{\xi}_{it}^{\dagger} \bigg) \, , 
\end{align*}
where $\hat{p}_{itt^{\prime}}^{f} \coloneqq \hat{\mathbf{f}}_{t}^{\prime} \widehat{\boldsymbol{\Phi}}_{i}^{- 1} \hat{\mathbf{f}}_{t^{\prime}}$, $\hat{\xi}_{it}^{\dagger} \coloneqq \hat{\boldsymbol{\lambda}}_{i}^{\prime} \widehat{\boldsymbol{\Psi}}_{t} \widehat{\boldsymbol{\Phi}}_{i} \hat{\mathbf{f}}_{t}^{\phantom{\prime}}$, $\widehat{\boldsymbol{\Phi}}_{i} \coloneqq \sum_{t \in \mathcal{T}_{i}} \hat{\mathbf{f}}_{t}^{\phantom{\prime}} \hat{\mathbf{f}}_{t}^{\prime}$, $\widehat{\boldsymbol{\Psi}}_{t} \coloneqq \sum_{i \in \mathcal{I}_{t}} \hat{\boldsymbol{\lambda}}_{i}^{\phantom{\prime}} \hat{\boldsymbol{\lambda}}_{i}^{\prime}$, $\widehat{\mathbf{A}}^{\cdot} \coloneqq (\hat{\mathbf{a}}_{1}^{\cdot}, \ldots, \hat{\mathbf{a}}_{T}^{\cdot})^{\prime}$, $\widehat{\mathbf{C}}^{\cdot} \coloneqq (\hat{\mathbf{c}}_{1}^{\cdot}, \ldots, \hat{\mathbf{c}}_{N}^{\cdot})^{\prime}$,
\begin{align}
    \hat{x}_{it, k}^{\lambda f} \coloneqq& \, x_{it, k} - \hat{\boldsymbol{\lambda}}_{i}^{\prime} \hat{\mathbf{a}}_{t}^{\ast\ast} - \hat{\mathbf{f}}_{t}^{\prime} \hat{\mathbf{c}}_{i}^{\ast\ast} \, , \quad (\widehat{\mathbf{A}}^{\ast\ast}, \widehat{\mathbf{C}}^{\ast\ast}) \in \, \underset{\mathbf{A} \in \mathbb{R}^{T \times R}, \, \mathbf{C} \in \mathbb{R}^{N \times R}}{\argmin} \sum_{(i, t) \in \mathcal{D}} (x_{it, k} - \hat{\boldsymbol{\lambda}}_{i}^{\prime} \mathbf{a}_{t}^{\phantom{\prime}} - \hat{\mathbf{f}}_{t}^{\prime} \mathbf{c}_{i}^{\phantom{\prime}})^{2} \, , \label{eq:residuals1} \\
    \hat{x}_{it, k}^{\lambda} \coloneqq& \, x_{it, k} - \hat{\boldsymbol{\lambda}}_{i}^{\prime} \hat{\mathbf{a}}_{t}^{\ast} \, , \quad \widehat{\mathbf{A}}^{\ast} \in \, \underset{\mathbf{A} \in \mathbb{R}^{T \times R}}{\argmin} \sum_{(i, t) \in \mathcal{D}} (x_{it, k} - \hat{\boldsymbol{\lambda}}_{i}^{\prime} \mathbf{a}_{t}^{\phantom{\prime}})^{2} \, , \label{eq:residuals2}  \\
    \hat{x}_{it, k}^{f} \coloneqq& \, x_{it, k} - \hat{\mathbf{f}}_{t}^{\prime} \hat{\mathbf{c}}_{i}^{\ast} \, , \quad \widehat{\mathbf{C}}^{\ast} \in \, \underset{\mathbf{C} \in \mathbb{R}^{N \times R}}{\argmin} \sum_{(i, t) \in \mathcal{D}} (x_{it, k} - \hat{\mathbf{f}}_{t}^{\prime} \mathbf{c}_{i}^{\phantom{\prime}})^{2} \, . \label{eq:residuals3}
\end{align}
A debiased estimator for $\boldsymbol{\beta}$ is then constructed as
\begin{equation}
    \label{eq:debiased_estimator}
    \tilde{\boldsymbol{\beta}} \coloneqq \hat{\boldsymbol{\beta}} + \frac{N}{n} \, \widehat{\mathbf{W}}^{- 1} \widehat{\mathbf{B}}_{1} + \frac{T}{n} \, \widehat{\mathbf{W}}^{- 1} \widehat{\mathbf{B}}_{2} + \frac{N}{n} \, \widehat{\mathbf{W}}^{- 1} \widehat{\mathbf{B}}_{3} \, ,
\end{equation}
such that
\begin{equation}
    \label{eq:asympotic_distribution_debiased}
    \sqrt{n} (\tilde{\boldsymbol{\beta}} - \boldsymbol{\beta}) \overset{d}{\rightarrow} \N(0, \mathbf{W}^{- 1} \boldsymbol{\Omega} \mathbf{W}^{- 1}) \, .
\end{equation}

The factor $\lvert\mathcal{T}_{i}\rvert / (\lvert\mathcal{T}_{i}\rvert - j)$ in $\widehat{\mathbf{B}}_{1}$ is a finite-sample adjustment proposed by \textcite{fw2016}. Following \textcite{fw2018}, we use $\sqrt{n}$ rather than $\sqrt{NT}$ as the normalizing factor in \eqref{eq:asympotic_distribution_debiased} to improve finite-sample approximation. The uncorrected estimator $\hat{\boldsymbol{\beta}}$ is obtained using the algorithms of Section \ref{sec:estimation}.

To construct $\tilde{\boldsymbol{\beta}}$, we require a computationally feasible method for the residuals defined in \eqref{eq:residuals1}, \eqref{eq:residuals2}, and \eqref{eq:residuals3}. Consider an arbitrary $n$-dimensional vector $\mathbf{v}$. The minimization over $\mathbf{A}$ in \eqref{eq:residuals2},
\begin{equation*}
    \widehat{\mathbf{A}}^{\ast} \in \, \underset{\mathbf{A} \in \mathbb{R}^{T \times R}}{\argmin} \sum_{(i, t) \in \mathcal{D}} (v_{it} - \hat{\boldsymbol{\lambda}}_{i}^{\prime} \mathbf{a}_{t}^{\phantom{\prime}})^{2} \, ,
\end{equation*}
is separable across $t$. For each $t \in \{1, \ldots, T\}$, the solution reduces to a cross-sectional regression:
\begin{equation}
    \label{eq:ls_solution2}
    \hat{\mathbf{a}}_{t}^{\ast} = \Big(\sum_{i \in \mathcal{I}_{t}} \hat{\boldsymbol{\lambda}}_{i}^{\phantom{\prime}} \hat{\boldsymbol{\lambda}}_{i}^{\prime}\Big)^{- 1} \sum_{i \in \mathcal{I}_{t}} \hat{\boldsymbol{\lambda}}_{i} v_{it} = \widehat{\Psi}_{t}^{- 1} \sum_{i \in \mathcal{I}_{t}} \hat{\boldsymbol{\lambda}}_{i} v_{it} \, .
\end{equation}
The corresponding residuals are
\begin{equation}
    \label{eq:residuals2_closedform}
    \hat{v}_{it}^{\lambda} = v_{it} - \hat{\boldsymbol{\lambda}}_{i}^{\prime} \widehat{\Psi}_{t}^{- 1} \sum_{i^{\prime} \in \mathcal{I}_{t}} \hat{\boldsymbol{\lambda}}_{i^{\prime}} v_{i^{\prime}t} \, .
\end{equation}
The same argument applies to the minimization problem in \eqref{eq:residuals3}.\footnote{The separability of both minimization problems is also exploited in the two-step least squares debiasing procedure of \textcites{chlz2019}{chlz2023} (see Steps 3 and 4 of Algorithm \ref{alg:mc_redebias}).} For each $i \in \{1, \ldots, N\}$, the solution reduces to a time-series regression, with residuals
\begin{equation}
    \label{eq:residuals3_closedform}
    \hat{v}_{it}^{f} = v_{it} - \hat{\mathbf{f}}_{t}^{\prime} \widehat{\Phi}_{i}^{- 1} \sum_{t^{\prime} \in \mathcal{T}_{i}} \hat{\mathbf{f}}_{t^{\prime}} v_{it^{\prime}} \, .
\end{equation}
For the minimization problem in \eqref{eq:residuals1}, no closed-form expressions analogous to \eqref{eq:residuals2_closedform} and \eqref{eq:residuals3_closedform} are available.

We propose a novel algorithm based on the \textit{Method of Alternating Projections} (MAP, see \cites{vn1949}{vn1950}{h1962}) as a computationally feasible method to compute the residuals \eqref{eq:residuals1}.\footnote{Our algorithm adapts \textcite{s2018}, who introduced MAP as a powerful tool for demeaning variables in the optimization of fixed effects estimators for nonlinear models with multi-way fixed effects, such as binary choice models with individual and time effects. \textcite{s2020} (Chapter 3) and \textcite{cs2019} noted the usefulness of this approach for unbalanced panel data. MAP is particularly appealing because it computes residuals from complex regressions (including unbalanced, weighted, and multi-way fixed effects) by alternating between one-way fixed effects demeaning steps.} Let $\hat{\mathbf{v}}^{\lambda}$ be the $n$-dimensional vector with entries \eqref{eq:residuals2_closedform}, and $\hat{\mathbf{v}}^{f}$ the $n$-dimensional vector with entries \eqref{eq:residuals3_closedform}. We define two orthogonal projection operators, $\mathcal{M}_{\hat{\lambda}}(\mathbf{v})$ and $\mathcal{M}_{\hat{f}}(\mathbf{v})$, such that $\mathcal{M}_{\hat{\lambda}}(\mathbf{v}) = \hat{\mathbf{v}}^{\lambda}$ and $\mathcal{M}_{\hat{f}}(\mathbf{v}) = \hat{\mathbf{v}}^{f}$.
\begin{algorithm} \label{alg:map}
	MAP algorithm

    \noindent Given $\mathbf{v}$, $\widehat{\boldsymbol{\Lambda}}$, and $\widehat{\mathbf{F}}$, where $\mathbf{v}$ is an $n$-dimensional vector with entries $v_{it}$. Initialize $\hat{\mathbf{v}}^{\lambda f} = \mathbf{v}$ and repeat the following steps until convergence.
	\begin{description}
		\item[Step 1.] Update $\hat{\mathbf{v}}^{\lambda f} = \mathcal{M}_{\hat{\lambda}}(\hat{\mathbf{v}}^{\lambda f})$.
		\item[Step 2.] Update $\hat{\mathbf{v}}^{\lambda f} = \mathcal{M}_{\hat{f}}(\hat{\mathbf{v}}^{\lambda f})$.
	\end{description}
    Return $\hat{\mathbf{v}}^{\lambda f}$ after convergence.
\end{algorithm}

Algorithm \ref{alg:map} iterates between orthogonal projections onto two closed subspaces and converges strongly to the projection $\hat{\mathbf{v}}^{\lambda f}$, i.e., to the residuals \eqref{eq:residuals1}, as established by \textcites{vn1949}{vn1950}. The linear rate of convergence was first proved by \textcite{a1950}. Acceleration techniques are discussed in \textcite{er2011}, among others.

We now summarize how the components of this section are combined to conduct inference on $\boldsymbol{\beta}$ using the debiased interactive fixed effects estimator for unbalanced panels.
\begin{algorithm} \label{alg:inference}
    Inference on $\boldsymbol{\beta}$ using $\tilde{\boldsymbol{\beta}}$ for unbalanced panels
    
    \noindent Given $\mathbf{y}$, $\mathbf{X}$, $\mathcal{D}$, and $R$, where $\mathbf{y}$ is an $n$-dimensional vector with elements $y_{it}$ and $\mathbf{X}$ is an $n \times K$ matrix with rows $\mathbf{x}_{it}$. Conduct the following steps.
    \begin{description}
		\item[Step 1.] Use $\boldsymbol{\beta}^{\star}$, defined in \eqref{eq:nnm_estimator}, as the starting guess for the subsequent steps.
        \item[Step 2.] Choose a matrix completion procedure (Algorithm \ref{alg:mc_em} or Algorithm \ref{alg:mc_redebias}). If using Algorithm \ref{alg:mc_redebias}, select some $\nu > 0$. Selection strategies for $\nu$ are discussed in Remark 3.
		\item[Step 3.] Obtain $\hat{\boldsymbol{\beta}}$ by minimizing \eqref{eq:ife_objfunction2} using $\boldsymbol{\beta}^{\star}$ as the starting guess. The completed matrix $\boldsymbol{\Gamma}^{\ast}(\boldsymbol{\beta})$ in \eqref{eq:ife_objfunction2} is obtained using the procedure chosen in Step 2.
        \item[Step 4.] Obtain $\widehat{\boldsymbol{\Lambda}}$ and $\widehat{\mathbf{F}}$ by decomposing $\widehat{\boldsymbol{\Gamma}}^{\ast} = \boldsymbol{\Gamma}^{\ast}(\hat{\boldsymbol{\beta}})$, where $\widehat{\boldsymbol{\Gamma}}^{\ast}$ is obtained using the procedure chosen in Step 2. $\widehat{\mathbf{F}}$ equals the first $R$ eigenvectors of $\widehat{\boldsymbol{\Gamma}}^{\ast\prime} \widehat{\boldsymbol{\Gamma}}^{\ast}$ multiplied by $\sqrt{T}$, and $\widehat{\boldsymbol{\Lambda}} = \widehat{\boldsymbol{\Gamma}}^{\ast} \widehat{\mathbf{F}} / T$.
        \item[Step 5.] Obtain $\widehat{\mathbf{W}}$, $\widehat{\boldsymbol{\Omega}}$, $\widehat{\mathbf{B}}_{1}$, $\widehat{\mathbf{B}}_{2}$, and $\widehat{\mathbf{B}}_{3}$ to construct $\tilde{\boldsymbol{\beta}}$, defined in \eqref{eq:debiased_estimator}, and the corresponding covariance matrix $\widehat{\boldsymbol{V}} \coloneqq \widehat{\mathbf{W}}^{- 1} \widehat{\boldsymbol{\Omega}} \, \widehat{\mathbf{W}}^{- 1}$. The residuals \eqref{eq:residuals2} and \eqref{eq:residuals3} are obtained via \eqref{eq:residuals2_closedform} and \eqref{eq:residuals3_closedform}, respectively. The residuals \eqref{eq:residuals1} are obtained using Algorithm \ref{alg:map}.
        \item[Step 6.] Use $\tilde{\boldsymbol{\beta}}$ and $\widehat{\mathbf{V}}$ to construct a debiased test statistic, such as a debiased Wald statistic for testing linear restrictions on $\boldsymbol{\beta}$.
	\end{description}
\end{algorithm}

If certain bias terms are not required, for example, when all regressors are strictly exogenous (so that $\mathbf{B}_{1} = \mathbf{0}_{K}$), the corresponding estimates, $\widehat{\mathbf{B}}_{1}$ in the example, can be omitted from Step 5. Since choosing an appropriate bandwidth $L$ for $\widehat{\mathbf{B}}_{1}$ is non-trivial, \textcites{fw2016}{fw2018} recommend a sensitivity analysis reporting estimates across different values of $L$.

Algorithm \ref{alg:inference} applies to balanced panels as well, by removing Step 2 and replacing the completed matrix $\boldsymbol{\Gamma}^{\ast}(\boldsymbol{\beta})$ with $\boldsymbol{\Gamma}(\boldsymbol{\beta})$.

\section{Estimating the Number of Factors}
\label{sec:nof}

\textcites{b2009}{mw2017} derived their results under the assumption that the number of factors is known. In practice, this assumption is often very unlikely unless economic theory provides a clear prediction about the number of factors. Even in that case, it may be necessary to support the theoretical prediction with additional empirical evidence. We therefore need a reliable method to estimate the number of factors. We denote the true number of factors by $R^{0}$.

For pure factor models, i.e., \eqref{eq:ifemodel} without additional regressors, there is an extensive literature on estimating the number of factors (see, among others, \cite{be1992}, \cite{bn2002}, \cite{hl2007}, \cite{abc2010}, \cite{o2010}, \cite{ah2013}, and \cite{do2019}). As pointed out by \textcite{b2009},
\begin{equation*}
	y_{it} - \mathbf{x}_{it}^{\prime} \hat{\boldsymbol{\beta}} = \boldsymbol{\lambda}_{i}^{\prime} \mathbf{f}_{t}^{\phantom{\prime}} + e_{it} - \mathbf{x}_{it} (\hat{\boldsymbol{\beta}} - \boldsymbol{\beta})
\end{equation*}
is essentially a pure factor model. Thus, given an estimator for $\boldsymbol{\beta}$ such that the estimation error $\mathbf{x}_{it} (\hat{\boldsymbol{\beta}} - \boldsymbol{\beta})$ is asymptotically negligible, the number of factors can be estimated consistently using methods developed for pure factor models (see {\citereset\cite{b2009}} Remark 5 and the corresponding appendix). Since \textcite{mw2015} show that the interactive fixed effects estimator is at least $\sqrt{\min(N, T)}$-consistent for any $R \geq R^{0}$, the initial estimate of $\boldsymbol{\beta}$ should be based on a sufficiently large value of $R$.

We consider the estimators of \textcites{bn2002}{o2010}{ah2013}{do2019}. Specifically, we apply them to $\widehat{\boldsymbol{\Gamma}}$, where $\boldsymbol{\beta}$ is estimated using $R = \overline{R}$ and $\overline{R}$ is a known upper bound on the number of factors. \textcite{bn2002} proposes model selection criteria that minimize the sum of squared residuals plus a penalty for the number of estimated parameters. \textcites{o2010}{ah2013}{do2019} segment the eigenvalue spectrum of the sample covariance of $\widehat{\boldsymbol{\Gamma}}$ to identify a cut-off between the common factors and the noise from the idiosyncratic error term. \textcite{o2010} proposes the edge distribution estimator (ED), based on differences of consecutive eigenvalues. \textcite{ah2013} proposes using ratios (ER) and growth rates (GR) instead of differences. \textcite{be1992} proposes a specific version of parallel analysis (PA), which compares eigenvalues to those obtained from independent data to identify a cut-off between common factors and noise. Independent data are constructed by permuting each column of $\widehat{\boldsymbol{\Gamma}}$, which preserves the marginal variances while destroying the correlation pattern induced by the common factors. Theoretical justification for PA was recently provided by \textcite{d2020}.

For unbalanced panels, we follow \textcite{jms2021} and apply the estimators to $\mathcal{P}_{\mathcal{D}}^{\phantom{\perp}}(\widehat{\boldsymbol{\Gamma}}^{\ast}) / (1 - \psi) = \mathcal{P}_{\mathcal{D}}^{\phantom{\perp}}(\boldsymbol{\Gamma}(\hat{\boldsymbol{\beta}})) / (1 - \psi)$ rather than $\widehat{\boldsymbol{\Gamma}}^{\ast}$, where $\psi \coloneqq 1 - n / (NT)$ is the share of missing observations.

\section{Simulation Experiments}
\label{sec:simulation}

We use Monte Carlo simulations to analyze the finite-sample properties of the debiased estimator $\tilde{\beta}$, defined in \eqref{eq:debiased_estimator}, in the presence of missing data. Specifically, we compare relative biases (\textit{Bias}), average ratios of standard errors to standard deviations (\textit{Ratio}), and empirical sizes of $z$-tests with a 5\% nominal size (\textit{Size}) across different shares of missing data ($\psi$) and relative to the balanced panel case. We use Algorithm \ref{alg:mc_em} as matrix completion procedure for unbalanced panels. Because the number of factors is typically unknown, we also compare different estimators for the number of factors. Specifically, we consider the estimators of \textcites{bn2002}{o2010}{ah2013}{do2019}. Of the information criteria introduced by \textcite{bn2002}, we focus on $\text{IC}_{2}$ and $\text{BIC}_{3}$, which are also used in \textcite{o2010} and \textcite{ah2013}. Performance is assessed by comparing the average estimated number of factors.

We follow \textcite{mw2017} and consider an AR(1) model with $R = 1$ factor,
\begin{equation*}
    y_{it} =  \beta \, y_{i(t - 1)} + \lambda_{i} f_{t} + e_{it} \, .
\end{equation*}
The idiosyncratic error term $e_{it}$ is homoskedastic with fat tails. Specifically, $e_{it}$ is drawn independently and identically from the $t$-distribution with five degrees of freedom. The factor structure is constructed from $\lambda_{i} \sim \iid \N(1, 1)$ and $f_{t} = \rho \, f_{t-1} + u_{t}$, where $u_{t} \sim \iid \N(0, (1 - \rho^2) \sigma^2)$ and $\rho = \sigma = 0.5$. We discard the first $1{,}000$ time periods to ensure that the simulated data are drawn from the stationary distribution of the model. All random variables are redrawn in each replication, and all results are based on $1{,}000$ replications.

We consider three shares of missing data, $\psi \in \{0, 0.2, 0.4\}$, with $\psi = 0$ corresponding to a balanced panel. The total sample size satisfies $n = NT(1 - \psi)$. As implied by the results in Section \ref{sec:model_consistency}, the biases shrink with $\overline{N}$ and $\overline{T}$. To ensure comparability across values of $\psi$, we therefore select $N$ and $T$ so that both $\overline{N}$ and $\overline{T}$ remain constant, setting $N = \overline{N} / (1 - \psi)$ and $T = \overline{T} / (1 - \psi)$. We consider panels with $\overline{N} = 100$ and $\overline{T} \in \{5, 10, 20, 40, 80\}$, and AR(1) models with $\beta = 0.3$ and $\beta = 0.9$.

\begin{figure}[!htbp]
	\centering
	\caption{Missing Data Pattern for Different $\psi$ -- $\overline{N} = 100$ and $\overline{T} = 20$}
	\includegraphics[width=.9\textwidth]{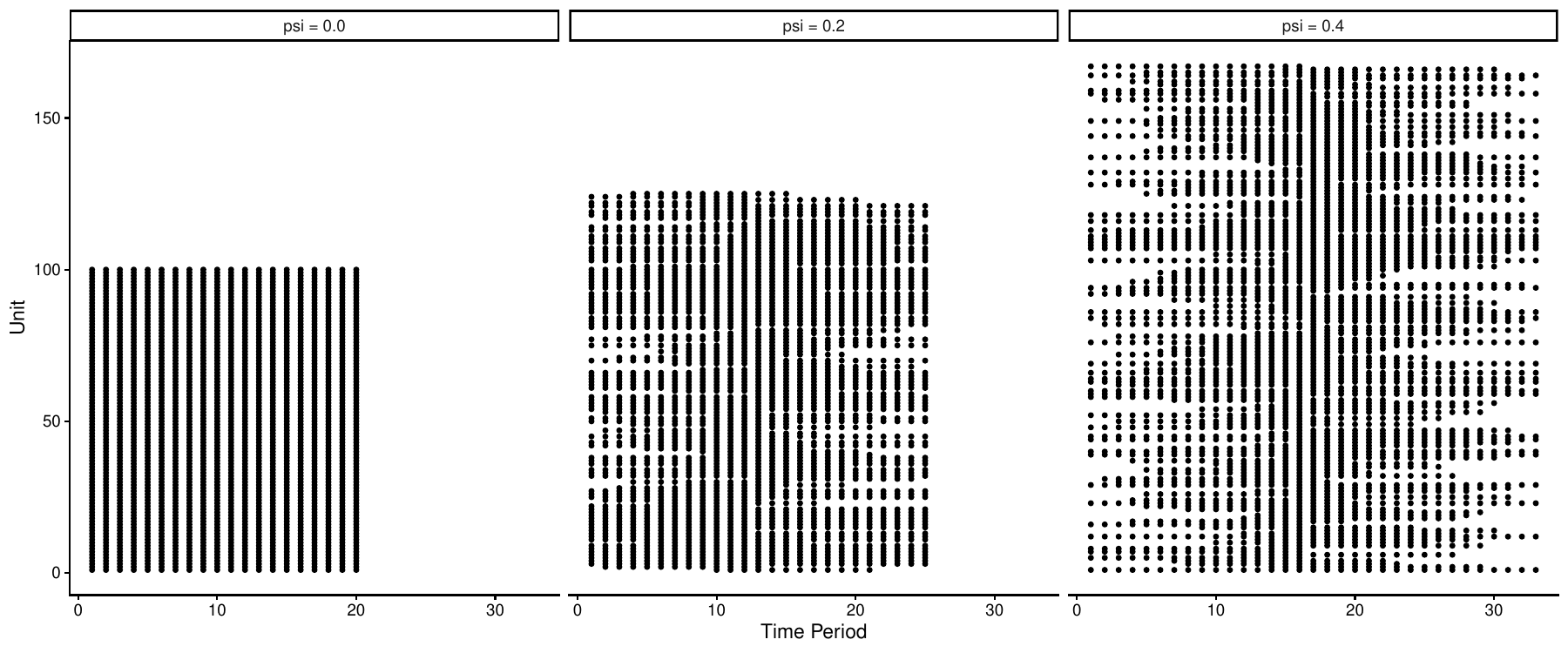}
	\label{fig:missing}
\end{figure}

Figure \ref{fig:missing} illustrates the missing data pattern for $\overline{N} = 100$ and $\overline{T} = 20$ across different values of $\psi$.
The pattern is taken from \textcite{cs2019}. All units are divided into two types. \textit{Type 1} consists of $N_{1} = 2 \psi N$ units observed over $T_{1} = T / 2$ consecutive time periods. The remaining $N_{2} = N - N_{1}$ units are \textit{Type 2} and are observed over the entire time horizon, i.e., $T_{2} = T$. The initial period is drawn uniformly at random from $\{0, 1, \ldots, T - T_{1}\}$. All unbalanced data sets are generated from initially balanced panels. Whether unit $i$ is \textit{Type 1} or \textit{Type 2} is determined by the value of $\lambda_{i}$: units with the lowest values of $\lambda_{i}$ are assigned to \textit{Type 1}. Observations are therefore not missing completely at random but are conditionally missing at random. Note also that the missing probabilities are homogeneous across $i$ but heterogeneous across $t$: they are lowest for time periods at the beginning or end of the time series and highest for time periods near $T / 2$.

Table \ref{tab:finite_sample_properties} presents the simulation results for $\tilde{\beta}$.
\begin{table}[!htbp]
	\centering
	\begin{threeparttable}
		\caption{Finite Sample Properties of $\tilde{\beta}$}
		\label{tab:finite_sample_properties}
		\begin{tabular}{@{}*{2}{l}*{3}{c}@{}}
		      \toprule
 $\overline{T}$&$L$&\multicolumn{3}{c}{$\psi = 0.0 \; / \; \psi = 0.2 \; / \; \psi = 0.4$}\\
 \cmidrule(lr){3-5}
 &&Bias&Ratio&Size\\
 \midrule
&&\multicolumn{3}{c}{$\beta = 0.3$}\\
\cmidrule(lr){3-5}
  5 &   2 & -15.052 / -17.811 / -21.268 & 0.393 / 0.423 / 0.462 & 0.509 / 0.499 / 0.520 \\ 
   10 &   3 & -7.507 / -8.610 / -8.706 & 0.736 / 0.779 / 0.820 & 0.189 / 0.197 / 0.211 \\ 
   20 &   4 & -3.269 / -3.291 / -2.586 & 0.908 / 0.906 / 0.961 & 0.091 / 0.094 / 0.074 \\ 
   40 &   5 & -1.252 / -0.889 / -1.041 & 0.984 / 0.958 / 0.964 & 0.053 / 0.063 / 0.077 \\ 
   80 &   6 & -0.529 / -0.270 / -0.222 & 1.032 / 0.988 / 0.942 & 0.040 / 0.055 / 0.064 \\ 
   \midrule
&&\multicolumn{3}{c}{$\beta = 0.9$}\\
\cmidrule(lr){3-5}
  5 &   2 & -13.848 / -11.241 / -11.208 & 0.239 / 0.221 / 0.202 & 0.613 / 0.538 / 0.549 \\ 
   10 &   3 & -4.584 / -4.251 / -3.938 & 0.292 / 0.279 / 0.269 & 0.422 / 0.401 / 0.429 \\ 
   20 &   4 & -1.009 / -0.895 / -0.795 & 0.505 / 0.520 / 0.524 & 0.212 / 0.216 / 0.210 \\ 
   40 &   5 & -0.210 / -0.191 / -0.178 & 0.809 / 0.851 / 0.898 & 0.102 / 0.085 / 0.081 \\ 
   80 &   6 & -0.067 / -0.087 / -0.058 & 0.965 / 0.955 / 0.961 & 0.049 / 0.063 / 0.067 \\ 
   \bottomrule
		\end{tabular}
		\begin{tablenotes}
			\footnotesize
			\item \emph{Note:} $\overline{N} = 100$ and $L$ is a bandwidth parameter; $\psi$ denotes the share of missing observations; Bias refers to relative biases in percentage, Ratio denotes the average ratios of standard errors to standard deviations, and Size is the empirical size of $z$-tests with 5\% nominal size; results are based on $1{,}000$ replications.
		\end{tablenotes}
	\end{threeparttable}
\end{table}
Although $e_{it}$ is homoskedastic and the missing probabilities are homogeneous across $i$, we do not exploit this information. Instead, we apply the debiased estimator and its covariance matrix estimator exactly as described in Section \ref{sec:bc_inference}, correcting for all three bias terms and using a covariance estimator that is robust to arbitrary heteroskedasticity. This approach yields a more realistic assessment of finite-sample performance in practice, where the true data-generating process is unknown and heteroskedasticity-robust inference is standard. The bandwidth parameter $L$ is taken from Table 1 of \textcite{mw2017}. For both $\beta = 0.3$ and $\beta = 0.9$, the biases, ratios, and sizes are similar to those in the balanced case, regardless of the share of missing data. Overall, the finite-sample performance of $\tilde{\beta}$ in unbalanced panels is well predicted by our theory.

Table \ref{tab:number_of_factors} presents the simulation results for the various estimators of the number of factors, $\widehat{R}$.
\begin{table}[!htbp]
	\centering
	\begin{threeparttable}
		\caption{Average of $\widehat{R}$}
		\label{tab:number_of_factors}
		\begin{tabular}{@{}*{2}{l}*{3}{c}@{}}
			  \toprule
 $\overline{T}$&$\overline{R}$&\multicolumn{3}{c}{$\psi = 0.0 \; / \; \psi = 0.2 \; / \; \psi = 0.4$}\\
 \cmidrule(lr){3-5}
 &&$\text{IC}_{2}$&$\text{BIC}_{3}$&ER\\
 \midrule
&&\multicolumn{3}{c}{$\beta = 0.3$}\\
\cmidrule(lr){3-5}
  5 &   2 & 2.000 / 1.937 / 1.795 & 0.652 / 0.480 / 0.456 & 0.919 / 0.879 / 0.844 \\ 
   10 &   5 & 4.463 / 2.650 / 2.266 & 2.005 / 1.554 / 1.100 & 1.004 / 0.931 / 0.932 \\ 
   20 &  10 & 2.318 / 2.488 / 1.489 & 2.896 / 1.798 / 1.189 & 0.823 / 0.979 / 0.994 \\ 
   40 &  10 & 1.007 / 1.007 / 1.034 & 1.005 / 1.004 / 1.008 & 0.996 / 1.002 / 1.001 \\ 
   80 &  10 & 1.001 / 1.001 / 1.004 & 1.001 / 1.001 / 1.001 & 1.001 / 1.001 / 1.000 \\ 
   \midrule
&&\multicolumn{3}{c}{$\beta = 0.9$}\\
\cmidrule(lr){3-5}
  5 &   2 & 1.988 / 1.853 / 1.708 & 1.228 / 1.027 / 1.023 & 0.863 / 0.914 / 0.975 \\ 
   10 &   5 & 3.425 / 3.266 / 3.313 & 3.247 / 2.620 / 2.367 & 1.143 / 1.170 / 1.235 \\ 
   20 &  10 & 6.046 / 5.755 / 5.303 & 5.701 / 4.675 / 4.096 & 1.400 / 1.297 / 1.215 \\ 
   40 &  10 & 6.218 / 5.967 / 6.567 & 4.745 / 4.411 / 4.676 & 1.353 / 1.142 / 1.048 \\ 
   80 &  10 & 3.727 / 3.251 / 3.729 & 2.980 / 2.367 / 2.512 & 1.015 / 1.002 / 1.004 \\ 
   \midrule
   &&GR&ED&PA\\
 \midrule
&&\multicolumn{3}{c}{$\beta = 0.3$}\\
\cmidrule(lr){3-5}
  5 &   2 & 0.857 / 0.896 / 0.871 & 0.627 / 0.723 / 0.901 & 0.761 / 0.815 / 0.971 \\ 
   10 &   5 & 0.920 / 0.978 / 0.963 & 0.752 / 0.970 / 1.052 & 1.079 / 1.266 / 1.248 \\ 
   20 &  10 & 0.872 / 0.987 / 0.996 & 1.001 / 1.078 / 1.082 & 1.339 / 1.350 / 1.162 \\ 
   40 &  10 & 0.996 / 1.003 / 1.001 & 1.091 / 1.085 / 1.151 & 1.045 / 1.013 / 1.094 \\ 
   80 &  10 & 1.001 / 1.001 / 1.000 & 1.089 / 1.112 / 1.514 & 1.001 / 1.000 / 1.034 \\ 
   \midrule
&&\multicolumn{3}{c}{$\beta = 0.9$}\\
\cmidrule(lr){3-5}
  5 &   2 & 0.992 / 0.972 / 1.051 & 1.283 / 1.198 / 1.332 & 0.692 / 0.777 / 1.004 \\ 
   10 &   5 & 1.401 / 1.436 / 1.568 & 2.616 / 2.447 / 2.266 & 1.105 / 1.280 / 1.529 \\ 
   20 &  10 & 1.719 / 1.598 / 1.431 & 2.628 / 2.400 / 2.094 & 1.636 / 1.776 / 1.890 \\ 
   40 &  10 & 1.667 / 1.363 / 1.148 & 2.625 / 2.242 / 1.848 & 2.419 / 2.712 / 3.045 \\ 
   80 &  10 & 1.021 / 1.002 / 1.005 & 1.897 / 1.419 / 1.402 & 3.196 / 3.189 / 3.286 \\ 
   \bottomrule
	    \end{tabular}
	\begin{tablenotes}
	\footnotesize
	\item \emph{Note:} $\overline{N} = 100$; $\psi$ denotes the share of missing observations; $\text{IC}_2$ and $\text{BIC}_{3}$ denote the information criteria of \textcite{bn2002}, ER and GR are the estimators of \textcite{ah2013}, ED is the estimator of \textcite{o2010}, and PA is the parallel analysis described in \textcite{do2019}. The true number of factors is one. The initial estimator for $\beta$ uses $R = \overline{R}$ factors. Results are based on $1{,}000$ replications.
	\end{tablenotes}
	\end{threeparttable}
\end{table}
The initial estimator uses $R = \overline{R}$, with $\overline{R} = 2$ for $\overline{T} = 5$, $\overline{R} = 5$ for $\overline{T} = 10$, and $\overline{R} = 10$ for $\overline{T} \in \{20, 40, 80\}$.\footnote{Our choice of $\overline{R}$ differs from studies such as \textcites{bn2002}{o2010}{ah2013}, which hold $\overline{R}$ fixed regardless of the sample size.} For $\psi > 0$, we apply the estimators to $\mathcal{P}_{\mathcal{D}}^{\phantom{\perp}}(\boldsymbol{\Gamma}(\hat{\beta}_{\bar{R}})) / (1 - \psi)$ as suggested by \textcite{jms2021}, where $\hat{\beta}_{\bar{R}}$ denotes the initial estimator with $R = \overline{R}$. For ER and GR, we use the mock eigenvalue of \textcite{ah2013} to allow for the possibility of selecting zero factors. For sufficiently large $\overline{T}$, all estimators perform similarly across different shares of missing data. However, performance differs between $\beta = 0.3$ and $\beta = 0.9$. In the low-persistence setting with sufficiently large $\overline{T}$, all estimators recover the correct number of factors, $R = 1$, nearly perfectly. In the high-persistence setting, only ER and GR achieve good performance. The results also suggest that the estimation error in $\boldsymbol{\beta}$ is asymptotically negligible, and they support the conjecture of \textcite{mw2015} that their main results extend beyond the case of independent and identically normally distributed errors.

Tables 7 and 8 in the Online Supplement \ref{os:additional_results} report the simulation results for $\tilde{\beta}$ and $\widehat{R}$ using Algorithm \ref{alg:mc_redebias} in place of Algorithm \ref{alg:mc_em} as the matrix completion procedure. The tuning parameter $\nu$ is selected via the plug-in approach of \textcites{chlz2019}{chlz2023}. The results are virtually identical to those reported here. Additional simulation results for a static panel data model with one regressor, two factors, and other missing data patterns can be found in the Online Supplement \ref{os:additional_simulation}.

\section{Empirical Example}
\label{sec:empirical_example}

The effect of democracy on economic growth remains a highly debated topic among economists. \textcite{anrr2019} provide evidence that democratization has a substantial positive impact on GDP per capita. Using annual data from 175 countries observed between 1960 and 2010, their main findings suggest a long-run effect of about 20\%. The dataset they construct is well suited for our purposes: it is naturally unbalanced, spans a long time horizon, and contains several unobserved common shocks triggered by technological progress and financial crises.\footnote{The data are part of the \href{https://www.journals.uchicago.edu/doi/suppl/10.1086/700936}{replication package} provided by the authors.}

The sample consists of $6{,}934$ observations, of which $3{,}558$ are classified as democratic. Of the 175 countries, 88 transition between democracy and non-democracy or vice versa. Average GDP, measured in year-2000 dollars, is $8{,}150$ for democratic and $2{,}074$ for non-democratic countries. A total of 71 countries are observed over the entire time horizon; on average, the dataset covers 136 countries and 40 years. The fraction and pattern of missing data are comparable to the configuration of our simulation study with $\psi = 0.2$.

\begin{figure}[!htbp]
	\centering
	\caption{GDP per capita before and after Democratization}
	\includegraphics[width=.9\textwidth]{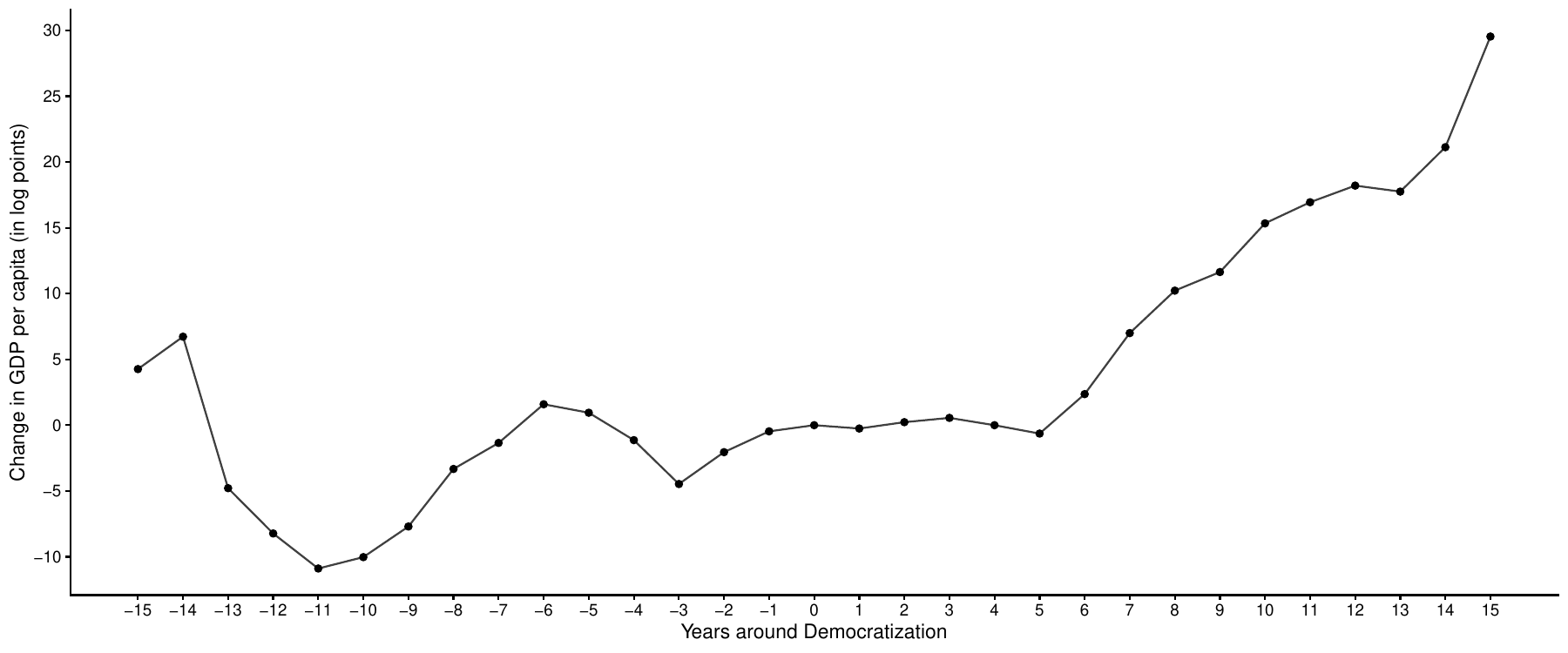}
	\begin{minipage}{.9\textwidth}
		\footnotesize
		\emph{Note:} Difference in average natural logarithm of GDP per capita between transitioning countries and those remaining non-democratic. The series is normalized to zero at the year of transition.
	\end{minipage}
	\label{fig:motivation}
\end{figure}

Figure \ref{fig:motivation} illustrates the evolution of GDP per capita around democratization for transitioning countries relative to persistently non-democratic ones.
The pre-transition dip and subsequent recovery suggest that the timing of democratization is endogenous to past GDP shocks. Failing to account for these dynamics by including sufficient lags of GDP could bias estimates, incorrectly attributing natural mean reversion following a crisis to the effect of democracy. To address this, \textcite{anrr2019} adopt the following dynamic panel specification:
\begin{equation*}
    y_{it} =  \theta \, D_{it} + \sum_{j = 1}^{p} \gamma_{j} \, y_{i(t - j)} + \alpha_{i} + \delta_{t} + u_{it} \, ,
\end{equation*}
where $y_{it}$ is the natural logarithm of GDP per capita of country $i$ at time $t$, $D_{it}$ is a democracy indicator, $\alpha_{i}$ and $\delta_{t}$ denote country and year fixed effects, and $u_{it}$ is an idiosyncratic error term. $\boldsymbol{\beta} = (\theta, \boldsymbol{\gamma}^{\prime})^{\prime}$ are the parameters of interest. This specification also allows us to distinguish between the short-run effect, $\theta$, and the long-run effect of democratization, $\phi(\boldsymbol{\beta}) \coloneqq \theta / (1 - \sum_{j = 1}^{p} \gamma_{j})$.

In contrast to \textcite{anrr2019}, we further decompose the error term into a factor structure $\boldsymbol{\lambda}_{i}^{\prime} \mathbf{f}_{t}^{\phantom{\prime}}$ and a residual idiosyncratic component $e_{it}$, i.e., $u_{it} = \boldsymbol{\lambda}_{i}^{\prime} \mathbf{f}_{t}^{\phantom{\prime}} + e_{it}$. This decomposition captures unobserved common shocks ($\mathbf{f}_{t}$) that simultaneously affect GDP growth and democratization in heterogeneous ways ($\boldsymbol{\lambda}_{i}$). Following \textcite{anrr2019}, we report results for $p \in \{1, 2, 4\}$, noting that $p = 4$ is the authors' preferred specification for modeling the GDP dynamics that follow a transition. As in the simulation study of Section \ref{sec:simulation}, we use Algorithm \ref{alg:mc_em} as the matrix completion procedure.

To reduce the number of parameters during optimization, we project out the country and time fixed effects before estimating $\boldsymbol{\beta}$:
\begin{equation*}
    \dot{y}_{it} = \theta \, \dot{D}_{it} + \sum_{j = 1}^{p} \gamma_{j} \, \dot{y}_{i(t - j)} + \dot{\boldsymbol{\lambda}}_{i}^{\prime} \dot{\mathbf{f}}_{t}^{\phantom{\prime}} + \dot{e}_{it} \, ,
\end{equation*}
where a dot denotes variables after projecting out country and time fixed effects. For example,
\begin{equation*}
    \dot{y}_{it} \coloneqq \, y_{it} - \hat{a}_{t} - \hat{c}_{i} \, , \quad (\hat{\mathbf{a}}, \hat{\mathbf{c}}) \in \, \underset{\mathbf{a} \in \mathbb{R}^{T}, \, \mathbf{c} \in \mathbb{R}^{N}}{\argmin} \sum_{(i, t) \in \mathcal{D}} (y_{it} - a_{t} - c_{i})^{2} \, .
\end{equation*}
The remaining residuals, $\dot{D}_{it}$, $\dot{y}_{i(t - 1)}$, $\dot{y}_{i(t - 2)}$, $\dot{y}_{i(t - 3)}$, and $\dot{y}_{i(t - 4)}$, are defined analogously. These residuals can be computed using the MAP algorithm (Algorithm \ref{alg:map}) presented in Section \ref{sec:bc_inference} with $R = 1$, $\hat{\lambda}_{i1} = 1$ for all $i \in \{1, \ldots, N\}$, and $\hat{f}_{t1} = 1$ for all $t \in \{1, \ldots, T\}$.\footnote{In this case, Algorithm \ref{alg:map} reduces to the algorithm proposed by \textcite{g2013_paper}, which is also used in popular fixed effects estimation software such as $\textit{lfe}$ \parencite{g2013_software} and \textit{reghdfe} \parencite{c2016}.}

For valid inference, the true number of factors must be known, or at least consistently overestimated. Since the true number is unknown, we proceed as follows. We estimate each specification with $R = 5$ to obtain $\mathcal{P}_{\mathcal{D}}^{\phantom{\perp}}(\boldsymbol{\Gamma}( \hat{\boldsymbol{\beta}})) / (1 - \psi)$, where $[\boldsymbol{\Gamma}( \hat{\boldsymbol{\beta}})]_{it} \coloneqq \hat{\theta} \, \dot{D}_{it} + \sum_{j = 1}^{p} \hat{\gamma}_{j} \, \dot{y}_{it - j}$. We then apply the estimators of \textcites{be1992}{bn2002}{o2010}{ah2013} to estimate the number of factors. For ER and GR, we use the mock eigenvalue of \textcite{ah2013} to accommodate the possibility of zero factors.

Table \ref{tab:emp_nof} summarizes the results.
\begin{table}[!htbp]
	\centering
	\begin{threeparttable}
		\caption{Estimated Number of Factors}
		\label{tab:emp_nof}
		\begin{tabular}{@{}*{1}{l}*{6}{c}@{}}
        \toprule
            Specification & $\text{IC}_{2}$ & $\text{BIC}_{3}$ & ER & GR & ED & PA \\ 
  \midrule
$p = 1$ &   5 &   2 &   1 &   1 &   2 &   3 \\ 
  $p = 2$ &   1 &   0 &   1 &   1 &   1 &   1 \\ 
  $p = 4$ &   1 &   0 &   0 &   0 &   1 &   1 \\ 
   \bottomrule
		\end{tabular}
		\begin{tablenotes}
			\footnotesize
			\item \emph{Note:} $\text{IC}_2$ and $\text{BIC}_{3}$ denote the information criteria of \textcite{bn2002}, ER and GR are the estimators of \textcite{ah2013}, ED is the estimator of \textcite{o2010}, and PA is the parallel analysis described in \textcite{do2019}. Estimators applied to $\mathcal{P}_{\mathcal{D}}^{\phantom{\perp}}(\boldsymbol{\Gamma}( \hat{\boldsymbol{\beta}})) / (1 - \psi)$. The initial estimator for $\boldsymbol{\beta}$ uses $R = 5$.
		\end{tablenotes}
	\end{threeparttable}
\end{table}
The estimates are nearly identical for $p > 1$, i.e., for specifications with flexible dynamics. For $p = 2$, most estimators select one common factor; for $p = 4$, half do so. The ER and GR results for $p = 4$ should be interpreted with caution, as they depend on the definition of the mock eigenvalue, which has multiple formulations (see {\citereset\cite{ah2013}}). Estimates for $p = 1$ vary substantially across estimators, ranging from one to five common factors, which may reflect insufficiently specified dynamics.

\begin{figure}[!htbp]
	\centering
	\caption{Largest Singular Values in Descending Order}
	\includegraphics[width=.9\textwidth]{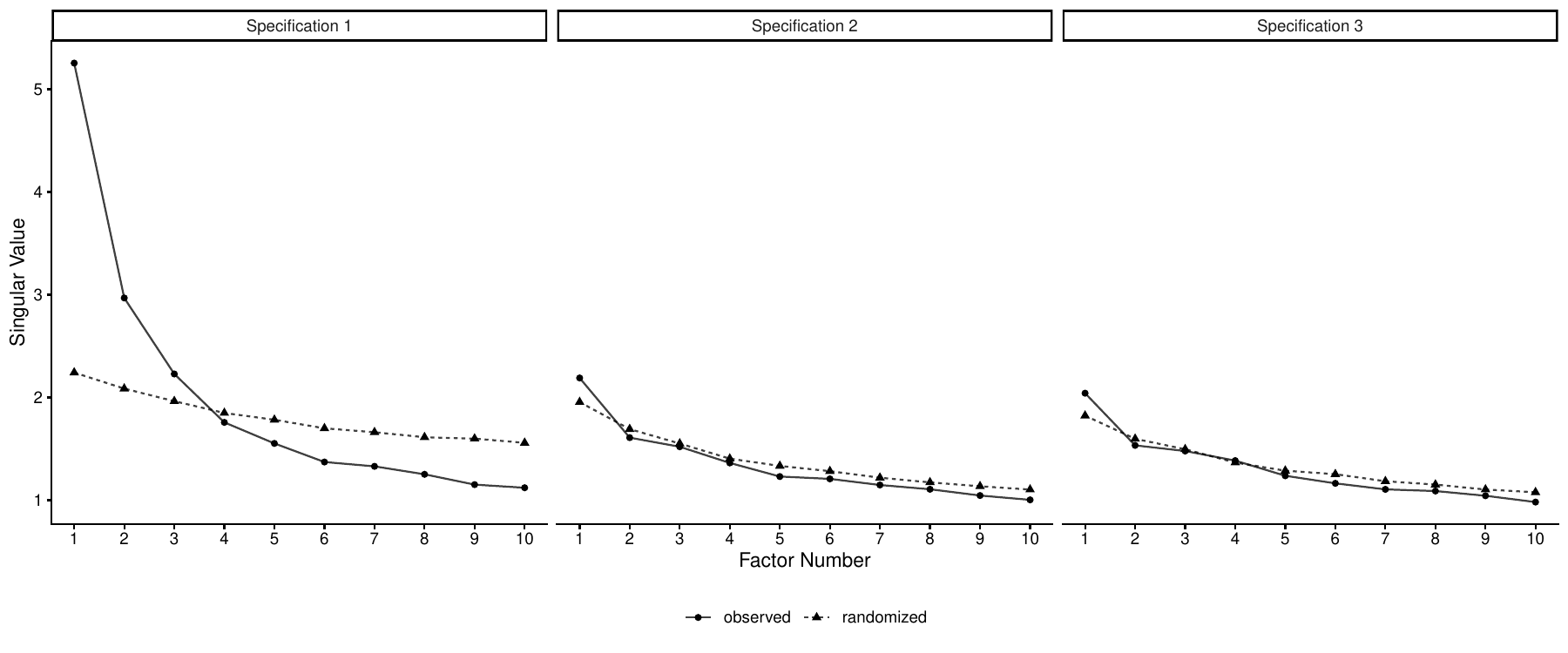}
	\begin{minipage}{.9\textwidth}
		\footnotesize
		\emph{Note:} Singular values for permuted data are based on 199 replications;  initial estimator for $\boldsymbol{\beta}$ uses $R = 5$.
	\end{minipage}
	\label{fig:nof}
\end{figure}

Figure \ref{fig:nof} displays the singular values of the pure factor models alongside those of the permuted versions.\footnote{More precisely, we randomly shuffle each column of $\mathcal{P}_{\mathcal{D}}^{\phantom{\perp}}(\boldsymbol{\Gamma}(\hat{\boldsymbol{\beta}}))$ and compute the maximum singular value across 199 randomized samples, multiplied by 1.05. This scaling factor is suggested by \textcite{do2019}, whose reasoning is that a factor only marginally exceeding what would be expected from pure noise should not be included in the model.}
We focus on the flexible dynamic specifications with $p > 1$. The gap between the first and second singular values explains why most estimators that decompose the eigenvalue spectrum select one common factor. Comparing the spectra with those of the permuted data, however, reveals that this common factor has explanatory power beyond what noise alone would generate, even if it accounts for only a small share of total variance. In light of the finding by \textcite{mw2015} that overestimating the number of factors is preferable to underestimating it, $R = 1$ is our preferred choice for $p > 1$.

Table \ref{tab:emp_coefs} summarizes our results.
\begin{table}[!htbp]
	\centering
	\begin{threeparttable}
		\caption{Effect of Democracy on Logarithmic GDP per Capita $(\times 100)$}
		\label{tab:emp_coefs}
		\begin{tabular}{@{}*{1}{l}*{6}{c}@{}}
			    \toprule
 &FE&AB&HHK&\multicolumn{3}{c}{IFE}\\
 \cmidrule(lr){5-7}
 &&&&$R = 1$&$R = 2$&$R = 3$\\
 \midrule
&\multicolumn{6}{c}{Specification 1 - $p = 1$}\\
\cmidrule(lr){2-7}
Democracy & 0.977 & 0.959 & 0.781 & 0.767 & 0.768 & 0.833 \\ 
   & (0.245) & (0.477) & (0.455) & (0.235) & (0.223) & (0.228) \\ 
  Persistence of & 0.980 & 0.946 & 0.938 & 0.960 & 0.973 & 0.968 \\ 
  \quad GDP process & (0.004) & (0.009) & (0.011) & (0.005) & (0.003) & (0.003) \\ 
  Long-run effect & 49.909 & 17.608 & 12.644 & 19.209 & 28.125 & 25.930 \\ 
  \quad of democracy & (19.761) & (10.609) & (8.282) & (6.991) & (9.233) & (8.035) \\ 
   \midrule
&\multicolumn{6}{c}{Specification 2 - $p = 2$}\\
\cmidrule(lr){2-7}
Democracy & 0.608 & 0.797 & 0.582 & 0.546 & 0.555 & 0.559 \\ 
   & (0.237) & (0.417) & (0.387) & (0.235) & (0.219) & (0.218) \\ 
  Persistence of & 0.973 & 0.946 & 0.941 & 0.956 & 0.968 & 0.967 \\ 
  \quad GDP process & (0.004) & (0.009) & (0.010) & (0.005) & (0.003) & (0.003) \\ 
  Long-run effect & 22.314 & 14.882 & 9.929 & 12.418 & 17.355 & 16.743 \\ 
  \quad of democracy & (10.459) & (9.152) & (7.258) & (5.979) & (7.306) & (6.956) \\ 
   \midrule
&\multicolumn{6}{c}{Specification 3 - $p = 4$}\\
\cmidrule(lr){2-7}
Democracy & 0.725 & 0.875 & 1.178 & 0.519 & 0.606 & 0.638 \\ 
   & (0.236) & (0.374) & (0.370) & (0.227) & (0.221) & (0.220) \\ 
  Persistence of & 0.967 & 0.947 & 0.953 & 0.958 & 0.964 & 0.966 \\ 
  \quad GDP process & (0.004) & (0.009) & (0.009) & (0.004) & (0.003) & (0.003) \\ 
  Long-run effect & 22.221 & 16.448 & 25.032 & 12.334 & 17.026 & 18.523 \\ 
  \quad of democracy & (8.708) & (8.436) & (10.581) & (5.780) & (6.626) & (6.853) \\ 
   \bottomrule
		\end{tabular}
		\begin{tablenotes}
			\footnotesize
			\item \emph{Note:} FE, AB, HHK, and IFE denote the debiased fixed effects estimator, the Arellano-Bond estimator, the Hahn-Hausman-Kuersteiner estimator, and the debiased IFE estimator. Standard errors in parentheses. FE and IFE use bandwidth $L = 5$ for the estimation of the asymptotic biases. The results of AB and HHK are taken from Table 2 in \textcite{anrr2019}.
		\end{tablenotes}
	\end{threeparttable}
\end{table}
Following \textcite{anrr2019}, we report results for the fixed effects estimator (FE), the Arellano-Bond estimator (AB, see \cite{ab1991}), and the Hahn-Hausman-Kuersteiner estimator (HHK, see \cite{hhk2004}). However, rather than the uncorrected FE estimator used by \textcite{anrr2019}, we report results from a debiased estimator with bandwidth $L = 5$ to correct for feedback bias.\footnote{Let $\hat{\boldsymbol{\beta}}_{\text{FE}}$ denote the uncorrected FE estimator and let $\dot{\mathbf{x}}_{it} \coloneqq (\dot{D}_{it}, \dot{y}_{i(t-1)}, \ldots)^{\prime}$. Then, the debiased FE estimator is constructed as
\begin{equation*}
    \tilde{\boldsymbol{\beta}}_{\text{FE}} \coloneqq \hat{\boldsymbol{\beta}}_{\text{FE}} + \bigg(\sum_{(i, t) \in \mathcal{D}} \dot{\mathbf{x}}_{it} (\dot{\mathbf{x}}_{it})^{\prime} \bigg)^{- 1} \, \bigg(\sum_{j = 1}^{L} \sum_{t = j + 1}^{T} \sum_{i \in \mathcal{I}_{t} \cap \mathcal{I}_{t - j}} (\lvert\mathcal{T}_{i}\rvert - j)^{- 1} \, \dot{\mathbf{x}}_{it} \hat{u}_{i(t - j)} \bigg) \, .
\end{equation*}
A similar estimator was also used by \textcite{ccf2019} for the same empirical illustration on a balanced subset of the data. Our debiased FE estimator follows the formulation of \textcite{fw2018}, augmented with the finite-sample adjustment described in Section 2.3. The key difference between \textcite{ccf2019} and \textcite{fw2018} is that the latter uses residualized regressors in place of raw regressors. We adopt the formulation of \textcite{fw2018} as it is more closely aligned with the bias expressions underlying our debiased IFE estimator.} In addition, we report results for three debiased interactive fixed effects estimators (IFE) with $R \in \{1, 2, 3\}$. We correct for both feedback bias and biases induced by heteroskedasticity, with bandwidth $L = 5$. We report estimates and standard errors for the short- and long-run effects of democratization and the persistence of GDP processes.

All estimators indicate strong and significant GDP persistence across all specifications. The democracy coefficients from FE and IFE are significant at the 5\% level throughout, whereas those from AB and HHK are significant only for $p = 4$. Focusing on the preferred specification $p = 4$, the estimators used by the authors imply short-run effects of democratization between 0.725\% and 1.178\%, and long-run effects between 16.448\% and 25.032\%. After controlling for additional time-varying unobserved heterogeneity, however, both effects are substantially smaller. Our preferred specification, IFE with $R = 1$, yields short- and long-run estimates of 0.519\% and 12.334\%, respectively.\footnote{Additional sensitivity checks are provided in Appendix \ref{app:empirical}. In particular, all IFE estimates are remarkably stable across bandwidth choices $L \in \{1, \ldots, 8\}$, and across different values of $R$, with the exception of $p = 1$.}

In summary, we find further support for the ``democracy does cause growth'' hypothesis of \textcite{anrr2019}. Controlling for time-varying unobserved heterogeneity via the interactive fixed effects estimator yields results that are qualitatively similar to those of the original authors. In the preferred specification $p = 4$, comparing HHK to IFE with $R = 1$ shows that both the short-run and long-run effects of democratization are roughly halved.

\section{Other Related Estimators}
\label{sec:other_estimators}

Although our analysis focuses on the interactive fixed effects (IFE) estimator of \textcite{b2009}, we briefly discuss three related estimators for which our findings and algorithms may also prove useful. First, in the presence of endogenous regressors, \textcites{mw2017}{msw2018} propose a minimum distance estimator in the spirit of \textcites{ch2006}{ch2008}. Second, because the IFE objective function is generally nonconvex, \textcite{mw2026} proposes an alternative estimator that replaces the potentially difficult nonconvex optimization problem with a convex one. Third, \textcite{cfw2021} propose an estimator for nonlinear parametric single-index models with interactive effects, such as logit, probit, ordered probit, and Poisson models.
\vspace{0.5em}

\noindent\underline{\textit{Minimum Distance Estimator.}} Suppose that $\mathbf{x}_{it}$ can be decomposed into $K_{1}$ endogenous and $K_{2}$ exogenous regressors, so that $K = K_{1} + K_{2}$. We use superscripts to distinguish between endogenous and exogenous regressors. Let $\mathbf{z}_{it} = (z_{1, it}, \ldots, z_{M, it})^{\prime}$ be a vector of excluded exogenous instruments, where $M \geq K_{1}$. \textcite{mw2017} suggest the following minimum distance estimator. In the first step, an estimator for $\boldsymbol{\beta}^{\text{end}}$ is obtained by
\begin{equation*}
	\hat{\boldsymbol{\beta}}^{\text{end}} \in \underset{\boldsymbol{\beta}^{\text{end}} \in \mathbb{R}^{K_{1}}}{\argmin} \hat{\boldsymbol{\pi}}(\boldsymbol{\beta}^{\text{end}})^{\prime} \, \boldsymbol{\Sigma} \, \hat{\boldsymbol{\pi}}(\boldsymbol{\beta}^{\text{end}}) \, ,
\end{equation*}
where $\hat{\boldsymbol{\pi}}(\boldsymbol{\beta}^{\text{end}})$ is the IFE estimator of 
\begin{equation*}
	y_{it} - \mathbf{x}_{it}^{\text{end} \, \prime} \boldsymbol{\beta}^{\text{end}}  = \mathbf{x}_{it}^{\text{exo} \, \prime} \boldsymbol{\beta}^{\text{exo}} + \mathbf{z}_{it}^{\prime} \boldsymbol{\pi} + \boldsymbol{\lambda}_{i}^{\prime} \mathbf{f}_{t}^{\phantom{\prime}} + e_{it}
\end{equation*}
and $\boldsymbol{\Sigma}$ is a positive definite $M \times M$ weighting matrix. At the true value of $\boldsymbol{\beta}^{\text{end}}$, the instrumental variable moment conditions imply $\boldsymbol{\pi} = \mathbf{0}_{M}$. In the second step, $\hat{\boldsymbol{\beta}}^{\text{exo}}$ is the IFE estimator of 
\begin{equation*}
	y_{it} - \mathbf{x}_{it}^{\text{end} \, \prime} \hat{\boldsymbol{\beta}}^{\text{end}}  = \mathbf{x}_{it}^{\text{exo} \, \prime} \boldsymbol{\beta}^{\text{exo}} + \boldsymbol{\lambda}_{i}^{\prime} \mathbf{f}_{t}^{\phantom{\prime}} + e_{it} \, .
\end{equation*}
	
The properties of the minimum distance estimator are studied in \textcite{msw2018}, where the authors extend the random coefficient demand model of \textcite{blp1995} to include interactive fixed effects, thereby accounting for unobserved product-market-specific heterogeneity, such as perceived utility from advertising at the product-market level. Under assumptions similar to those of \textcite{mw2017}, the authors establish consistency and derive the asymptotic distribution of the estimator. \textcite{lmw2012} apply the same estimator to address measurement error in the dependent variable of dynamic interactive fixed effects models.
\vspace{0.5em}

\noindent\underline{\textit{Nuclear Norm Regularized Estimator.}} \textcite{mw2026} show that the rank constraint on the factor structure renders the optimization problem nonconvex. They propose two alternative estimators based on a convex relaxation of this constraint. The nuclear norm minimizing estimator was already presented in \eqref{eq:nnm_estimator}; the second estimator uses nuclear norm regularization. \textcite{mw2026} establish consistency for both estimators, but only at the rate $\sqrt{\min(N, T)}$. To recover the properties of the IFE estimator, they suggest estimating the number of factors from $\mathcal{P}_{\mathcal{D}}^{\phantom{\perp}}(\boldsymbol{\Gamma}(\hat{\boldsymbol{\beta}}^{\star}))$ and then applying an iterative post-estimation routine. After a finite number of iterations, the estimator attains the same limiting distribution as the IFE estimator.
\begin{algorithm}
Post-Estimation Routine after Nuclear Norm Regularized Estimation
		
\noindent Given $\hat{\boldsymbol{\beta}}^{\star}$ and $R$. Choose a matrix completion procedure (Algorithm \ref{alg:mc_em} or Algorithm \ref{alg:mc_redebias}), initialize $\hat{\boldsymbol{\beta}} = \hat{\boldsymbol{\beta}}^{\star}$, and repeat the following steps a finite number of times (e.g., five times).
\begin{description}
    \item[Step 1.] If using Algorithm \ref{alg:mc_redebias}, select some $\nu > 0$. Selection strategies for $\nu$ are discussed in Remark 3.
    \item[Step 2.] Obtain $\widehat{\boldsymbol{\Lambda}}$ and $\widehat{\mathbf{F}}$ by decomposing $\widehat{\boldsymbol{\Gamma}}^{\ast} = \boldsymbol{\Gamma}^{\ast}(\hat{\boldsymbol{\beta}})$, where $\widehat{\boldsymbol{\Gamma}}^{\ast}$ is obtained using the procedure chosen in Step 1. $\widehat{\mathbf{F}}$ equals the first $R$ eigenvectors of $\widehat{\boldsymbol{\Gamma}}^{\ast\prime} \widehat{\boldsymbol{\Gamma}}^{\ast}$ multiplied by $\sqrt{T}$, and $\widehat{\boldsymbol{\Lambda}} = \widehat{\boldsymbol{\Gamma}}^{\ast} \widehat{\mathbf{F}} / T$.
    \item[Step 3.] Use Algorithm \ref{alg:map} to obtain $\hat{\mathbf{x}}_{k}^{\lambda f}$ for all $k \in \{1, \ldots, K\}$, where $\mathbf{x}_{k}$ is an $n$-dimensional vector with elements $x_{it, k}$.
	\item[Step 4.] Update $\hat{\boldsymbol{\beta}} = \big((\widehat{\mathbf{X}}^{\lambda f})^{\prime} \widehat{\mathbf{X}}^{\lambda f}\big)^{- 1} (\widehat{\mathbf{X}}^{\lambda f})^{\prime} \mathbf{y}$, where $\widehat{\mathbf{X}}^{\lambda f} = (\hat{\mathbf{x}}_{1}^{\lambda f}, \ldots, \hat{\mathbf{x}}_{K}^{\lambda f})$.
\end{description}
\end{algorithm}
\vspace{0.5em}

\noindent\underline{\textit{Estimator for Nonlinear Factor Models.}} Suppose the outcome variable is generated by
\begin{equation*}
    y_{it} \mid \mathbf{x}_{it}, \boldsymbol{\beta}, \boldsymbol{\Lambda}, \mathbf{F} \sim f(\cdot \mid \mathbf{x}_{it}^{\prime} \boldsymbol{\beta} + \boldsymbol{\lambda}_{i}^{\prime} \mathbf{f}_{t}^{\phantom{\prime}}) \, ,
\end{equation*}
where $f(\cdot)$ is a known density, such as the logistic density. To maximize the corresponding log-likelihood, \textcite{cfw2021} propose an EM-type optimization algorithm.

\begin{algorithm}
EM-type Log-Likelihood Maximization
		
\noindent Given $R$. Choose a matrix completion procedure (Algorithm \ref{alg:mc_em} or Algorithm \ref{alg:mc_redebias}), initialize $\hat{\boldsymbol{\beta}} = (\hat{\boldsymbol{\theta}}^{\prime}, \hat{\boldsymbol{\lambda}}_{1}^{\prime}, \ldots, \hat{\boldsymbol{\lambda}}_{N}^{\prime}, \hat{\mathbf{f}}_{1}^{\prime}, \ldots, \hat{\mathbf{f}}_{T}^{\prime})^{\prime}$, and repeat the following steps until convergence.
\begin{description}
    \item[Step 1.] Set $[\boldsymbol{\Gamma}(\hat{\boldsymbol{\beta}})]_{it} = r_{it}(\hat{\boldsymbol{\beta}})$, where $r_{it}(\hat{\boldsymbol{\beta}}) = z_{it}(\hat{\boldsymbol{\beta}}) - \partial_{z} l_{it}(z_{it}(\hat{\boldsymbol{\beta}})) / \partial_{z^{2}} l_{it}(z_{it}(\hat{\boldsymbol{\beta}}))$, $z_{it}(\hat{\boldsymbol{\beta}}) = \mathbf{x}_{it}^{\prime} \boldsymbol{\theta} + \boldsymbol{\lambda}_{i}^{\prime} \mathbf{f}_{t}^{\phantom{\prime}}$, and $\partial_{z^{j}} l_{it}(z_{it}(\hat{\boldsymbol{\beta}}))$ is the $j$-th partial derivative of the log-likelihood contribution $l_{it}$ with respect to the linear index $z_{it}(\hat{\boldsymbol{\beta}})$. $\boldsymbol{\Gamma}(\hat{\boldsymbol{\beta}})$ has missing entries corresponding to unobserved index pairs.
    \item[Step 2.] If using Algorithm \ref{alg:mc_redebias}, select some $\nu > 0$. Selection strategies for $\nu$ are discussed in Remark 3.
    \item[Step 3.] Obtain $\widehat{\boldsymbol{\Lambda}}$ and $\widehat{\mathbf{F}}$ by decomposing $\widehat{\boldsymbol{\Gamma}}^{\ast} = \boldsymbol{\Gamma}^{\ast}(\hat{\boldsymbol{\beta}})$, where $\widehat{\boldsymbol{\Gamma}}^{\ast}$ is obtained using the procedure chosen in the beginning. $\widehat{\mathbf{F}}$ equals the first $R$ eigenvectors of $\widehat{\boldsymbol{\Gamma}}^{\ast\prime} \widehat{\boldsymbol{\Gamma}}^{\ast}$ multiplied by $\sqrt{T}$, and $\widehat{\boldsymbol{\Lambda}} = \widehat{\boldsymbol{\Gamma}}^{\ast} \widehat{\mathbf{F}} / T$.
    \item[Step 4.] Use Algorithm \ref{alg:map} to obtain $\hat{\mathbf{x}}_{k}^{\lambda f}$ for all $k \in \{1, \ldots, K\}$, where $\mathbf{x}_{k}$ is an $n$-dimensional vector with elements $x_{it, k}$.
	\item[Step 5.] Update $\hat{\boldsymbol{\theta}} = \big((\widehat{\mathbf{X}}^{\lambda f})^{\prime} \widehat{\mathbf{X}}^{\lambda f}\big)^{- 1} (\widehat{\mathbf{X}}^{\lambda f})^{\prime} \mathbf{r}(\hat{\boldsymbol{\beta}})$, where $\widehat{\mathbf{X}}^{\lambda f} = (\hat{\mathbf{x}}_{1}^{\lambda f}, \ldots, \hat{\mathbf{x}}_{K}^{\lambda f})$.
\end{description}
\end{algorithm}

\section{Concluding Remarks}
\label{sec:conclusion}

The assumption that unobserved heterogeneity is constant over time is often too restrictive. In panels that span a long time horizon, such as macroeconomic country panels, it is implausible that a global shock affects all units equally. Interactive fixed effects estimators offer researchers a flexible way to accommodate this form of heterogeneity (see, among others, \cite{hnr1988}, \cite{p2006}, and \cite{b2009}). These panels are, however, often naturally unbalanced. Although \textcite{b2009} proposed an estimation algorithm for this case, the practical aspects of inference remained unclear. Drawing on insights from \textcite{fw2018} and extending \textcite{mw2017}, we derive the asymptotic distribution of \textcite{b2009}'s interactive fixed effects estimator for unbalanced panels, thereby establishing a foundation for inference in this practically relevant setting. We also develop a novel algorithm to compute the residualized variables required for estimating the bias terms and the covariance matrix.

Our findings and algorithms may further prove useful for related estimators, including the minimum distance estimator of \textcites{mw2017}{msw2018}, the nuclear norm estimator of \textcite{mw2026}, and the estimator for nonlinear factor models of \textcite{cfw2021}.

\clearpage
\printbibliography


\clearpage
\appendix


\section{Appendix}

\subsection{Derivation of Asymptotic Distribution}
\label{app:derivation_distribution}

We derive the results of Section \ref{sec:model_consistency} starting from the quadratic expansion for balanced panels in \textcite{mw2015}, which also underlies the asymptotic distribution in \textcite{mw2017} (see Theorem 4.1, Corollary 4.2, and Theorem 4.3).

The derivation proceeds in four parts. In \textit{Part 1}, we present the estimator and its asymptotic distribution for balanced panels. Because \textcites{mw2015}{mw2017} use $N \times T$ matrix notation for regressors and errors, we rewrite their results in summation form. In \textit{Part 2}, using the attrition indicator $\delta_{it} = \ind\{(i, t) \in \mathcal{D}\}$, we express \eqref{eq:ife_objfunction} as a profile objective for a balanced panel with outcomes and regressors premultiplied by $\delta_{it}$, yielding a balanced-panel representation of unbalanced data. In \textit{Part 3}, we show that, under Assumption \ref{ass:missing_data_stochastic} and additional high-level regularity conditions, we obtain the asymptotic distribution stated in Section \ref{sec:model_consistency}. In \textit{Part 4}, we verify these additional conditions using the assumptions of \textcite{mw2017} and Assumption \ref{ass:missing_data_stochastic}.

For clarity, we present the derivation for $K = 1$ (a single regressor); the extension to $K > 1$ is straightforward.

\vspace{0.5em}
\noindent\textit{Part 1.} The common parameter $\beta$ in balanced panels can be estimated as follows:
\begin{equation*}
    \hat{b} \coloneqq \underset{b \in \mathbb{R}}{\argmin} \mathbb{Q}\left(b\right) \, , \quad \mathbb{Q}\left(b\right) \coloneqq \underset{\boldsymbol{\Lambda}, \mathbf{F}}{\min} \; \frac{1}{NT} \sum_{i = 1}^{N} \sum_{t = 1}^{T} \left(y_{it} - x_{it} b - \boldsymbol{\lambda}_{i}^{\prime} \mathbf{f}_{t}^{\phantom{\prime}}\right)^{2} \, .
\end{equation*}
The least squares objective function reduces to the following eigenvalue problem:
\begin{equation*}
    \frac{1}{NT} \sum_{r = R + 1}^{T} \mu_{r} \big(\boldsymbol{\Gamma}(b)^{\prime} \boldsymbol{\Gamma}(b)\big) \, .
\end{equation*}
Because no closed-form expression for the derivatives of eigenvalues exists, standard Taylor expansion techniques are infeasible. Instead, \textcites{mw2015}{mw2017} obtain a quadratic expansion of the profile objective function using perturbation theory.

Let $\mathbb{U}$ be mean zero and
\begin{equation*}
    \text{Var}[\mathbb{U} \mid \mathcal{C}] = \EX[\mathbb{U}^{2} \mid \mathcal{C}] = \mathbb{V} + o_{P}(1) \, .
\end{equation*}
In addition, let $\theta_{it}^{\dagger} \coloneqq [\boldsymbol{\Theta}^{\dagger}]_{it} = \boldsymbol{\lambda}_{i}^{\prime} (\boldsymbol{\Lambda}^{\prime} \boldsymbol{\Lambda})^{- 1} (\mathbf{F}^{\prime} \mathbf{F})^{- 1} \mathbf{f}_{t}$, where $\boldsymbol{\Theta} \coloneqq \boldsymbol{\Lambda} \mathbf{F}^{\prime}$ is an $N \times T$ matrix of interactive effects. The quadratic approximation of \textcite{mw2015} then yields the following asymptotic expansion:
\begin{equation*}
    \sqrt{NT} \, \mathbb{W} (\hat{b} - \beta) + \kappa \, \mathbb{B}_{1} + \kappa^{- 1} \, \mathbb{B}_{2} + \kappa \, \mathbb{B}_{3} = \mathbb{U} + o_{P}(1) \, ,
\end{equation*}
where
\begin{align*} 
    &\mathbb{W} \coloneqq \frac{1}{NT} \sum_{i = 1}^{N} \sum_{t = 1}^{T} \big(r_{it}^{\lambda f}\big)^{2}\, , \quad \mathbb{V} \coloneqq \frac{1}{NT} \sum_{i = 1}^{N} \sum_{t = 1}^{T} \big(r_{it}^{\lambda f} e_{it}\big)^{2}  \, , \quad \mathbb{B}_{1} \coloneqq \frac{1}{N} \sum_{i = 1}^{N} \sum_{t = 1}^{T} p_{it}^{f} e_{it} \, , \\
    &\mathbb{B}_{2} \coloneqq \frac{1}{T} \sum_{i = 1}^{N} \bigg(\sum_{t = 1}^{T} \EX\left[e_{it}^{2} \mid \mathcal{C}\right] \bigg) \bigg(\sum_{t = 1}^{T} r_{it}^{\lambda} \theta_{it}^{\dagger} \bigg) \, , \\
    &\mathbb{B}_{3} \coloneqq \frac{1}{N} \sum_{t = 1}^{T} \bigg(\sum_{i = 1}^{N} \EX\left[e_{it}^{2} \mid \mathcal{C}\right] \bigg) \bigg(\sum_{i = 1}^{N} r_{it}^{f} \theta_{it}^{\dagger} \bigg) \, ,
\end{align*}
with 
\begin{align*}
    p_{it}^{\lambda f} \coloneqq& \, \boldsymbol{\lambda}_{i}^{\prime} \mathbf{a}_{t}^{\lambda f} + \mathbf{f}_{t}^{\prime} \mathbf{c}_{i}^{\lambda f} \, , \quad (\mathbf{A}^{\lambda f}, \mathbf{C}^{\lambda f}) \in \, \underset{\mathbf{A} \in \mathbb{R}^{T \times R}, \, \mathbf{C} \in \mathbb{R}^{N \times R}}{\argmin} \sum_{i = 1}^{N} \sum_{t = 1}^{T} (x_{it} - \boldsymbol{\lambda}_{i}^{\prime} \mathbf{a}_{t}^{\phantom{\prime}} - \mathbf{f}_{t}^{\prime} \mathbf{c}_{i}^{\phantom{\prime}})^{2} \, ,  \\
    p_{it}^{\lambda} \coloneqq& \, \boldsymbol{\lambda}_{i}^{\prime} \mathbf{a}_{t}^{\lambda} \, , \quad \mathbf{A}^{\lambda} \in \, \underset{\mathbf{A} \in \mathbb{R}^{T \times R}}{\argmin} \sum_{i = 1}^{N} \sum_{t = 1}^{T} (x_{it} - \boldsymbol{\lambda}_{i}^{\prime} \mathbf{a}_{t}^{\phantom{\prime}})^{2} \, ,  \\
    p_{it}^{f} \coloneqq& \, \mathbf{f}_{t}^{\prime} \mathbf{c}_{i}^{f} \, , \quad \mathbf{C}^{f} \in \, \underset{\mathbf{C} \in \mathbb{R}^{N \times R}}{\argmin} \sum_{i = 1}^{N} \sum_{t = 1}^{T} (x_{it} - \mathbf{f}_{t}^{\prime} \mathbf{c}_{i}^{\phantom{\prime}})^{2} \, , 
\end{align*}
$r_{it}^{\lambda f} \coloneqq x_{it} - p_{it}^{\lambda f}$, $r_{it}^{\lambda} \coloneqq x_{it} - p_{it}^{\lambda}$, and $r_{it}^{f} \coloneqq x_{it} - p_{it}^{f}$.

Two points are noteworthy. First, to compare our representation with \textcites{mw2015}{mw2017}, note that $\mathbf{r}^{\lambda f} = \MX_{\boldsymbol{\Lambda}} \mathbf{X} \MX_{\mathbf{F}}$, $\mathbf{r}^{\lambda} = \MX_{\boldsymbol{\Lambda}} \mathbf{X}$, and $\mathbf{r}^{f} = \mathbf{X} \MX_{\mathbf{F}}$, where $\MX_{\boldsymbol{\Lambda}} \coloneqq \eye_{N} - \boldsymbol{\Lambda} (\boldsymbol{\Lambda}^{\prime} \boldsymbol{\Lambda})^{\dagger} \boldsymbol{\Lambda}^{\prime}$, $\MX_{\mathbf{F}} \coloneqq \eye_{T} - \mathbf{F} (\mathbf{F}^{\prime} \mathbf{F})^{\dagger} \mathbf{F}^{\prime}$, $(\cdot)^{\dagger}$ is the Moore-Penrose inverse, and $\mathbf{X}$ is an $N \times T$ matrix with $[\mathbf{X}]_{it} = x_{it}$. Second, the expansion already exploits the fact that $x_{it} e_{it}$ and $e_{it}$ are uncorrelated across $i$ and $t$ for all $i, t, N, T$, conditional on $\mathcal{C}$. Under Assumption \ref{ass:missing_data_stochastic}, the same holds for $\delta_{it} x_{it} e_{it}$ and $\delta_{it} e_{it}$, which justifies the corresponding simplifications.

\vspace{0.5em}
\noindent\textit{Part 2.} Using $\delta_{it}^{q} = \delta_{it}$ for any $q > 0$, the profile objective function \eqref{eq:ife_objfunction} can be rewritten as
\begin{align*}
    Q\left(b\right) =& \, \underset{\boldsymbol{\Lambda}, \mathbf{F}}{\min} \; \frac{1}{NT} \sum_{(i, t) \in \mathcal{D}} \left(y_{it} - x_{it} b - \boldsymbol{\lambda}_{i}^{\prime} \mathbf{f}_{t}^{\phantom{\prime}}\right)^{2} \\
    =& \, \underset{\boldsymbol{\Lambda}, \mathbf{F}}{\min} \; \frac{1}{NT} \sum_{i = 1}^{N} \sum_{t = 1}^{T} \left(\delta_{it} y_{it} - \delta_{it} x_{it} b - \delta_{it} \boldsymbol{\lambda}_{i}^{\prime} \mathbf{f}_{t}^{\phantom{\prime}}\right)^{2} \, .
\end{align*}

The quadratic expansion can be interpreted as a projection-based approach for constructing the profile estimator of the common parameters. The second-order terms ($\mathbb{B}_{2}$ and $\mathbb{B}_{3}$) depend on the pseudo-inverse of the interactive effects. To apply this expansion to unbalanced panels, the projections and the pseudo-inverse must therefore be appropriately adapted. The adapted projections are
\begin{align*}
    &(\mathbf{A}^{\lambda f}, \mathbf{C}^{\lambda f}) \in \, \underset{\mathbf{A} \in \mathbb{R}^{T \times R}, \, \mathbf{C} \in \mathbb{R}^{N \times R}}{\argmin} \sum_{i = 1}^{N} \sum_{t = 1}^{T} (\delta_{it} x_{it} - \delta_{it} \boldsymbol{\lambda}_{i}^{\prime} \mathbf{a}_{t}^{\phantom{\prime}} - \delta_{it} \mathbf{f}_{t}^{\prime} \mathbf{c}_{i}^{\phantom{\prime}})^{2} \, ,  \\
    &\mathbf{A}^{\lambda} \in \, \underset{\mathbf{A} \in \mathbb{R}^{T \times R}}{\argmin} \sum_{i = 1}^{N} \sum_{t = 1}^{T} (\delta_{it} x_{it} - \delta_{it} \boldsymbol{\lambda}_{i}^{\prime} \mathbf{a}_{t}^{\phantom{\prime}})^{2} \, ,  \\
    &\mathbf{C}^{f} \in \, \underset{\mathbf{C} \in \mathbb{R}^{N \times R}}{\argmin} \sum_{i = 1}^{N} \sum_{t = 1}^{T} (\delta_{it} x_{it} - \delta_{it} \mathbf{f}_{t}^{\prime} \mathbf{c}_{i}^{\phantom{\prime}})^{2} \, ,
\end{align*}
and the adapted (approximate) pseudo-inverse is $\xi_{it}^{\dagger} \coloneqq \boldsymbol{\lambda}_{i}^{\prime} \boldsymbol{\Psi}_{t}^{- 1} \boldsymbol{\Phi}_{i}^{- 1} \mathbf{f}_{t}^{\phantom{\prime}}$, where $\boldsymbol{\Phi}_{i} \coloneqq \sum_{t = 1}^{T} \delta_{it} \, \mathbf{f}_{t}^{\phantom{\prime}} \mathbf{f}_{t}^{\prime}$ and $\boldsymbol{\Psi}_{t} \coloneqq \sum_{i = 1}^{N} \delta_{it} \, \boldsymbol{\lambda}_{i}^{\phantom{\prime}} \boldsymbol{\lambda}_{i}^{\prime}$.

We introduce additional notation. Let $\mathbf{D}_{\lambda} \coloneqq \boldsymbol{\Lambda} \otimes \eye_{T}$ and $\mathbf{D}_{f} \coloneqq \eye_{N} \otimes \, \mathbf{F}$ be $NT \times TR$ and $NT \times NR$ matrices, respectively, where $\otimes$ denotes the Kronecker product. Let $\mathbf{D}_{\lambda f} \coloneqq (\mathbf{D}_{\lambda}, \mathbf{D}_{f})$ and $\boldsymbol{\nabla} \coloneqq \diag(\delta_{11}, \ldots, \delta_{NT})$. The fitted values of the adapted projections can then be written as $\mathbf{p}^{\lambda f} = \boldsymbol{\nabla} \mathbb{L}^{\lambda f} \boldsymbol{\nabla} \mathbf{x}$, $\mathbf{p}^{\lambda} = \boldsymbol{\nabla} \mathbb{L}^{\lambda} \boldsymbol{\nabla} \mathbf{x}$, and $\mathbf{p}^{f} = \boldsymbol{\nabla} \mathbb{L}^{f} \boldsymbol{\nabla} \mathbf{x}$, where $\mathbb{L}^{\lambda f} \coloneqq \mathbf{D}_{\lambda f} \mathbf{H}_{\lambda f}^{- 1} \mathbf{D}_{\lambda f}^{\prime} / \sqrt{NT}$, $\mathbb{L}^{\lambda} \coloneqq \mathbf{D}_{\lambda} \mathbf{H}_{\lambda}^{- 1} \mathbf{D}_{\lambda}^{\prime} / \sqrt{NT}$, $\mathbb{L}^{f} \coloneqq \mathbf{D}_{f} \mathbf{H}_{f}^{- 1} \mathbf{D}_{f}^{\prime} / \sqrt{NT}$, $\mathbf{H}_{\lambda f} \coloneqq (\mathbf{D}_{\lambda f}^{\prime} \boldsymbol{\nabla} \mathbf{D}_{\lambda f} + \mathbf{V}_{\lambda f} \mathbf{V}_{\lambda f}^{\prime}) / \sqrt{NT}$, $\mathbf{H}_{\lambda} \coloneqq \mathbf{D}_{\lambda}^{\prime} \boldsymbol{\nabla} \mathbf{D}_{\lambda} / \sqrt{NT}$, $\mathbf{H}_{f} \coloneqq \mathbf{D}_{f}^{\prime} \boldsymbol{\nabla} \mathbf{D}_{f} / \sqrt{NT}$, and $\mathbf{V}_{\lambda f}$ is an $(N + T) R \times R^{2}$ matrix defined implicitly to impose the restriction
\begin{equation*}
    \bigg(\sum_{i = 1}^{N} \lambda_{i1} c_{i1} - \sum_{t = 1}^{T} a_{t1} f_{t1}, \ldots, \sum_{i = 1}^{N} \lambda_{iR} c_{iR} - \sum_{t = 1}^{T} a_{tR} f_{tR} \bigg) = \mathbf{0}_{R^{2}} 
\end{equation*}
that ensures uniqueness of the solution. $\mathbf{V}_{\lambda f}$ is defined analogously in \textcite{cfw2021}. The residuals are consequently $\mathbf{r}^{\lambda f} = \boldsymbol{\nabla}(\mathbf{x} - \mathbb{L}^{\lambda f} \boldsymbol{\nabla} \mathbf{x})$, $\mathbf{r}^{\lambda} = \boldsymbol{\nabla}(\mathbf{x} - \mathbb{L}^{\lambda} \boldsymbol{\nabla} \mathbf{x})$, and $\mathbf{r}^{f} = \boldsymbol{\nabla}(\mathbf{x} - \mathbb{L}^{f} \boldsymbol{\nabla} \mathbf{x})$. Finally, let $\mathbf{D}_{N} \coloneqq \mathbf{1}_{N} \otimes \eye_{T}$ and $\mathbf{D}_{T} \coloneqq \eye_{N} \otimes \, \mathbf{1}_{T}$ be $NT \times T$ and $NT \times N$ matrices, respectively.

\vspace{0.5em}
\noindent\textit{Part 3.} Conditional on $\mathcal{C}$, the projections and the pseudo-inverse are stochastic, as they depend on the attrition indicators through the inverses of $\mathbf{H}_{\cdot}$ ($\mathbf{H}_{\lambda f}$, $\mathbf{H}_{\lambda}$, or $\mathbf{H}_{f}$, where $\cdot$ serves as a placeholder), $\boldsymbol{\Phi}_{i}$, and $\boldsymbol{\Psi}_{t}$. To address this, we draw on ideas from \textcite{fw2016}. We adopt bar and tilde notation for conditional expectations and deviations therefrom: we write $\overline{\mathbf{H}}_{\cdot}$ for $\EX[\mathbf{H}_{\cdot} \mid \mathcal{C}]$ and $\widetilde{\mathbf{H}}_{\cdot}$ for $\mathbf{H}_{\cdot} - \overline{\mathbf{H}}_{\cdot}$. We then show that the approximation $\mathbf{H}_{\cdot}^{- 1} \approx \overline{\mathbf{H}}_{\cdot}^{- 1} - \overline{\mathbf{H}}_{\cdot}^{- 1} \widetilde{\mathbf{H}}_{\cdot} \, \overline{\mathbf{H}}_{\cdot}^{- 1}$ can be used to replace the stochastic projections in the quadratic expansion with projections that are deterministic conditional on $\mathcal{C}$. Specifically, we show that $\mathbb{L}^{\cdot} \boldsymbol{\nabla} \mathbf{x} \approx \overline{\mathbb{L}}^{\cdot} \overline{\boldsymbol{\nabla} \mathbf{x}}$, where $\overline{\mathbb{L}}^{\cdot} \coloneqq \mathbf{D}_{\cdot} \overline{\mathbf{H}}_{\cdot}^{- 1} \mathbf{D}_{\cdot}^{\prime} / \sqrt{NT}$. For the pseudo-inverse, we show that $\boldsymbol{\xi}^{\dagger} \approx \bar{\boldsymbol{\xi}}^{\dagger}$, where $\bar{\boldsymbol{\xi}}^{\dagger}$ is an $NT$ vector with elements $\bar{\xi}_{it}^{\dagger} = \boldsymbol{\lambda}_{i}^{\prime} \overline{\boldsymbol{\Psi}}_{t}^{- 1} \overline{\boldsymbol{\Phi}}_{i}^{- 1} \mathbf{f}_{t}^{\phantom{\prime}}$. These approximations are used to complete the derivation of the asymptotic distribution.

Let $\odot$ denote the Hadamard (element-wise) product. We impose the following regularity conditions.
\begin{assumption}[Regularity Conditions]
\label{ass:regul_con}
\hspace{0px}
\begin{enumerate}[i)]
    \item for $2 \leq q \leq 4$, $\norm{\mathbf{x}}_{q} = \mathcal{O}_{P}((NT)^{1 / q})$, $\norm{\mathbf{e}}_{q} = \mathcal{O}_{P}((NT)^{1 / q})$, $\norm{\boldsymbol{\lambda}_{i}}_{2}$ and $\norm{\mathbf{f}_{t}}_{2}$ are uniformly bounded over $i, t, N, T$;
    \item $\overline{\mathbf{H}}_{\lambda f} > 0$ wpa1, $\overline{\mathbf{H}}_{\lambda} > 0$ wpa1, $\overline{\mathbf{H}}_{f} > 0$ wpa1 $\overline{\boldsymbol{\Phi}}_{i} / T > 0$ and $\overline{\boldsymbol{\Psi}}_{t} / N > 0$ wpa1 uniformly over $i, t, N, T$;
    \item for $2 \leq q \leq 4$, $\norm{\overline{\boldsymbol{\nabla} \mathbf{x}}}_{q} = \mathcal{O}_{P}((NT)^{1 / q})$, $\norm{\overline{\boldsymbol{\nabla} \mathbf{e} \odot \mathbf{e}}}_{q} = \mathcal{O}_{P}((NT)^{1 / q})$, $\norm{\overline{\mathbf{H}}_{\lambda}^{- 1}}_{q} = \mathcal{O}_{P}(1)$, $\norm{\overline{\mathbf{H}}_{f}^{- 1}}_{q} = \mathcal{O}_{P}(1)$, $\norm{\overline{\mathbf{H}}_{\lambda f}^{- 1}}_{q} = \mathcal{O}_{P}(1)$, $\sup_{i, N} \norm{(\overline{\boldsymbol{\Phi}}_{i} / T)^{- 1}}_{q} = \mathcal{O}_{P}(1)$, $\sup_{t, T} \norm{(\overline{\boldsymbol{\Psi}}_{t} / N)^{- 1}}_{q} = \mathcal{O}_{P}(1)$;
    \item for $2 \leq q \leq 4$, $\norm{\widetilde{\mathbf{H}}_{\lambda f}}_{q} = o_{P}((NT)^{- 1 / (4 q)})$, $\norm{\widetilde{\mathbf{H}}_{\lambda} }_{q} = o_{P}((NT)^{- 1 / (4 q)})$, $\norm{\widetilde{\mathbf{H}}_{f}}_{q} = o_{P}((NT)^{- 1 / (4 q)})$, $\sup_{i, N} \norm{(\widetilde{\boldsymbol{\Phi}}_{i} / T)^{- 1}}_{q} = o_{P}(1)$, $\sup_{t, T} \norm{(\widetilde{\boldsymbol{\Psi}}_{t} / N)^{- 1}}_{q} = o_{P}(1)$;  
    \item for $2 \leq q \leq 4$, $\norm{\mathbf{D}_{\lambda f}}_{q} = \mathcal{O}_{P}((NT)^{1 / (2 q)})$, $\norm{\mathbf{D}_{\lambda}}_{q} = \mathcal{O}_{P}((NT)^{1 / (2 q)})$, \linebreak $\norm{\mathbf{D}_{f}}_{q} = \mathcal{O}_{P}((NT)^{1 / (2 q)})$, $\norm{\mathbf{D}_{\lambda f}^{\prime}}_{q} = \mathcal{O}_{P}((NT)^{1 / 2 - 1 / (2 q)})$, $\norm{\mathbf{D}_{\lambda}^{\prime}}_{q} = \mathcal{O}_{P}((NT)^{1 / 2 - 1 / (2 q)})$, $\norm{\mathbf{D}_{f}^{\prime}}_{q} = \mathcal{O}_{P}((NT)^{1 / 2 - 1 / (2 q)})$, $\norm{\mathbf{D}_{N}}_{q} = \mathcal{O}((NT)^{1 / (2 q)})$, $\norm{\mathbf{D}_{T}}_{q} = \mathcal{O}((NT)^{1 / (2 q)})$, \linebreak $\norm{\mathbf{D}_{N}^{\prime}}_{q} = \mathcal{O}((NT)^{1 / 2 - 1 / (2 q)})$, $\norm{\mathbf{D}_{T}^{\prime}}_{q} = \mathcal{O}((NT)^{1 / 2 - 1 / (2 q)})$;
    \item for $2 \leq q \leq 4$, $\norm{\mathbf{D}_{\lambda f}^{\prime} \boldsymbol{\nabla}\mathbf{e}}_{q} = \mathcal{O}_{P}((NT)^{1 / 4 + 1 / (2 q)})$, $\norm{\mathbf{D}_{\lambda}^{\prime} \boldsymbol{\nabla}\mathbf{e}}_{q} = \mathcal{O}_{P}((NT)^{1 / 4 + 1 / (2 q)})$, $\norm{\mathbf{D}_{f}^{\prime} \boldsymbol{\nabla}\mathbf{e}}_{q} = \mathcal{O}_{P}((NT)^{1 / 4 + 1 / (2 q)})$, $\norm{\mathbf{D}_{\lambda f}^{\prime} \widetilde{\boldsymbol{\nabla} \mathbf{x}}}_{q} = \mathcal{O}_{P}((NT)^{1 / 4 + 1 / (2 q)})$, \linebreak $\norm{\mathbf{D}_{\lambda}^{\prime} \widetilde{\boldsymbol{\nabla} \mathbf{x}}}_{q} = \mathcal{O}_{P}((NT)^{1 / 4 + 1 / (2 q)})$, $\norm{\mathbf{D}_{f}^{\prime} \widetilde{\boldsymbol{\nabla} \mathbf{x}}}_{q} = \mathcal{O}_{P}((NT)^{1 / 4 + 1 / (2 q)})$;
    \item $\lvert\widetilde{(\ddot{\mathbf{x}}^{\lambda f})^{\prime} \boldsymbol{\nabla} \ddot{\mathbf{x}}^{\lambda f}}\rvert / (NT) = o_{P}(1)$, $\lvert\widetilde{(\ddot{\mathbf{x}}^{\lambda f} \odot \mathbf{e})^{\prime} \boldsymbol{\nabla} (\ddot{\mathbf{x}}^{\lambda f} \odot \mathbf{e})}\rvert / (NT) = o_{P}(1)$, $\lvert\widetilde{\mathbf{e}^{\prime} \boldsymbol{\nabla} \overline{\mathbb{L}}^{f} \widetilde{\boldsymbol{\nabla} \ddot{\mathbf{x}}^{f}}}\rvert / N = o_{P}(1)$, $\norm{\mathbf{D}_{T}^{\prime} (\widetilde{\boldsymbol{\nabla} \ddot{\mathbf{x}}^{\lambda}} \odot \bar{\boldsymbol{\xi}}^{\dagger})}_{2} = o_{P}((NT)^{1 / 4})$, $\norm{\mathbf{D}_{N}^{\prime} (\widetilde{\boldsymbol{\nabla} \ddot{\mathbf{x}}^{f}} \odot \bar{\boldsymbol{\xi}}^{\dagger})}_{2} = o_{P}((NT)^{1 / 4})$;
    \item $\overline{(\ddot{\mathbf{x}}^{\lambda f})^{\prime} \boldsymbol{\nabla} \ddot{\mathbf{x}}^{\lambda f}} / (NT) > 0$ wpa1.
\end{enumerate}  
\end{assumption}

Under our assumptions, $\overline{\mathbf{H}}_{\cdot}$ is invertible. By Corollary 5.6.16 of \textcite{hj2012}, $\eye + \widetilde{\mathbf{H}}_{\cdot} \overline{\mathbf{H}}_{\cdot}^{- 1}$ is also invertible for sufficiently large $N$ and $T$, since $\norm{\widetilde{\mathbf{H}}_{\cdot} \overline{\mathbf{H}}_{\cdot}^{- 1}}_{q} \leq \norm{\widetilde{\mathbf{H}}_{\cdot}}_{q} \norm{\overline{\mathbf{H}}_{\cdot}^{- 1}}_{q} = o_{P}(1)$. Consequently, $\mathbf{H}_{\cdot}^{- 1}$ admits the following Neumann series representation:
\begin{equation*}
    \mathbf{H}_{\cdot}^{- 1} = \overline{\mathbf{H}}_{\cdot}^{- 1} \big(\eye + \widetilde{\mathbf{H}}_{\cdot} \overline{\mathbf{H}}_{\cdot}^{- 1} \big)^{- 1} =  \overline{\mathbf{H}}_{\cdot}^{- 1} \sum_{r = 0}^{\infty} (- \widetilde{\mathbf{H}}_{\cdot} \overline{\mathbf{H}}_{\cdot}^{- 1})^{r} \, \quad \text{wpa1} \, .
\end{equation*}
Let $\mathbf{U}_{\cdot} \coloneqq \sum_{r = 2}^{\infty} (- \widetilde{\mathbf{H}}_{\cdot} \overline{\mathbf{H}}_{\cdot}^{- 1})^{r}$ denote the truncation remainder, so that
\begin{equation*}
    \mathbf{H}_{\cdot}^{- 1} = \overline{\mathbf{H}}_{\cdot}^{- 1} - \overline{\mathbf{H}}_{\cdot}^{- 1} \widetilde{\mathbf{H}}_{\cdot} \overline{\mathbf{H}}_{\cdot}^{- 1} + \overline{\mathbf{H}}_{\cdot}^{- 1} \mathbf{U}_{\cdot} \, .
\end{equation*}
The remainder satisfies
\begin{equation*}
    \norm{\mathbf{U}_{\cdot}}_{q} \leq \norm{\widetilde{\mathbf{H}}_{\cdot}}_{q}^{2} \norm{\overline{\mathbf{H}}_{\cdot}^{- 1}}_{q}^{2} \big(1 - \norm{\widetilde{\mathbf{H}}_{\cdot}}_{q} \norm{\overline{\mathbf{H}}_{\cdot}^{- 1}}_{q}\big)^{- 1} \, .
\end{equation*}
Since $\big(1 - \norm{\widetilde{\mathbf{H}}_{\cdot}}_{q} \norm{\overline{\mathbf{H}}_{\cdot}^{- 1}}_{q}\big)^{- 1} = (1 - o_{P}(1))^{- 1} = \mathcal{O}_{P}(1)$, we obtain
\begin{align}
    &\norm{\mathbf{H}_{\cdot}^{- 1} - \overline{\mathbf{H}}_{\cdot}^{- 1}}_{q} = o_{P}\big((NT)^{- 1 / (4 q)}\big) \, , \label{eq:approx_inv_ip_hessian1} \\
    &\norm{\mathbf{H}_{\cdot}^{- 1} - \overline{\mathbf{H}}_{\cdot}^{- 1} + \overline{\mathbf{H}}_{\cdot}^{- 1} \widetilde{\mathbf{H}}_{\cdot} \overline{\mathbf{H}}_{\cdot}^{- 1}}_{q} = o_{P}\big((NT)^{- 1 / (2 q)}\big) \, . \label{eq:approx_inv_ip_hessian2}
\end{align}
By analogous arguments,
\begin{align}
    &\sup_{i, N} \, \norm{(\boldsymbol{\Phi}_{i} / T))^{- 1} - (\overline{\boldsymbol{\Phi}}_{i} / T)^{- 1}}_{q} = o_{P}(1) \label{eq:approx_inv_pseudo_inverse} \, , \\
    &\sup_{t, T} \, \norm{(\boldsymbol{\Psi}_{t} / N))^{- 1} - (\overline{\boldsymbol{\Psi}}_{t} / N)^{- 1}}_{q} = o_{P}(1) \nonumber \, .
\end{align}

Decomposing the projection
\begin{equation*}
    \mathbb{L}^{\cdot} \boldsymbol{\nabla} \mathbf{x} = \overline{\mathbb{L}}^{\cdot} \overline{\boldsymbol{\nabla} \mathbf{x}} + \mathbf{D}_{\cdot} (\mathbf{H}_{\cdot}^{- 1} - \overline{\mathbf{H}}_{\cdot}^{- 1}) \mathbf{D}_{\cdot}^{\prime} \boldsymbol{\nabla} \mathbf{x} / \sqrt{NT} + \mathbf{D}_{\cdot} \overline{\mathbf{H}}_{\cdot}^{- 1} \mathbf{D}_{\cdot}^{\prime} \widetilde{\boldsymbol{\nabla} \mathbf{x}} / \sqrt{NT} \, ,
\end{equation*}
and applying the triangle inequality, \eqref{eq:approx_inv_ip_hessian1}, and $\norm{\boldsymbol{\nabla}}_{q} = \norm{\boldsymbol{\delta}}_{\infty} \leq 1$, where $\boldsymbol{\delta} \coloneqq (\delta_{11}, \ldots, \delta_{NT})$, we obtain
\begin{align}
    &\norm{\boldsymbol{\nabla} (\mathbb{L}^{\cdot} \boldsymbol{\nabla} \mathbf{x} - \overline{\mathbb{L}}^{\cdot} \overline{\boldsymbol{\nabla} \mathbf{x}})}_{q} \leq \norm{\boldsymbol{\nabla}}_{q} \norm{\mathbf{D}_{\cdot}}_{q} \norm{\mathbf{D}_{\cdot}^{\prime}}_{q} \norm{\mathbf{H}_{\cdot}^{- 1} - \overline{\mathbf{H}}_{\cdot}^{- 1}}_{q} \norm{\boldsymbol{\nabla} \mathbf{x}}_{q} / \sqrt{NT} \, 
    + \nonumber\\
    &\qquad \norm{\boldsymbol{\nabla}}_{q} \norm{\mathbf{D}_{\cdot}}_{q} \norm{\overline{\mathbf{H}}_{\cdot}^{- 1}}_{q} \norm{\mathbf{D}_{\cdot}^{\prime} \widetilde{\boldsymbol{\nabla} \mathbf{x}}}_{q} / \sqrt{NT} = o_{P}\big((NT)^{3 / (4 q)}\big) \, . \label{eq:approx_projection}
\end{align}
By analogous arguments,
\begin{align*}
    &\norm{\boldsymbol{\xi}^{\dagger} - \bar{\boldsymbol{\xi}}^{\dagger}}_{q} \leq \bigg(\sum_{i = 1}^{N} \sum_{t = 1}^{T} \lvert \boldsymbol{\lambda}_{i}^{\prime} \big(\boldsymbol{\Psi}_{t}^{- 1} - \overline{\boldsymbol{\Psi}}_{t}^{- 1}\big) \overline{\boldsymbol{\Phi}}_{i}^{- 1} \mathbf{f}_{t} \rvert^{q}\bigg)^{\frac{1}{q}} + \bigg(\sum_{i = 1}^{N} \sum_{t = 1}^{T} \lvert \boldsymbol{\lambda}_{i}^{\prime} \overline{\boldsymbol{\Psi}}_{t}^{- 1} \big(\boldsymbol{\Phi}_{i}^{- 1} - \overline{\boldsymbol{\Phi}}_{i}^{- 1}\big) \mathbf{f}_{t} \rvert^{q}\bigg)^{\frac{1}{q}} \, + \\
    &\quad \bigg(\sum_{i = 1}^{N} \sum_{t = 1}^{T} \lvert \boldsymbol{\lambda}_{i}^{\prime} \big(\boldsymbol{\Psi}_{t}^{- 1} - \overline{\boldsymbol{\Psi}}_{t}^{- 1}\big) \big(\boldsymbol{\Phi}_{i}^{- 1} - \overline{\boldsymbol{\Phi}}_{i}^{- 1}\big) \mathbf{f}_{t} \rvert^{q}\bigg)^{\frac{1}{q}} \eqqcolon \norm{\Delta \boldsymbol{\xi}_{1}^{\dagger}}_{q} + \norm{\Delta \boldsymbol{\xi}_{2}^{\dagger}}_{q} + \norm{\Delta \boldsymbol{\xi}_{3}^{\dagger}}_{q} \, . 
\end{align*}
Then, by the Cauchy–Schwarz inequality and \eqref{eq:approx_inv_pseudo_inverse},
\begin{align*}
    &\norm{\Delta \boldsymbol{\xi}_{1}^{\dagger}}_{q} \leq \bigg(\sum_{i = 1}^{N} \norm{\boldsymbol{\lambda}_{i}}_{2}^{2 q}\bigg)^{\frac{1}{2 q}} \bigg(\sum_{t = 1}^{T} \norm{(\boldsymbol{\Psi}_{t} / N)^{- 1} - (\overline{\boldsymbol{\Psi}}_{t} / N)^{- 1}}_{2}^{2 q}\bigg)^{\frac{1}{2 q}} \bigg(\sum_{i = 1}^{N} \norm{(\overline{\boldsymbol{\Phi}}_{i} / T)^{- 1}}_{2}^{2 q}\bigg)^{\frac{1}{2 q}} \\
    &\quad \bigg(\sum_{t = 1}^{T} \norm{\mathbf{f}_{t}}_{2}^{2 q}\bigg)^{\frac{1}{2 q}} / NT = o_{P}\big((NT)^{- 1 + 1 / q}\big) \, .
\end{align*}
Analogously, $\norm{\Delta \boldsymbol{\xi}_{2}^{\dagger}}_{q} = o_{P}((NT)^{- 1 + 1 / q})$ and $\norm{\Delta \boldsymbol{\xi}_{3}^{\dagger}}_{q} = o_{P}((NT)^{- 1 + 1 / q})$. Hence,
\begin{equation}
    \label{eq:approx_pseudo_inverse}
     \norm{\boldsymbol{\xi}^{\dagger} - \bar{\boldsymbol{\xi}}^{\dagger}}_{q} = o_{P}\big((NT)^{- 1 + 1 / q}\big) \, .
\end{equation}
In addition,
\begin{equation}
    \label{eq:bound_projection}
    \norm{\ddot{\mathbf{x}}^{\cdot}}_{q} \leq \norm{\mathbf{x}}_{q} + \norm{\mathbf{D}_{\cdot}}_{q} \norm{\mathbf{D}_{\cdot}^{\prime}}_{q} \norm{\overline{\mathbf{H}}_{\cdot}^{- 1}}_{q} \norm{\overline{\boldsymbol{\nabla} \mathbf{x}}}_{q} / \sqrt{NT} = \mathcal{O}_{P}\big((NT)^{1 / q}\big)
\end{equation}
and
\begin{align}
    &\norm{\bar{\boldsymbol{\xi}}^{\dagger}}_{q} \leq \bigg(\sum_{i = 1}^{N} \norm{\boldsymbol{\lambda}_{i}}_{2}^{2 q}\bigg)^{\frac{1}{2 q}} \bigg(\sum_{t = 1}^{T} \norm{(\overline{\boldsymbol{\Psi}}_{t} / N)^{- 1}}_{2}^{2 q}\bigg)^{\frac{1}{2 q}} \bigg(\sum_{i = 1}^{N} \norm{(\overline{\boldsymbol{\Phi}}_{i} / T)^{- 1}}_{2}^{2 q}\bigg)^{\frac{1}{2 q}} \bigg(\sum_{t = 1}^{T} \norm{\mathbf{f}_{t}}_{2}^{2 q}\bigg)^{\frac{1}{2 q}} / (NT) \nonumber \\
    &\quad = \mathcal{O}_{P}\big((NT)^{- 1 + 1 / q}\big) \label{eq:bound_pseudo_inverse} \, .
\end{align}

Decomposing the terms of the quadratic approximation then yields
\begin{align*}
    &\mathbb{W} = \frac{(\mathbf{r}^{\lambda f})^{\prime} \mathbf{r}^{\lambda f}}{NT} = \frac{\overline{(\ddot{\mathbf{x}}^{\lambda f})^{\prime} \boldsymbol{\nabla} \ddot{\mathbf{x}}^{\lambda f}}}{NT} + \frac{\widetilde{(\ddot{\mathbf{x}}^{\lambda f})^{\prime} \boldsymbol{\nabla} \ddot{\mathbf{x}}^{\lambda f}}}{NT} - 2 \frac{(\ddot{\mathbf{x}}^{\lambda f})^{\prime} \boldsymbol{\nabla} (\mathbb{L}^{\lambda f} \boldsymbol{\nabla} \mathbf{x} - \overline{\mathbb{L}}^{\lambda f} \overline{\boldsymbol{\nabla} \mathbf{x}})}{NT} \, + \\
    &\qquad \frac{(\mathbb{L}^{\lambda f} \boldsymbol{\nabla} \mathbf{x} - \overline{\mathbb{L}}^{\lambda f} \overline{\boldsymbol{\nabla} \mathbf{x}})^{\prime} \boldsymbol{\nabla} (\mathbb{L}^{\lambda f} \boldsymbol{\nabla} \mathbf{x} - \overline{\mathbb{L}}^{\lambda f} \overline{\boldsymbol{\nabla} \mathbf{x}})}{NT} \eqqcolon \mathbb{W}_{1} + \ldots + \mathbb{W}_{4} \, , \\
    &\mathbb{V} = \frac{(\mathbf{r}^{\lambda f} \odot \mathbf{e})^{\prime} (\mathbf{r}^{\lambda f} \odot \mathbf{e})}{NT} = \frac{\overline{(\ddot{\mathbf{x}}^{\lambda f}  \odot \mathbf{e})^{\prime} \boldsymbol{\nabla} (\ddot{\mathbf{x}}^{\lambda f} \odot \mathbf{e})}}{NT} + \frac{\widetilde{(\ddot{\mathbf{x}}^{\lambda f} \odot \mathbf{e})^{\prime} \boldsymbol{\nabla} (\ddot{\mathbf{x}}^{\lambda f} \odot \mathbf{e})}}{NT} \, - \\
    &\qquad 2 \frac{(\ddot{\mathbf{x}}^{\lambda f} \odot \mathbf{e})^{\prime} \boldsymbol{\nabla} ((\mathbb{L}^{\lambda f} \boldsymbol{\nabla} \mathbf{x} - \overline{\mathbb{L}}^{\lambda f} \overline{\boldsymbol{\nabla} \mathbf{x}}) \odot \mathbf{e})}{NT} \, + \\
    &\qquad \frac{((\mathbb{L}^{\lambda f} \boldsymbol{\nabla} \mathbf{x} - \overline{\mathbb{L}}^{\lambda f} \overline{\boldsymbol{\nabla} \mathbf{x}}) \odot \mathbf{e})^{\prime} \boldsymbol{\nabla} ((\mathbb{L}^{\lambda f} \boldsymbol{\nabla} \mathbf{x} - \overline{\mathbb{L}}^{\lambda f} \overline{\boldsymbol{\nabla} \mathbf{x}}) \odot \mathbf{e})}{NT} \eqqcolon \mathbb{V}_{1} + \ldots + \mathbb{V}_{4} \, , \\
    &\mathbb{B}_{1} = \frac{\mathbf{e}^{\prime} \mathbf{p}^{f}}{N} = \frac{\mathbf{e}^{\prime} \boldsymbol{\nabla} \overline{\mathbb{L}}^{f} \overline{\boldsymbol{\nabla} \mathbf{x}}}{N} + \frac{\overline{\mathbf{e}^{\prime} \boldsymbol{\nabla} \overline{\mathbb{L}}^{f} \boldsymbol{\nabla} \ddot{\mathbf{x}}^{f}}}{N} + \frac{\widetilde{\mathbf{e}^{\prime} \boldsymbol{\nabla} \overline{\mathbb{L}}^{f} \widetilde{\boldsymbol{\nabla} \ddot{\mathbf{x}}^{f}}}}{N} \, - \\
    &\qquad \frac{\mathbf{e}^{\prime} \boldsymbol{\nabla} \mathbf{D}_{f} \overline{\mathbf{H}}_{f}^{- 1} \widetilde{\mathbf{H}}_{f} \overline{\mathbf{H}}_{f}^{- 1} \mathbf{D}_{f}^{\prime} \widetilde{\boldsymbol{\nabla} \mathbf{x}}}{N \sqrt{NT}} \, + \\
    &\qquad \frac{\mathbf{e}^{\prime} \boldsymbol{\nabla} \mathbf{D}_{f} \big(\mathbf{H}_{f}^{- 1} - \overline{\mathbf{H}}_{f}^{- 1} + \overline{\mathbf{H}}_{f}^{- 1} \widetilde{\mathbf{H}}_{f} \overline{\mathbf{H}}_{f}^{- 1}\big) \mathbf{D}_{f}^{\prime} \boldsymbol{\nabla} \mathbf{x}}{N \sqrt{NT}} \eqqcolon \mathbb{B}_{1, 1} + \ldots + \mathbb{B}_{1, 5} \, , \\
    &\mathbb{B}_{2} = \frac{\mathbf{D}_{T}^{\prime} \overline{\boldsymbol{\nabla} \mathbf{e} \odot \mathbf{e}} \odot \mathbf{D}_{T}^{\prime} (\mathbf{r}^{\lambda} \odot \boldsymbol{\xi}^{\dagger})}{T} = \frac{\mathbf{D}_{T}^{\prime} \overline{\boldsymbol{\nabla} \mathbf{e} \odot \mathbf{e}} \odot \mathbf{D}_{T}^{\prime} (\overline{\boldsymbol{\nabla} \ddot{\mathbf{x}}^{\lambda}} \odot \bar{\boldsymbol{\xi}}^{\dagger})}{T} \, + \\
    &\qquad \frac{\mathbf{D}_{T}^{\prime} \overline{\boldsymbol{\nabla} \mathbf{e} \odot \mathbf{e}} \odot \mathbf{D}_{T}^{\prime} (\widetilde{\boldsymbol{\nabla} \ddot{\mathbf{x}}^{\lambda}} \odot \bar{\boldsymbol{\xi}}^{\dagger})}{T} - \frac{\mathbf{D}_{T}^{\prime} \overline{\boldsymbol{\nabla} \mathbf{e} \odot \mathbf{e}} \odot \mathbf{D}_{T}^{\prime} (\boldsymbol{\nabla} (\mathbb{L}^{\lambda} \boldsymbol{\nabla} \mathbf{x} - \overline{\mathbb{L}}^{\lambda} \overline{\boldsymbol{\nabla} \mathbf{x}}) \odot \bar{\boldsymbol{\xi}}^{\dagger})}{T} \, + \\
    &\qquad \frac{\mathbf{D}_{T}^{\prime} \overline{\boldsymbol{\nabla} \mathbf{e} \odot \mathbf{e}} \odot \mathbf{D}_{T}^{\prime} (\boldsymbol{\nabla} \ddot{\mathbf{x}}^{\lambda} \odot (\boldsymbol{\xi}^{\dagger} - \bar{\boldsymbol{\xi}}^{\dagger}))}{T} \, - \\
    &\qquad \frac{\mathbf{D}_{T}^{\prime} \overline{\boldsymbol{\nabla} \mathbf{e} \odot \mathbf{e}} \odot \mathbf{D}_{T}^{\prime} (\boldsymbol{\nabla} (\mathbb{L}^{\lambda} \boldsymbol{\nabla} \mathbf{x} - \overline{\mathbb{L}}^{\lambda} \overline{\boldsymbol{\nabla} \mathbf{x}}) \odot (\boldsymbol{\xi}^{\dagger} - \bar{\boldsymbol{\xi}}^{\dagger}))}{T} \eqqcolon \mathbb{B}_{2, 1} + \ldots + \mathbb{B}_{2, 5} \, , \\
    &\mathbb{B}_{3} = \frac{\mathbf{D}_{N}^{\prime} \overline{\boldsymbol{\nabla} \mathbf{e} \odot \mathbf{e}} \odot \mathbf{D}_{N}^{\prime} (\mathbf{r}^{f} \odot \boldsymbol{\xi}^{\dagger})}{N} = \frac{\mathbf{D}_{N}^{\prime} \overline{\boldsymbol{\nabla} \mathbf{e} \odot \mathbf{e}} \odot \mathbf{D}_{N}^{\prime} (\overline{\boldsymbol{\nabla} \ddot{\mathbf{x}}^{f}} \odot \bar{\boldsymbol{\xi}}^{\dagger})}{N} \, + \\
    &\qquad \frac{\mathbf{D}_{N}^{\prime} \overline{\boldsymbol{\nabla} \mathbf{e} \odot \mathbf{e}} \odot \mathbf{D}_{N}^{\prime} (\widetilde{\boldsymbol{\nabla} \ddot{\mathbf{x}}^{f}} \odot \bar{\boldsymbol{\xi}}^{\dagger})}{N} - \frac{\mathbf{D}_{N}^{\prime} \overline{\boldsymbol{\nabla} \mathbf{e} \odot \mathbf{e}} \odot \mathbf{D}_{N}^{\prime} (\boldsymbol{\nabla} (\mathbb{L}^{f} \boldsymbol{\nabla} \mathbf{x} - \overline{\mathbb{L}}^{f} \overline{\boldsymbol{\nabla} \mathbf{x}}) \odot \bar{\boldsymbol{\xi}}^{\dagger})}{N} \, + \\
    &\qquad \frac{\mathbf{D}_{N}^{\prime} \overline{\boldsymbol{\nabla} \mathbf{e} \odot \mathbf{e}} \odot \mathbf{D}_{N}^{\prime} (\boldsymbol{\nabla} \ddot{\mathbf{x}}^{f} \odot (\boldsymbol{\xi}^{\dagger} - \bar{\boldsymbol{\xi}}^{\dagger}))}{N} \, - \\
    &\qquad \frac{\mathbf{D}_{N}^{\prime} \overline{\boldsymbol{\nabla} \mathbf{e} \odot \mathbf{e}} \odot \mathbf{D}_{N}^{\prime} (\boldsymbol{\nabla} (\mathbb{L}^{f} \boldsymbol{\nabla} \mathbf{x} - \overline{\mathbb{L}}^{f} \overline{\boldsymbol{\nabla} \mathbf{x}}) \odot (\boldsymbol{\xi}^{\dagger} - \bar{\boldsymbol{\xi}}^{\dagger}))}{N} \eqqcolon \mathbb{B}_{3, 1} + \ldots + \mathbb{B}_{3, 5} \, .
\end{align*}
Furthermore, given our assumptions and intermediate results \eqref{eq:approx_inv_ip_hessian1}--\eqref{eq:bound_pseudo_inverse},
\begin{align*}
    &\lvert\mathbb{W}_{2}\rvert = o_{P}(1) \, , \\
    &\lvert\mathbb{W}_{3}\rvert \leq \norm{\ddot{\mathbf{x}}^{\lambda f}}_{2} \norm{\boldsymbol{\nabla} (\mathbb{L}^{\lambda f} \boldsymbol{\nabla} \mathbf{x} - \overline{\mathbb{L}}^{\lambda f} \overline{\boldsymbol{\nabla} \mathbf{x}})}_{2} / NT = o_{P}(1) \, , \\
    &\lvert\mathbb{W}_{4}\rvert \leq \norm{\boldsymbol{\nabla} (\mathbb{L}^{\lambda f} \boldsymbol{\nabla} \mathbf{x} - \overline{\mathbb{L}}^{\lambda f} \overline{\boldsymbol{\nabla} \mathbf{x}})}_{2}^{2} / NT = o_{P}(1) \, , \\
    &\lvert\mathbb{V}_{2}\rvert = o_{P}(1) \, , \\
    &\lvert\mathbb{V}_{3}\rvert \leq 2 \norm{\ddot{\mathbf{x}}^{\lambda f}}_{4} \norm{\mathbf{e}}_{4}^{2} \norm{\boldsymbol{\nabla} (\mathbb{L}^{\lambda f} \boldsymbol{\nabla} \mathbf{x} - \overline{\mathbb{L}}^{\lambda f} \overline{\boldsymbol{\nabla} \mathbf{x}})}_{4} = o_{P}(1) \, , \\
    &\lvert\mathbb{V}_{4}\rvert \leq \norm{\mathbf{e}}_{4}^{2} \norm{\boldsymbol{\nabla} (\mathbb{L}^{\lambda f} \boldsymbol{\nabla} \mathbf{x} - \overline{\mathbb{L}}^{\lambda f} \overline{\boldsymbol{\nabla} \mathbf{x}})}_{4}^{2} = o_{P}(1) \, , \\
    &\lvert\mathbb{B}_{1, 3}\rvert = o_{P}(1) \, , \\
    &\lvert\mathbb{B}_{1, 4}\rvert \leq \norm{\mathbf{D}_{f}^{\prime} \widetilde{\boldsymbol{\nabla} \mathbf{x}}}_{2} \norm{\mathbf{D}_{f}^{\prime} \boldsymbol{\nabla} \mathbf{e}}_{2} \norm{\overline{\mathbf{H}}_{f}^{- 1}}_{2}^{2} \norm{\widetilde{\mathbf{H}}_{f}}_{2} / (N \sqrt{NT}) = o_{P}(1) \, , \\
    &\lvert\mathbb{B}_{1, 5}\rvert \leq \norm{\mathbf{D}_{f}}_{2} \norm{\mathbf{x}}_{2} \norm{\mathbf{D}_{f}^{\prime} \boldsymbol{\nabla} \mathbf{e}}_{2} \norm{\mathbf{H}_{f}^{- 1} - \overline{\mathbf{H}}_{f}^{- 1} + \overline{\mathbf{H}}_{f}^{- 1} \widetilde{\mathbf{H}}_{f} \overline{\mathbf{H}}_{f}^{- 1}}_{2} / (N \sqrt{NT}) = o_{P}(1) \, , \\
    &\lvert\mathbb{B}_{2, 2}\rvert \leq \norm{\mathbf{D}_{T}}_{2} \norm{\overline{\boldsymbol{\nabla} \mathbf{e} \odot \mathbf{e}}}_{2} \norm{\mathbf{D}_{T}^{\prime} (\widetilde{\boldsymbol{\nabla} \ddot{\mathbf{x}}^{\lambda}} \odot \bar{\boldsymbol{\xi}}^{\dagger})}_{2} / T = o_{P}(1) \, , \\
    &\lvert\mathbb{B}_{2, 3}\rvert \leq \norm{\mathbf{D}_{T}}_{2}^{2} \norm{\overline{\boldsymbol{\nabla} \mathbf{e} \odot \mathbf{e}}}_{2} \norm{\boldsymbol{\nabla} (\mathbb{L}^{\lambda} \boldsymbol{\nabla} \mathbf{x} - \overline{\mathbb{L}}^{\lambda} \overline{\boldsymbol{\nabla} \mathbf{x}})}_{4} \norm{\bar{\boldsymbol{\xi}}^{\dagger}}_{4} / T = o_{P}(1) \, , \\
    &\lvert\mathbb{B}_{2, 4}\rvert \leq \norm{\mathbf{D}_{T}}_{2}^{2} \norm{\overline{\boldsymbol{\nabla} \mathbf{e} \odot \mathbf{e}}}_{2} \norm{\ddot{\mathbf{x}}^{\lambda}}_{4} \norm{\boldsymbol{\xi}^{\dagger} - \bar{\boldsymbol{\xi}}^{\dagger}}_{4} / T  = o_{P}(1) \, , \\
    &\lvert\mathbb{B}_{2, 5}\rvert \leq \norm{\mathbf{D}_{T}}_{2}^{2} \norm{\overline{\boldsymbol{\nabla} \mathbf{e} \odot \mathbf{e}}}_{2} \norm{\boldsymbol{\nabla} (\mathbb{L}^{\lambda} \boldsymbol{\nabla} \mathbf{x} - \overline{\mathbb{L}}^{\lambda} \overline{\boldsymbol{\nabla} \mathbf{x}})}_{4} \norm{\boldsymbol{\xi}^{\dagger} - \bar{\boldsymbol{\xi}}^{\dagger}}_{4} / T = o_{P}(1) \, , \\
    &\lvert\mathbb{B}_{3, 2}\rvert \leq \norm{\mathbf{D}_{N}}_{2} \norm{\overline{\boldsymbol{\nabla} \mathbf{e} \odot \mathbf{e}}}_{2} \norm{\mathbf{D}_{N}^{\prime} (\widetilde{\boldsymbol{\nabla} \ddot{\mathbf{x}}^{f}} \odot \bar{\boldsymbol{\xi}}^{\dagger})}_{2} / N = o_{P}(1) \, , \\
    &\lvert\mathbb{B}_{3, 3}\rvert \leq \norm{\mathbf{D}_{N}}_{2}^{2} \norm{\overline{\boldsymbol{\nabla} \mathbf{e} \odot \mathbf{e}}}_{2} \norm{\boldsymbol{\nabla} (\mathbb{L}^{f} \boldsymbol{\nabla} \mathbf{x} - \overline{\mathbb{L}}^{f} \overline{\boldsymbol{\nabla} \mathbf{x}})}_{4} \norm{\bar{\boldsymbol{\xi}}^{\dagger}}_{4} / N = o_{P}(1) \, , \\
    &\lvert\mathbb{B}_{3, 4}\rvert \leq \norm{\mathbf{D}_{N}}_{2}^{2} \norm{\overline{\boldsymbol{\nabla} \mathbf{e} \odot \mathbf{e}}}_{2} \norm{\ddot{\mathbf{x}}^{f}}_{4} \norm{\boldsymbol{\xi}^{\dagger} - \bar{\boldsymbol{\xi}}^{\dagger}}_{4} / N = o_{P}(1) \, , \\
    &\lvert\mathbb{B}_{3, 5}\rvert \leq \norm{\mathbf{D}_{N}}_{2}^{2} \norm{\overline{\boldsymbol{\nabla} \mathbf{e} \odot \mathbf{e}}}_{2} \norm{\boldsymbol{\nabla} (\mathbb{L}^{f} \boldsymbol{\nabla} \mathbf{x} - \overline{\mathbb{L}}^{f} \overline{\boldsymbol{\nabla} \mathbf{x}})}_{4} \norm{\boldsymbol{\xi}^{\dagger} - \bar{\boldsymbol{\xi}}^{\dagger}}_{4} / N = o_{P}(1) \, ,
\end{align*}
and $\EX[\mathbb{B}_{1, 1} \mid \mathcal{C}] = 0$. In addition, $\mathbb{W}_{1}$ is invertible wpa1. The results of Section \ref{sec:model_consistency} follow immediately.

\vspace{0.5em}
\noindent\textit{Part 4.} We verify the additional regularity conditions of Assumption \ref{ass:regul_con} using the assumptions of \textcite{mw2017} and Assumption \ref{ass:missing_data_stochastic}.

\vspace{0.5em}
\noindent i) follows directly from the assumptions of \textcite{mw2017}.

\vspace{0.5em}
\noindent ii) is a consequence of $\EX[\delta_{it} \mid \mathcal{C}] \geq c_{\min} > 0$ a.\,s.\ uniformly over $i, t, N, T$. Recall $\overline{\mathbf{H}}_{\lambda f} = (\mathbf{D}_{\lambda f}^{\prime} \overline{\boldsymbol{\nabla}} \mathbf{D}_{\lambda f} + \mathbf{V}_{\lambda f} \mathbf{V}_{\lambda f}^{\prime}) / \sqrt{NT}$, where $\overline{\boldsymbol{\nabla}} = \diag(\EX[\delta_{11} \mid \mathcal{C}], \ldots, \EX[\delta_{NT} \mid \mathcal{C}])$. By the Courant–Fischer–Weyl min-max principle,
\begin{align*}
    \mu_{\min}\big(\overline{\mathbf{H}}_{\lambda f}\big) =& \, \min_{\norm{v}_{2} = 1} v^{\prime} \bigg\{\frac{\mathbf{D}_{\lambda f}^{\prime} \overline{\boldsymbol{\nabla}} \mathbf{D}_{\lambda f} + \mathbf{V}_{\lambda f} \mathbf{V}_{\lambda f}^{\prime}}{\sqrt{NT}}\bigg\} v \\
    =& \, \min_{\norm{v}_{2} = 1} v^{\prime} \bigg\{\frac{\mathbf{D}_{\lambda f}^{\prime} \overline{\boldsymbol{\nabla}} \mathbf{D}_{\lambda f} + c_{\min} \mathbf{V}_{\lambda f} \mathbf{V}_{\lambda f}^{\prime} + (1 - c_{\min}) \mathbf{V}_{\lambda f} \mathbf{V}_{\lambda f}^{\prime}}{\sqrt{NT}}\bigg\} v \, .
\end{align*}
By Weyl's inequality (see, e.g., \textcite{hj2012} Theorem 4.3.1),
\begin{align*}
    \mu_{\min}\big(\overline{\mathbf{H}}_{\lambda f}\big) \geq& \, \mu_{\min}\bigg(\frac{\mathbf{D}_{\lambda f}^{\prime} \overline{\boldsymbol{\nabla}} \mathbf{D}_{\lambda f} + c_{\min} \mathbf{V}_{\lambda f} \mathbf{V}_{\lambda f}^{\prime}}{\sqrt{NT}}\bigg) + \frac{(1 - c_{\min})}{\sqrt{NT}} \, \mu_{\min}(\mathbf{V}_{\lambda f} \mathbf{V}_{\lambda f}^{\prime}) \\
    =& \, \mu_{\min}\bigg(\frac{\mathbf{D}_{\lambda f}^{\prime} \overline{\boldsymbol{\nabla}} \mathbf{D}_{\lambda f} + c_{\min} \mathbf{V}_{\lambda f} \mathbf{V}_{\lambda f}^{\prime}}{\sqrt{NT}}\bigg) \, ,
\end{align*}
where we use $(1 - c_{\min}) \geq 0$ and $\lambda_{\min}(\mathbf{V}_{\lambda f} \mathbf{V}_{\lambda f}^{\prime}) = 0$. Hence,
\begin{align*}
    \mu_{\min}\big(\overline{\mathbf{H}}_{\lambda f}\big) \geq& \, c_{\min} \, \min_{\norm{v}_{2} = 1} v^{\prime} \bigg\{\frac{\mathbf{D}_{\lambda f}^{\prime} \mathbf{D}_{\lambda f} + \mathbf{V}_{\lambda f} \mathbf{V}_{\lambda f}^{\prime}}{\sqrt{NT}}\bigg\} v \\
    =& \, c_{\min} \, \mu_{\min}\bigg(\frac{\mathbf{D}_{\lambda f}^{\prime} \mathbf{D}_{\lambda f} + \mathbf{V}_{\lambda f} \mathbf{V}_{\lambda f}^{\prime}}{\sqrt{NT}}\bigg) > 0 \; \text{wpa1} \, ,
\end{align*}
where the last inequality follows from
\begin{equation*}
    \mu_{\min}((\mathbf{D}_{\lambda f}^{\prime} \mathbf{D}_{\lambda f} + \mathbf{V}_{\lambda f} \mathbf{V}_{\lambda f}^{\prime})/\sqrt{NT}) > 0 \; \text{wpa1}
\end{equation*}
by the assumptions of \textcite{mw2017}. Lower bounds for $\mu_{\min}(\overline{\mathbf{H}}_{\lambda})$ and $\mu_{\min}(\overline{\mathbf{H}}_{f})$ follow analogously. By similar arguments,
\begin{align*}
    &\inf_{i, N} \mu_{\min}(\overline{\boldsymbol{\Phi}}_{i} / T) \geq c_{\min} \, \mu_{\min}(\mathbf{F}^{\prime} \mathbf{F} / T)  > 0 \; \text{wpa1} \, , \\
    &\inf_{t, T} \mu_{\min}(\overline{\boldsymbol{\Psi}}_{t} / N) \geq c_{\min} \, \mu_{\min}(\boldsymbol{\Lambda}^{\prime} \boldsymbol{\Lambda} / N) > 0 \; \text{wpa1} \, .
\end{align*}

\vspace{0.5em}
\noindent iii) follows from $\EX[\delta_{it} \mid \mathcal{C}] \geq c_{\min} > 0$ a.\,s.\ uniformly over $i, t, N, T$ and the fact that $\delta_{it}$ is a binary indicator. In particular,
\begin{equation*}
    \norm{\overline{\nabla \mathbf{x}}}_{q} = \bigg(\sum_{i = 1}^{N} \sum_{t = 1}^{T} \lvert\EX[\delta_{it} \mid \mathcal{C}]\rvert^{q} \lvert\EX[x_{it} \mid \mathcal{C}]\rvert^{q} \bigg)^{\frac{1}{q}} \leq \bigg(\sum_{i = 1}^{N} \sum_{t = 1}^{T} \EX[\lvert x_{it} \mid \mathcal{C}\rvert^{q}] \bigg)^{\frac{1}{q}} = \mathcal{O}_{P}\big((NT)^{1 / q}\big) \, .
\end{equation*}
The bound for $\norm{\overline{\nabla \mathbf{e} \odot \mathbf{e}}}_{q}$ follows analogously. By the definition of the operator norm,
\begin{equation*}
    \norm{\overline{\mathbf{H}}_{\lambda f}^{- 1}}_{q} = \, \bigg\{\min_{\norm{v}_{q} = 1} \, \norm{\overline{\mathbf{H}}_{\lambda f} \, v}_{q}\bigg\}^{- 1} \leq c_{\min}^{- 1} \, \norm{((\mathbf{D}_{\lambda f}^{\prime} \mathbf{D}_{\lambda f} + \mathbf{V}_{\lambda f} \mathbf{V}_{\lambda f}^{\prime}/\sqrt{NT}))^{- 1}}_{q} = \mathcal{O}_{P}(1) \, ,
\end{equation*}
where the last inequality follows from
\begin{equation*}
    \norm{((\mathbf{D}_{\lambda f}^{\prime} \mathbf{D}_{\lambda f} + \mathbf{V}_{\lambda f} \mathbf{V}_{\lambda f}^{\prime}/\sqrt{NT}))^{- 1}}_{q} = \mathcal{O}_{P}(1)
\end{equation*}
by the assumptions of \textcite{mw2017}. All remaining bounds follow by analogous arguments.

\vspace{0.5em}
\noindent iv) follows from the techniques in the supplement of \textcite{fw2016} together with $\sum_{t^{\prime} = 1}^{T} \EX[\delta_{it^{\prime}} \delta_{it} \mid \mathcal{C}] - \EX[\delta_{it^{\prime}} \mid \mathcal{C}] \EX[\delta_{it} \mid \mathcal{C}] \leq c_{\max} < \infty$ a.\,s. uniformly over $i, t, N, T$. Recall,
\begin{equation*}
    \widetilde{\mathbf{H}}_{\lambda f} = \begin{pmatrix}
        \widetilde{\mathbf{H}}_{\lambda} & \mathbf{D}_{\lambda}^{\prime} \widetilde{\boldsymbol{\nabla}} \mathbf{D}_{f} / \sqrt{NT} \\
        \mathbf{D}_{f}^{\prime} \widetilde{\boldsymbol{\nabla}} \mathbf{D}_{\lambda} / \sqrt{NT}& \widetilde{\mathbf{H}}_{f}
    \end{pmatrix} = \begin{pmatrix}
        \mathbf{D}_{\lambda}^{\prime} \widetilde{\boldsymbol{\nabla}} \mathbf{D}_{\lambda} / \sqrt{NT} & \mathbf{D}_{\lambda}^{\prime} \widetilde{\boldsymbol{\nabla}} \mathbf{D}_{f} / \sqrt{NT} \\
        \mathbf{D}_{f}^{\prime} \widetilde{\boldsymbol{\nabla}} \mathbf{D}_{\lambda} / \sqrt{NT}& \mathbf{D}_{f}^{\prime} \widetilde{\boldsymbol{\nabla}} \mathbf{D}_{f} / \sqrt{NT}
    \end{pmatrix} \, ,
\end{equation*}
where $\widetilde{\boldsymbol{\nabla}} = \diag(\delta_{11} - \EX[\delta_{11} \mid \mathcal{C}], \ldots, \delta_{NT} - \EX[\delta_{NT} \mid \mathcal{C}])$. By the definition of the matrix $q$-norm, $\norm{\widetilde{\mathbf{H}}_{\lambda f}}_{q} \leq \norm{\widetilde{\mathbf{H}}_{\lambda}}_{q} + \norm{\widetilde{\mathbf{H}}_{f}}_{q} + \norm{\mathbf{D}_{\lambda}^{\prime} \widetilde{\boldsymbol{\nabla}} \mathbf{D}_{f} / \sqrt{NT}}_{q} + \norm{\mathbf{D}_{f}^{\prime} \widetilde{\boldsymbol{\nabla}} \mathbf{D}_{\lambda} / \sqrt{NT}}_{q}$. Applying Lemma S.4 of \textcite{fw2016},
\begin{align*}
    &\norm{\widetilde{\mathbf{H}}_{\lambda f}}_{q} \leq \norm{\widetilde{\mathbf{H}}_{\lambda}}_{\infty} + \norm{\widetilde{\mathbf{H}}_{f}}_{\infty} + \norm{\mathbf{D}_{f}^{\prime} \widetilde{\boldsymbol{\nabla}} \mathbf{D}_{\lambda} / \sqrt{NT}}_{2}^{\frac{2}{q}} \norm{\mathbf{D}_{f}^{\prime} \widetilde{\boldsymbol{\nabla}} \mathbf{D}_{\lambda} / \sqrt{NT}}_{1}^{1 - \frac{2}{q}} \, + \\
    &\qquad \norm{\mathbf{D}_{f}^{\prime} \widetilde{\boldsymbol{\nabla}} \mathbf{D}_{\lambda} / \sqrt{NT}}_{2}^{\frac{2}{q}} \norm{\mathbf{D}_{f}^{\prime} \widetilde{\boldsymbol{\nabla}} \mathbf{D}_{\lambda} / \sqrt{NT}}_{\infty}^{1 - \frac{2}{q}} \, .
\end{align*}
$\widetilde{\mathbf{H}}_{\lambda}$ and $\widetilde{\mathbf{H}}_{f}$ are block diagonal (with $R \times R$ blocks), or can be permuted to become so. Hence, $\norm{\widetilde{\mathbf{H}}_{\lambda}}_{\infty} = \mathcal{O}_{P}((NT)^{- 3 / 16})$ and $\norm{\widetilde{\mathbf{H}}_{f}}_{\infty} = \mathcal{O}_{P}((NT)^{- 3 / 16})$ follow from a suitable moment bound combined with Assumption 7 of \textcite{mw2017} ($\norm{\boldsymbol{\lambda}_{i}}_{2}$ and $\norm{\mathbf{f}_{t}}_{2}$ are uniformly bounded over $i, t, N, T$). $\mathbf{D}_{f}^{\prime} \widetilde{\boldsymbol{\nabla}} \mathbf{D}_{\lambda} / \sqrt{NT}$ is a $NR \times TR$ matrix whose entries have mean zero; it can be viewed as an $R \times R$ block matrix, where each block is an $N \times T$ matrix with independent coordinate rows conditional on $\mathcal{C}$. By Lemma S.6 of \textcite{fw2016}, $\norm{\mathbf{D}_{f}^{\prime} \widetilde{\boldsymbol{\nabla}} \mathbf{D}_{\lambda} / \sqrt{NT}}_{2} = \mathcal{O}_{P}((NT)^{- 3 / 16})$. Moreover, $\norm{\mathbf{D}_{f}^{\prime} \widetilde{\boldsymbol{\nabla}} \mathbf{D}_{\lambda} / \sqrt{NT}}_{1} = o_{P}((NT)^{1 / 16})$ and $\norm{\mathbf{D}_{f}^{\prime} \widetilde{\boldsymbol{\nabla}} \mathbf{D}_{\lambda} / \sqrt{NT}}_{\infty} = o_{P}((NT)^{1 / 16})$ follow from our assumptions. Consequently, $\norm{\widetilde{\mathbf{H}}_{\lambda f}}_{q} = o_{P}((NT)^{- 1 / (4 q)})$, $\norm{\widetilde{\mathbf{H}}_{\lambda}}_{q} = o_{P}((NT)^{- 1 / (4 q)})$, and $\norm{\widetilde{\mathbf{H}}_{f}}_{q} = o_{P}((NT)^{- 1 / (4 q)})$ for $2 \leq q \leq 4$. In addition, for each $i, N$, $\widetilde{\boldsymbol{\Phi}}_{i} / T = \tfrac{1}{T} \sum_{t = 1}^{T} \big(\delta_{i^{\prime} t} \delta_{it} -  \EX[\delta_{i^{\prime} t} \delta_{it} \mid \mathcal{C}]\big) \mathbf{f}_{t} \mathbf{f}_{t}^{\prime}$ is a symmetric $R \times R$ matrix with mean-zero entries. By similar arguments (and the union bound), $\sup_{i, N} \norm{\widetilde{\boldsymbol{\Phi}}_{i} / T}_{q} \leq \sup_{i, N} \norm{\widetilde{\boldsymbol{\Phi}}_{i} / T}_{\infty} = o_{P}(1)$. Analogously, $\sup_{t, T} \norm{\widetilde{\boldsymbol{\Psi}}_{t} / N}_{q} \leq \sup_{t, T} \norm{\widetilde{\boldsymbol{\Psi}}_{t} / N}_{\infty} = o_{P}(1)$.

\vspace{0.5em}
\noindent v) follows immediately from Assumption 7 of \textcite{mw2017}, Lemma S.4 of \textcite{fw2016}, and the inequality $\norm{A \otimes B}_{q} \leq \norm{A}_{q} \norm{B}_{q}$.

\vspace{0.5em}
\noindent vi) and vii) follow from suitable moment bounds, $\delta_{it} \in \{0, 1\}$, and Assumptions \ref{ass:missing_data_stochastic} i)--iii).

\vspace{0.5em}
\noindent viii) follows by the same arguments as ii).

\subsection{Numerical Comparison of Matrix Completion Algorithms}
\label{app:num_comp_mc}

We use a small-scale simulation to compare Algorithms \ref{alg:mc_em} and \ref{alg:mc_redebias} in terms of speed and predictive accuracy. We consider a two-factor ($R = 2$) model,
\begin{equation*}
    y_{it} = \boldsymbol{\lambda}_{i}^{\prime} \mathbf{f}_{t}^{\phantom{\prime}} + e_{it} \, ,
\end{equation*}
where the idiosyncratic error $e_{it}$ is heteroskedastic across cross-sectional units. Specifically, $e_{it} = (u_{it} - 5) / \sqrt{5}$ if $i$ is odd and $e_{it} = 2 \, (u_{it} - 5) / \sqrt{10}$ otherwise, where $u_{it}$ is drawn independently from the $\chi^{2}$-distribution with five degrees of freedom. The factor structure is constructed from $\lambda_{ir} \sim \iid \N(1, 1)$ and $f_{tr} = \rho \, f_{(t-1)r} + v_{tr}$, where $v_{tr} \sim \iid \N(0, (1 - \rho^2) \sigma^2)$ and $\rho = \sigma = 0.5$. We discard the first $1{,}000$ time periods to ensure that the simulated data are drawn from the stationary distribution of the model. All random variables are redrawn in each replication, and all results are based on $1{,}000$ replications.

The missing data pattern follows Section \ref{sec:simulation}: observations are conditionally missing at random, with missing probabilities that are homogeneous across $i$ but heterogeneous across $t$. The total sample size is $n = NT(1 - \psi)$, with $N = \overline{N} / (1 - \psi)$ and $T = \overline{T} / (1 - \psi)$. We set $\psi = 0.4$ and consider panels with $N = T \in \{100, 200, 300\}$, corresponding to $\overline{N} = \overline{T} \in \{60, 120, 180\}$.

We apply Algorithms \ref{alg:mc_em} (EM) and \ref{alg:mc_redebias} to complete the matrix $\mathbf{Y}$, where $\mathbf{Y}$ is an $N \times T$ matrix with elements $[\mathbf{Y}]_{it} = y_{it}$ and missing entries at unobserved index pairs. Let $\mathbf{Y}^{\ast}$ denote the completed matrix. Then $\widehat{\mathbf{F}}$ equals the first $R$ eigenvectors of $\mathbf{Y}^{\ast\prime} \mathbf{Y}^{\ast}$ multiplied by $\sqrt{T}$, and $\widehat{\boldsymbol{\Lambda}} = \mathbf{Y}^{\ast} \widehat{\mathbf{F}} / T$.

For Algorithm \ref{alg:mc_redebias}, we consider two approaches to select the tuning parameter $\nu$: a cross-validation approach (CV) proposed by \textcite{abdik2021} and a plug-in approach (PI) proposed by \textcites{chlz2019}{chlz2023}. For CV, we consider a decreasing sequence of 30 candidate values starting from $\sigma_{\max}$, the largest singular value of $\mathcal{P}_{D}(\mathbf{Y})$; the cross-validation error (average squared prediction error) is computed using five random subsets. For PI, we set $\nu$ equal to the 0.95th quantile of $2.2 \, \norm{\mathcal{P}_{\mathcal{D}}(\mathbf{Z})}_{2}$, where $\mathbf{Z}$ is an $N \times T$ matrix with elements $[\mathbf{Z}]_{it} = z_{it}$ and $z_{it} \iid \N(0, \sigma_{e}^{2})$. We estimate this quantile by simulation ($1{,}000$ replications), using the sample variance of $\mathbf{y}$ as an initial estimate of $\sigma_{e}^{2}$. The maximum number of iterations is 15. For further details, we refer to the replication package.

We compare computational speed across the three approaches. For Algorithm \ref{alg:mc_redebias}, speed depends on the choice of $\nu$: larger values yield faster convergence. Note that EM is a special case of Algorithm \ref{alg:mc_redebias} with $\nu = 0$, i.e., without regularization or debiasing. To facilitate comparison, we also report the average selected $\nu$, where $\nu_{\text{CV}}$ and $\nu_{\text{PI}}$ denote the values selected by CV and PI, respectively. We report computation time relative to EM to abstract from the unit of measurement. For each replication, computation time is recorded as the median over multiple calls. Results for $N = 100$ are omitted because the runtimes are too short to yield reliable estimates.

Table \ref{tab:tuning_time} reports the tuning parameter values and relative computation times.
\begin{table}[!htbp]
	\centering
	\begin{threeparttable}
		\caption{Tuning Parameter Selection and Computation Time}
		\label{tab:tuning_time}
		\begin{tabular}{@{}*{2}{l}*{4}{c}@{}}
		        \toprule
 $N = T$&$\overline{N}=\overline{T}$&$\nu_{\text{CV}}$&$\nu_{\text{PI}}$&Time CV / Time EM&Time PI / Time EM \\
 \midrule
200 & 120 & 9.2487 & 48.5255 & 0.5791 & 0.3434 \\ 
  300 & 180 & 8.7290 & 58.7705 & 0.6550 & 0.3899 \\ 
   \bottomrule
		\end{tabular}
		\begin{tablenotes}
			\footnotesize
			\item \emph{Note:} $\nu_{\text{CV}}$ and $\nu_{\text{PI}}$ refer to the average value of $\nu$ selected by CV and PI, respectively. Time CV / Time EM and Time PI / Time EM refer to computation time of CV and PI relative to EM, respectively. $\psi = 0.4$. Results are based on $1{,}000$ replications.
		\end{tablenotes}
	\end{threeparttable}
\end{table}
CV consistently selects a smaller tuning parameter than PI. The value selected by CV remains approximately 9 as the sample size grows, whereas PI selects larger values with increasing sample size. Recall from Remark 3 that the theoretical requirement is $\nu > c_{\nu} \sqrt{N}$ for some $c_{\nu} > 0$ \parencite{chlz2019, chlz2023}. In our experiments, PI yields $c_{\nu} \approx 3.5$ in both settings, while CV yields $c_{\nu} \approx 0.6$ for $N = 200$ and $c_{\nu} \approx 0.5$ for $N = 300$. Thus, $c_{\nu}$ is stable for PI but declines for CV as the sample size increases. Because $\nu_{\text{PI}} > \nu_{\text{CV}}$, PI also converges fastest, as expected: PI requires roughly 35\% of the computation time of EM, while CV requires roughly 60\%. Consistent with \textcite{fll2021}, we find that Algorithm \ref{alg:mc_redebias} outperforms Algorithm \ref{alg:mc_em} in terms of computational speed.

We also compare predictive performance. Specifically, we compare the true factor structure ($\boldsymbol{\lambda}_{i}^{\prime} \mathbf{f}_{t}^{\phantom{\prime}}$) with its estimates obtained after matrix completion. This comparison is natural because IFE estimation in unbalanced panels rests on imputing missing entries via the estimated factor structure; see the decomposition in \eqref{eq:ifemodel_decomposition}. We evaluate performance using the mean absolute deviation (Bias) and the root mean squared error (RMSE), computed separately for all index pairs, observed index pairs, and unobserved index pairs. For example, letting $\hat{\boldsymbol{\lambda}}_{i}^{\prime} \hat{\mathbf{f}}_{t}^{\phantom{\prime}}$ denote the estimated factor structure, the three bias measures are
\begin{align*}
    (NT)^{- 1} \sum_{i = 1}^{N} \sum_{t = 1}^{T} \lvert \hat{\boldsymbol{\lambda}}_{i}^{\prime} \hat{\mathbf{f}}_{t}^{\phantom{\prime}} - \boldsymbol{\lambda}_{i}^{\prime} \mathbf{f}_{t}^{\phantom{\prime}}\rvert, \;
    \lvert \mathcal{D} \rvert^{- 1} \sum_{(i, t) \in \mathcal{D}} \lvert \hat{\boldsymbol{\lambda}}_{i}^{\prime} \hat{\mathbf{f}}_{t}^{\phantom{\prime}} - \boldsymbol{\lambda}_{i}^{\prime} \mathbf{f}_{t}^{\phantom{\prime}}\rvert, \; \text{and} \;
    (NT - \lvert \mathcal{D} \rvert)^{- 1} \sum_{(i, t) \notin \mathcal{D}} \lvert \hat{\boldsymbol{\lambda}}_{i}^{\prime} \hat{\mathbf{f}}_{t}^{\phantom{\prime}} - \boldsymbol{\lambda}_{i}^{\prime} \mathbf{f}_{t}^{\phantom{\prime}}\rvert \, .
\end{align*}
The RMSE measures are computed analogously.

Table \ref{tab:prediction_performance} reports the predictive performance of the three approaches.
\begin{table}[!htbp]
	\centering
	\begin{threeparttable}
		\caption{Predictive Performance of Different Matrix Completion Approaches}
		\label{tab:prediction_performance}
		\begin{tabular}{@{}*{2}{l}*{6}{c}@{}}
		        \toprule
 $N = T$&$\overline{N}=\overline{T}$&\multicolumn{3}{c}{Bias}&\multicolumn{3}{c}{RMSE}\\
 \cmidrule(lr){3-8}
 &&EM&CV&PI&EM&CV&PI\\
 \midrule
&&\multicolumn{6}{c}{All Entries}\\
\cmidrule(lr){3-8}
100 &  60 & 0.4013 & 0.3802 & 0.3727 & 0.5643 & 0.5182 & 0.5053 \\ 
  200 & 120 & 0.2606 & 0.2566 & 0.2531 & 0.3541 & 0.3468 & 0.3411 \\ 
  300 & 180 & 0.2080 & 0.2061 & 0.2040 & 0.2812 & 0.2778 & 0.2743 \\ 
   \midrule
&&\multicolumn{6}{c}{Observed Entries}\\
\cmidrule(lr){3-8}
100 &  60 & 0.3733 & 0.3670 & 0.3640 & 0.5058 & 0.4941 & 0.4891 \\ 
  200 & 120 & 0.2497 & 0.2485 & 0.2472 & 0.3343 & 0.3322 & 0.3300 \\ 
  300 & 180 & 0.2004 & 0.1998 & 0.1990 & 0.2676 & 0.2666 & 0.2653 \\ 
   \midrule
&&\multicolumn{6}{c}{Missing Entries}\\
\cmidrule(lr){3-8}
100 &  60 & 0.4433 & 0.3999 & 0.3857 & 0.6380 & 0.5516 & 0.5280 \\ 
  200 & 120 & 0.2770 & 0.2688 & 0.2621 & 0.3815 & 0.3674 & 0.3567 \\ 
  300 & 180 & 0.2194 & 0.2156 & 0.2114 & 0.3003 & 0.2937 & 0.2872 \\ 
   \bottomrule
		\end{tabular}
		\begin{tablenotes}
			\footnotesize
			\item \emph{Note:} Bias denotes the mean absolute deviation from the true factor structure. RMSE denotes the root mean squared error. All Entries, Observed Entries, and Missing Entries refer to performance measures for all estimates of $\boldsymbol{\lambda}_{i}^{\prime} \mathbf{f}_{t}^{\phantom{\prime}}$, estimates for observed index pairs only, and estimates for unobserved index pairs only, respectively. $\psi = 0.4$. Results are based on $1{,}000$ replications.
		\end{tablenotes}
	\end{threeparttable}
\end{table}
Both CV and PI outperform EM, with the largest differences occurring at $N = 100$ and for missing entries (precisely the values of primary interest). At $N = 300$, the performance of all three approaches is nearly identical. Based on these results, we rank the approaches as follows: 1.\ PI, 2.\ CV, 3.\ EM.

This ranking is noteworthy. PI rests on the assumption that $e_{it}$ is iid.\ normal, which is violated here: errors are heteroskedastic across cross-sectional units and drawn from a right-skewed distribution. Despite this misspecification, PI performs best. Our findings therefore differ from those of \textcite{fll2021}, who find no difference in predictive performance between Algorithm \ref{alg:mc_em} and Algorithm \ref{alg:mc_redebias}. The discrepancy may reflect differences in the data-generating process, the missing data pattern, and the estimand considered.

We close with two remarks on the implications for IFE estimation. First, the algorithms above are compared in isolation using a pure factor model with $R = 2$. In the IFE estimator, matrix completion is called repeatedly during optimization. Although Algorithm \ref{alg:mc_redebias} offers substantial speed gains in isolation, we do not observe these gains in our IFE estimation procedure: the optimizer requires considerably more function evaluations to converge when Algorithm \ref{alg:mc_redebias} is used as the inner routine. This may differ for other IFE estimation procedures not considered here (see Remark 4). Second, although Table \ref{tab:prediction_performance} reveals meaningful differences in predictive performance, these differences do not translate into relevant differences in IFE inference. Tables 7 and 8 in the Online Supplement \ref{os:additional_results} report simulation results using Algorithm \ref{alg:mc_redebias} in place of Algorithm \ref{alg:mc_em}; the results are virtually identical to those in Section \ref{sec:simulation}.

\subsection{Empirical Example -- Sensitivity Checks}
\label{app:empirical}
We consider two sensitivity checks. First, we examine sensitivity of the results to different bandwidth choices $L \in \{1, \ldots, 8\}$, as recommended by \textcites{fw2016}{fw2018}. Second, we report estimates for $R \in \{1, \ldots, 5\}$. As shown in \textcite{mw2015}, the inclusion of redundant common factors should affect only the precision of the IFE estimator, once all relevant common factors have been controlled for.

\begin{figure}[!htbp]
	\centering
	\caption{Sensitivity to Different Bandwidth Choices}
	\includegraphics[width=.9\textwidth]{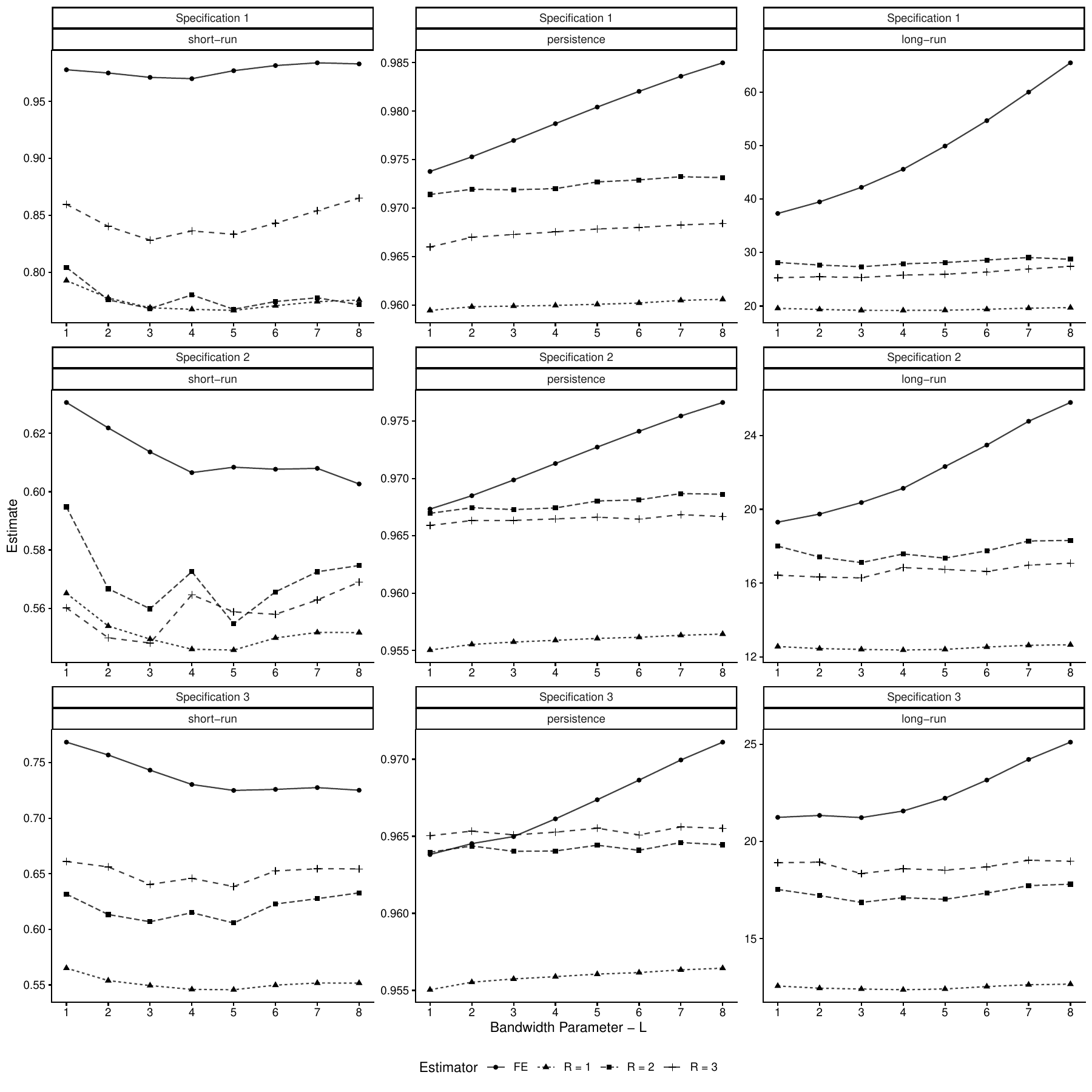}
	\begin{minipage}{.9\textwidth}
		\footnotesize
		\emph{Note:} Effect of democracy on logarithmic GDP per capita $(\times 100)$; FE denotes the debiased fixed effects estimator; $R = 1$, $R = 2$, and $R = 3$ denote  debiased IFE estimators with one, two, and three numbers of factors, respectively; Bandwidth choices $L \in \{1, \ldots, 8\}$.
	\end{minipage}
	\label{fig:sensitivity_bw}
\end{figure}

Figure \ref{fig:sensitivity_bw} shows the sensitivity of our results to different bandwidth choices.
The IFE estimates are remarkably stable across bandwidth choices, regardless of $R$. The fixed effects estimates are more sensitive: in particular, the estimated persistence of the GDP process increases substantially with $L$, which also inflates the long-run effects. For example, for $p = 1$, the implied long-run effects range from 37.305\% to 65.480\%.

\begin{figure}[!htbp]
	\centering
	\caption{Sensitivity to Number of Factors}
	\includegraphics[width=.9\textwidth]{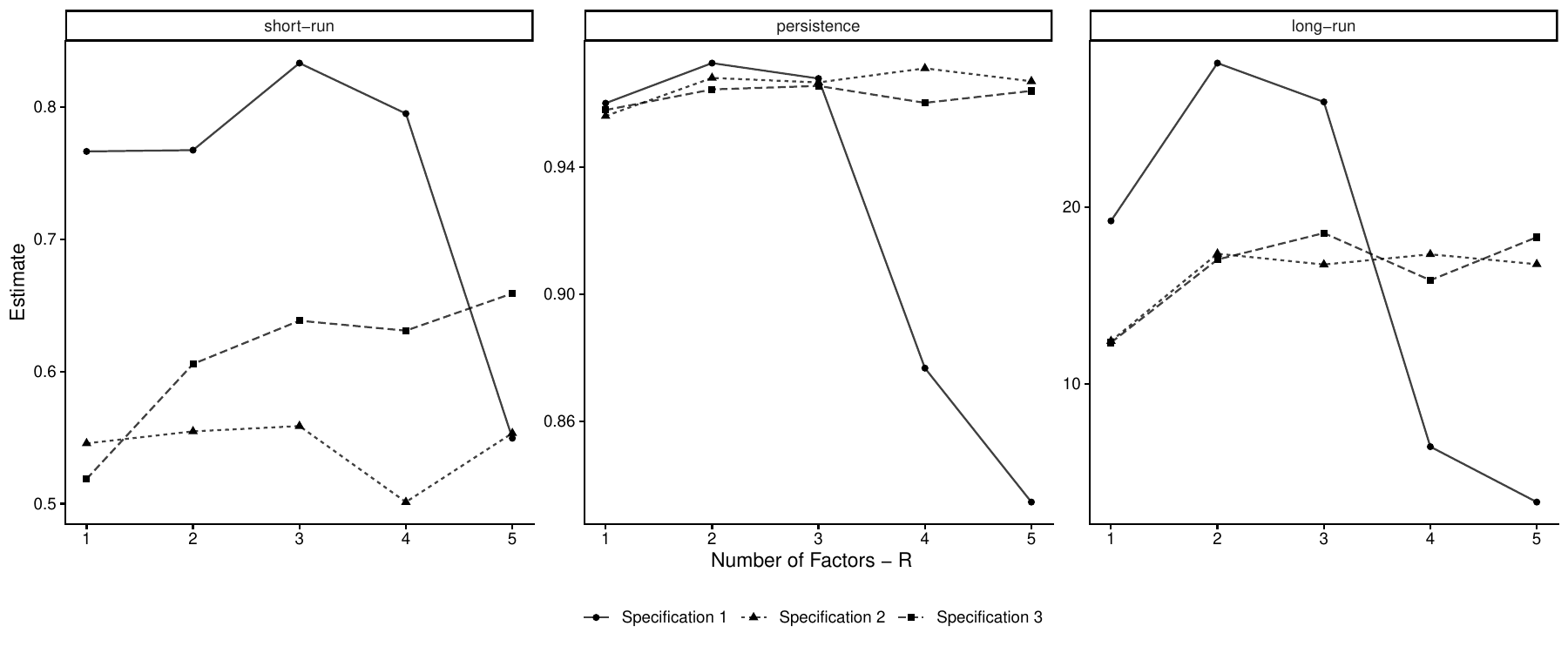}
	\begin{minipage}{.9\textwidth}
		\footnotesize
		\emph{Note:} Effect of democracy on logarithmic GDP per capita $(\times 100)$; results based on debiased IFE estimator for $R \in \{1, \ldots, 5\}$ with $L = 5$.
	\end{minipage}
	\label{fig:sensitivity_nof}
\end{figure}

Figure \ref{fig:sensitivity_nof} shows the sensitivity of the IFE estimates to different numbers of factors.
The estimates are quite stable across different values of $R$, with the exception of $p = 1$. For that specification, substantial drops in the estimated short-run effect and GDP persistence are observed after adding three and four factors, respectively.

\clearpage

\section{Online Supplement (Not for Publication)}

\subsection{Simulation Results for Regularized Matrix Completion}
\label{os:additional_results}

\begin{table}[!htbp]
	\centering
	\begin{threeparttable}
		\caption{Finite Sample Properties of $\tilde{\beta}$}
		\label{tab:finite_sample_properties_supp}
		\begin{tabular}{@{}*{2}{l}*{3}{c}@{}}
			    \toprule
 $\overline{T}$&$L$&\multicolumn{3}{c}{$\psi = 0.0 \; / \; \psi = 0.2 \; / \; \psi = 0.4$}\\
 \cmidrule(lr){3-5}
 &&Bias&Ratio&Size\\
 \midrule
&&\multicolumn{3}{c}{$\beta = 0.3$}\\
\cmidrule(lr){3-5}
  5 &   2 & -15.052 / -17.196 / -19.634 & 0.393 / 0.438 / 0.488 & 0.509 / 0.476 / 0.461 \\ 
   10 &   3 & -7.507 / -8.539 / -7.579 & 0.736 / 0.810 / 0.857 & 0.189 / 0.184 / 0.169 \\ 
   20 &   4 & -3.269 / -3.219 / -2.004 & 0.908 / 0.907 / 0.983 & 0.091 / 0.090 / 0.062 \\ 
   40 &   5 & -1.252 / -0.863 / -0.690 & 0.984 / 0.958 / 0.968 & 0.053 / 0.063 / 0.064 \\ 
   80 &   6 & -0.529 / -0.256 /  0.019 & 1.032 / 0.988 / 0.945 & 0.040 / 0.054 / 0.061 \\ 
   \midrule
&&\multicolumn{3}{c}{$\beta = 0.9$}\\
\cmidrule(lr){3-5}
  5 &   2 & -13.848 / -11.499 / -11.084 & 0.239 / 0.224 / 0.218 & 0.613 / 0.548 / 0.544 \\ 
   10 &   3 & -4.584 / -4.316 / -3.664 & 0.292 / 0.283 / 0.285 & 0.422 / 0.398 / 0.398 \\ 
   20 &   4 & -1.009 / -0.880 / -0.726 & 0.505 / 0.526 / 0.549 & 0.212 / 0.215 / 0.194 \\ 
   40 &   5 & -0.210 / -0.195 / -0.159 & 0.809 / 0.865 / 0.901 & 0.102 / 0.083 / 0.076 \\ 
   80 &   6 & -0.067 / -0.087 / -0.041 & 0.965 / 0.954 / 0.962 & 0.049 / 0.063 / 0.065 \\ 
   \bottomrule
		\end{tabular}
		\begin{tablenotes}
			\footnotesize
			\item \emph{Note:} $\overline{N} = 100$ and $L$ is a bandwidth parameter; $\psi$ denotes the share of missing observations; Bias refers to relative biases in percentage, Ratio denotes the average ratios of standard errors to standard deviations, and Size is the empirical size of $z$-tests with 5\% nominal size; results are based on $1{,}000$ replications.
		\end{tablenotes}
	\end{threeparttable}
\end{table}
\begin{table}[!htbp]
	\centering
	\begin{threeparttable}
		\caption{Average of $\widehat{R}$}
		\label{tab:number_of_factors_supp}
		\begin{tabular}{@{}*{2}{l}*{3}{c}@{}}
            \toprule
 $\overline{T}$&$\overline{R}$&\multicolumn{3}{c}{$\psi = 0.0 \; / \; \psi = 0.2 \; / \; \psi = 0.4$}\\
 \cmidrule(lr){3-5}
 &&$\text{IC}_{2}$&$\text{BIC}_{3}$&ER\\
 \midrule
&&\multicolumn{3}{c}{$\beta = 0.3$}\\
\cmidrule(lr){3-5}
  5 &   2 & 2.000 / 1.926 / 1.756 & 0.652 / 0.449 / 0.432 & 0.919 / 0.859 / 0.858 \\ 
   10 &   5 & 4.463 / 2.296 / 2.038 & 2.005 / 1.369 / 1.050 & 1.004 / 0.922 / 0.928 \\ 
   20 &  10 & 2.318 / 2.235 / 1.561 & 2.896 / 1.647 / 1.246 & 0.823 / 0.979 / 0.995 \\ 
   40 &  10 & 1.007 / 1.007 / 1.020 & 1.005 / 1.002 / 1.004 & 0.996 / 1.001 / 1.002 \\ 
   80 &  10 & 1.001 / 1.001 / 1.004 & 1.001 / 1.001 / 1.001 & 1.001 / 1.001 / 1.000 \\ 
   \midrule
&&\multicolumn{3}{c}{$\beta = 0.9$}\\
\cmidrule(lr){3-5}
  5 &   2 & 1.988 / 1.863 / 1.715 & 1.228 / 1.002 / 0.958 & 0.863 / 0.917 / 0.969 \\ 
   10 &   5 & 3.425 / 3.237 / 3.274 & 3.247 / 2.590 / 2.336 & 1.143 / 1.184 / 1.239 \\ 
   20 &  10 & 6.046 / 5.377 / 5.221 & 5.701 / 4.370 / 3.976 & 1.400 / 1.270 / 1.214 \\ 
   40 &  10 & 6.218 / 5.855 / 6.002 & 4.745 / 4.304 / 4.164 & 1.353 / 1.154 / 1.052 \\ 
   80 &  10 & 3.727 / 2.755 / 2.446 & 2.980 / 2.019 / 1.697 & 1.015 / 1.001 / 1.004 \\
   \midrule
   &&GR&ED&PA\\
 \midrule
&&\multicolumn{3}{c}{$\beta = 0.3$}\\
\cmidrule(lr){3-5}
  5 &   2 & 0.857 / 0.865 / 0.893 & 0.627 / 0.700 / 0.908 & 0.746 / 0.788 / 0.947 \\ 
   10 &   5 & 0.920 / 0.954 / 0.959 & 0.752 / 0.952 / 1.088 & 1.094 / 1.179 / 1.207 \\ 
   20 &  10 & 0.872 / 0.991 / 0.998 & 1.001 / 1.103 / 1.104 & 1.350 / 1.296 / 1.209 \\ 
   40 &  10 & 0.996 / 1.003 / 1.003 & 1.091 / 1.091 / 1.171 & 1.040 / 1.010 / 1.039 \\ 
   80 &  10 & 1.001 / 1.001 / 1.000 & 1.089 / 1.116 / 1.487 & 1.001 / 1.000 / 1.038 \\ 
   \midrule
&&\multicolumn{3}{c}{$\beta = 0.9$}\\
\cmidrule(lr){3-5}
  5 &   2 & 0.992 / 0.979 / 1.037 & 1.283 / 1.188 / 1.312 & 0.689 / 0.772 / 0.985 \\ 
   10 &   5 & 1.401 / 1.452 / 1.573 & 2.616 / 2.433 / 2.253 & 1.108 / 1.282 / 1.522 \\ 
   20 &  10 & 1.719 / 1.532 / 1.412 & 2.628 / 2.269 / 2.078 & 1.634 / 1.699 / 1.880 \\ 
   40 &  10 & 1.667 / 1.336 / 1.117 & 2.625 / 2.241 / 1.848 & 2.417 / 2.698 / 2.866 \\ 
   80 &  10 & 1.021 / 1.001 / 1.005 & 1.897 / 1.400 / 1.463 & 3.195 / 2.801 / 2.585 \\ 
   \bottomrule
	    \end{tabular}
	\begin{tablenotes}
	\footnotesize
	\item \emph{Note:} $\overline{N} = 100$; $\psi$ denotes the share of missing observations; $\text{IC}_2$ and $\text{BIC}_{3}$ denote the information criteria of \textcite{bn2002}, ER and GR are the estimators of \textcite{ah2013}, ED is the estimator of \textcite{o2010}, and PA is the parallel analysis described in \textcite{do2019}. The true number of factors is one. The initial estimator for $\beta$ uses $R = \overline{R}$ factors. Results are based on $1{,}000$ replications.
	\end{tablenotes}
	\end{threeparttable}
\end{table}

\clearpage

\subsection{Additional Simulation Experiments}
\label{os:additional_simulation}

We present additional Monte Carlo simulations to analyze the finite-sample properties of the debiased estimator $\tilde{\beta}$, defined in \eqref{eq:debiased_estimator}, in the presence of missing data. Specifically, we compare relative biases (\textit{Bias}), average ratios of standard errors to standard deviations (\textit{Ratio}), and empirical sizes of $z$-tests with a 5\% nominal size (\textit{Size}) across different shares of missing data ($\psi$) and relative to the balanced panel case. We use Algorithm \ref{alg:mc_em} as the matrix completion procedure for unbalanced panels. Because the number of factors is typically unknown, we also compare different estimators for this quantity. Specifically, we consider the estimators of \textcites{bn2002}{o2010}{ah2013}{do2019}. Of the information criteria introduced by \textcite{bn2002}, we focus on $\text{IC}_{2}$ and $\text{BIC}_{3}$, which are also used in \textcite{o2010} and \textcite{ah2013}. Performance is assessed by comparing the average estimated number of factors.

Following \textcite{mw2015}, we consider a static panel data model with one regressor and two factors:
\begin{eqnarray}
y_{it} &=&  \beta \, x_{it} + \sum_{r = 1}^{2} \lambda_{ir} f_{tr} + e_{it} \, , \nonumber \\
x_{it} &=&  1 + \sum_{r = 1}^{2} (\lambda_{ir} + \chi_{ir}) (f_{tr} + f_{t - 1,r}) + w_{it} \, , \nonumber
\end{eqnarray}
$i = 1, \ldots, N$, $t = 1, \ldots, T$, and $e_{it}$ is an idiosyncratic error term. The regressor $x_{it}$ is correlated with the common factors and their loadings. Throughout all experiments, $f_{tr}, w_{it} \sim \iid \N(0, 1)$ and $\lambda_{ir}, \chi_{ir} \sim \iid \N(1, 1)$.

We consider four configurations for the idiosyncratic error term: i) homoskedastic, ii) homoskedastic with fat tails, iii) heteroskedastic across units, and iv) heteroskedastic across units and over time. Specifically: i) $e_{it} \sim \iid \N(0, 4)$; ii) $e_{it} = \sqrt{12 / 5} \, u_{it}$, where $u_{it}$ follows a $t$-distribution with five degrees of freedom; iii) $e_{it} = u_{it}$, where $u_{it} \sim \iid \N(0, 2)$ if $i$ is odd and $u_{it} \sim \iid \N(0, 6)$ otherwise; iv) $e_{it} = u_{it} + v_{it}$, where $u_{it} \sim \iid \N(0, 1)$ if $i$ is odd and $u_{it} \sim \iid \N(0, 3)$ otherwise, and $v_{it} \sim \iid \N(0, 1)$ if $t$ is odd and $v_{it} \sim \iid \N(0, 3)$ otherwise.

\begin{figure}[!htbp]
	\centering
	\caption{Patterns of Randomly Missing Observations}
	\includegraphics[width=0.9\textwidth]{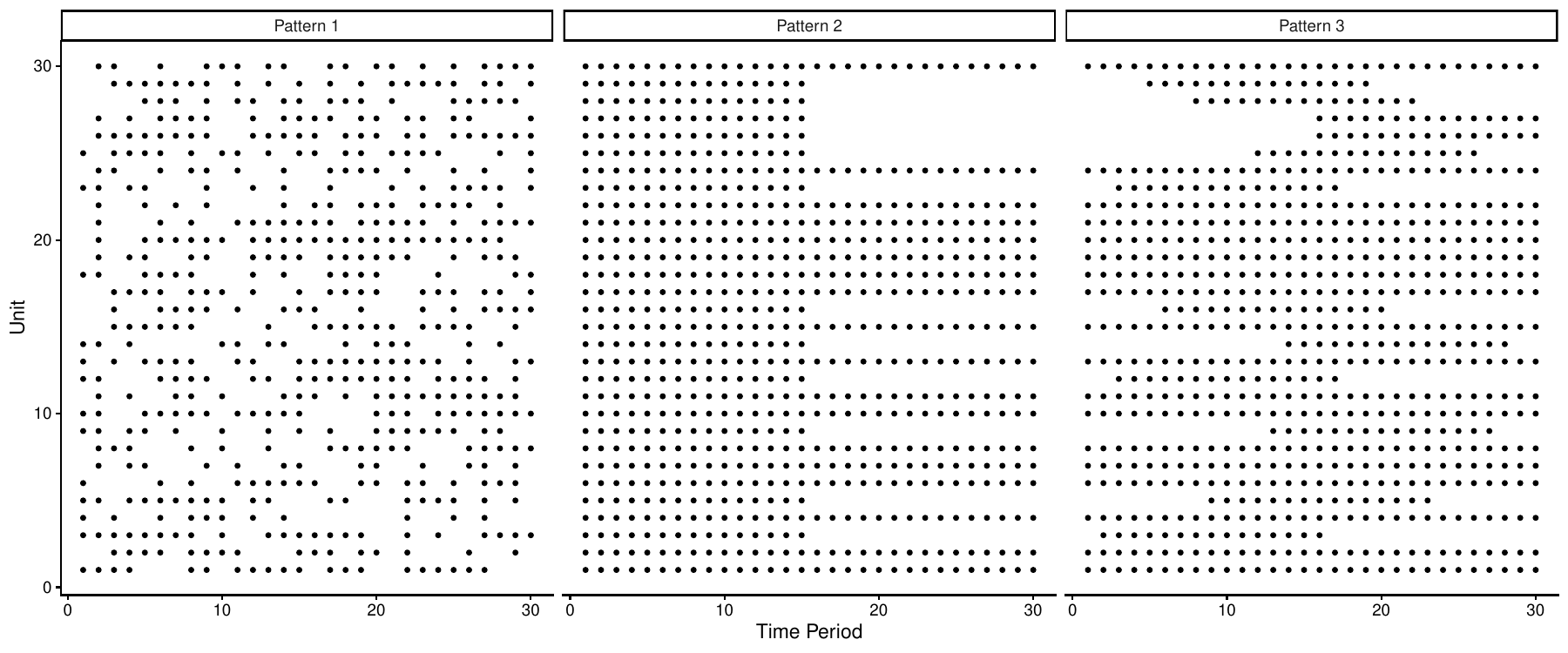}
	\label{fig:missing_supp}
\end{figure}

We consider three patterns in which a fraction $\psi \in \{0, 0.2, 0.4\}$ of observations is missing at random. The total sample size is $NT(1 - \psi)$. Figure \ref{fig:missing_supp} illustrates the three patterns.
In the first pattern, $NT\psi$ observations are dropped irregularly from the entire panel. This pattern is also studied by \textcite{bly2015} and mimics a setting in which survey respondents refuse or fail to answer certain questions. The other two patterns are borrowed from \textcite{cs2019} and reflect situations in which individuals either are or are not replaced after dropping out of a survey. To describe Patterns 2 and 3, we divide all individuals into two types. Type 1 consists of $N_{1} = 2 \psi N$ individuals observed for $T_{1} = T / 2$ time periods. The remaining $N_{2} = N - N_{1}$ individuals are of Type 2 and are observed over the entire time horizon ($T_{2} = T$). Patterns 2 and 3 differ only in the starting point of the time series for Type 1 individuals. In Pattern 2, all time series start at $t = 1$; in Pattern 3, the initial period is drawn with equal probability from $\{0, 1, \ldots, T - T_{1}\}$. All unbalanced datasets are generated from balanced panels by dropping observations according to the corresponding missing data pattern.

We consider panel datasets of different average sizes: $\overline{N} \in \{120, 240\}$ and $\overline{T} \in \{24, 48, 96\}$, where $N = \overline{N} / (1 - \psi)$ and $T = \overline{T} / (1 - \psi)$. All results are based on 500 replications and are summarized in Tables \ref{tab:prop_pattern1}--\ref{tab:nof_pattern3}.

First, we analyze the finite-sample properties of the IFE estimator. As in Section \ref{sec:simulation}, we apply the debiased estimator and its covariance matrix estimator exactly as described in Section \ref{sec:bc_inference} with $L = 0$, correcting for $\mathbf{B}_{2}$ and $\mathbf{B}_{3}$, and using a heteroskedasticity-robust covariance estimator. This approach yields a more realistic assessment of finite-sample performance in practice, where the true data-generating process is unknown and heteroskedasticity-robust inference is standard. The results are summarized in Tables \ref{tab:prop_pattern1}--\ref{tab:prop_pattern3}. Biases, ratios, and sizes are similar to those in the balanced case, irrespective of the fraction and pattern of missing data. As in Section \ref{sec:simulation}, the finite-sample performance of $\tilde{\beta}$ in unbalanced panels is well predicted by our theory.

\begin{table}[!htbp]
	\centering
	\begin{threeparttable}
		\caption{Properties of $\tilde{\beta}$ - Missing Data Pattern 1}
		\label{tab:prop_pattern1}
		\begin{tabular}{@{}*{2}{l}*{3}{c}@{}}
			\toprule
 $\overline{N}$&$\overline{T}$&\multicolumn{3}{c}{$\psi = 0.0 \; / \; \psi = 0.2 \; / \; \psi = 0.4$}\\
 \cmidrule(lr){3-5}
 &&Bias&Ratio&Size\\
 \midrule
&&\multicolumn{3}{c}{Homoskedastic}\\
\cmidrule(lr){3-5}
120 &  24 & 0.32 / 0.18 / 0.11 & 0.79 / 0.86 / 0.80 & 0.12 / 0.09 / 0.11 \\ 
  120 &  48 & 0.03 / 0.04 / 0.06 & 0.91 / 0.92 / 0.92 & 0.06 / 0.07 / 0.09 \\ 
  120 &  96 & -0.03 /  0.11 /  0.08 & 0.97 / 0.98 / 0.88 & 0.06 / 0.06 / 0.08 \\ 
  240 &  24 & 0.17 / 0.12 / 0.15 & 0.83 / 0.80 / 0.82 & 0.10 / 0.11 / 0.11 \\ 
  240 &  48 & 0.10 / 0.00 / 0.06 & 0.89 / 0.91 / 0.91 & 0.08 / 0.07 / 0.07 \\ 
  240 &  96 & 0.05 / 0.04 / 0.02 & 0.97 / 0.94 / 1.01 & 0.05 / 0.06 / 0.05 \\ 
   \midrule
&&\multicolumn{3}{c}{Homoskedastic with Fat Tails}\\
\cmidrule(lr){3-5}
120 &  24 & 0.32 / 0.35 / 0.32 & 0.81 / 0.71 / 0.77 & 0.12 / 0.13 / 0.12 \\ 
  120 &  48 & 0.16 / 0.08 / 0.07 & 0.87 / 0.90 / 0.91 & 0.08 / 0.08 / 0.07 \\ 
  120 &  96 & 0.06 / 0.06 / 0.06 & 0.91 / 0.92 / 0.88 & 0.07 / 0.06 / 0.08 \\ 
  240 &  24 & 0.21 / 0.16 / 0.12 & 0.87 / 0.90 / 0.86 & 0.10 / 0.08 / 0.09 \\ 
  240 &  48 & 0.06 / 0.07 / 0.05 & 0.93 / 0.94 / 0.93 & 0.06 / 0.06 / 0.06 \\ 
  240 &  96 & 0.02 / 0.03 / 0.04 & 0.94 / 0.91 / 0.95 & 0.06 / 0.09 / 0.07 \\ 
   \midrule
&&\multicolumn{3}{c}{Heteroskedastic across Units}\\
\cmidrule(lr){3-5}
120 &  24 & 0.14 / 0.28 / 0.29 & 0.84 / 0.82 / 0.82 & 0.10 / 0.11 / 0.12 \\ 
  120 &  48 & 0.10 / 0.03 / 0.10 & 0.88 / 0.93 / 0.88 & 0.07 / 0.07 / 0.09 \\ 
  120 &  96 & 0.10 / 0.10 / 0.08 & 0.96 / 0.99 / 0.93 & 0.07 / 0.07 / 0.09 \\ 
  240 &  24 & 0.20 / 0.02 / 0.22 & 0.92 / 0.83 / 0.85 & 0.08 / 0.11 / 0.10 \\ 
  240 &  48 & 0.06 / 0.07 / 0.11 & 0.95 / 0.99 / 0.94 & 0.07 / 0.05 / 0.05 \\ 
  240 &  96 & 0.03 / 0.02 / 0.01 & 0.92 / 0.93 / 1.00 & 0.07 / 0.07 / 0.04 \\ 
   \midrule
&&\multicolumn{3}{c}{Heteroskedastic across Units and over Time}\\
\cmidrule(lr){3-5}
120 &  24 & 0.10 / 0.23 / 0.31 & 0.81 / 0.79 / 0.82 & 0.13 / 0.14 / 0.12 \\ 
  120 &  48 & 0.07 / 0.08 / 0.21 & 0.87 / 0.91 / 0.91 & 0.09 / 0.07 / 0.09 \\ 
  120 &  96 & 0.07 / 0.01 / 0.08 & 0.94 / 0.91 / 0.94 & 0.07 / 0.07 / 0.07 \\ 
  240 &  24 & 0.26 / 0.23 / 0.15 & 0.85 / 0.84 / 0.82 & 0.10 / 0.11 / 0.10 \\ 
  240 &  48 & 0.07 / 0.05 / 0.02 & 0.91 / 0.95 / 0.94 & 0.07 / 0.06 / 0.06 \\ 
  240 &  96 & 0.04 / 0.03 / 0.00 & 0.93 / 0.96 / 0.92 & 0.07 / 0.06 / 0.06 \\ 
   \bottomrule
		\end{tabular}
		\begin{tablenotes}
			\footnotesize
			\item \emph{Note:} $\psi$ denotes the share of missing observations; Bias refers to relative biases in percentage, Ratio denotes the average ratios of standard errors to standard deviations, and Size is the empirical size of $z$-tests with 5\% nominal size; results are based on $500$ replications.
		\end{tablenotes}
	\end{threeparttable}
\end{table}

\begin{table}[!htbp]
	\centering
	\begin{threeparttable}
		\caption{Properties of $\tilde{\beta}$ - Missing Data Pattern 2}
		\label{tab:prop_pattern2}
		\begin{tabular}{@{}*{2}{l}*{3}{c}@{}}
			\toprule
 $\overline{N}$&$\overline{T}$&\multicolumn{3}{c}{$\psi = 0.0 \; / \; \psi = 0.2 \; / \; \psi = 0.4$}\\
 \cmidrule(lr){3-5}
 &&Bias&Ratio&Size\\
 \midrule
&&\multicolumn{3}{c}{Homoskedastic}\\
\cmidrule(lr){3-5}
120 &  24 & 0.32 / 0.21 / 0.31 & 0.79 / 0.82 / 0.83 & 0.12 / 0.11 / 0.11 \\ 
  120 &  48 & 0.03 / 0.08 / 0.13 & 0.91 / 0.91 / 0.87 & 0.06 / 0.08 / 0.09 \\ 
  120 &  96 & -0.03 /  0.07 /  0.11 & 0.97 / 0.93 / 0.95 & 0.06 / 0.08 / 0.07 \\ 
  240 &  24 & 0.17 / 0.12 / 0.05 & 0.83 / 0.84 / 0.83 & 0.10 / 0.11 / 0.10 \\ 
  240 &  48 & 0.10 / 0.06 / 0.05 & 0.89 / 0.94 / 0.95 & 0.08 / 0.08 / 0.05 \\ 
  240 &  96 & 0.05 / 0.04 / 0.03 & 0.97 / 0.97 / 0.96 & 0.05 / 0.05 / 0.06 \\ 
   \midrule
&&\multicolumn{3}{c}{Homoskedastic with Fat Tails}\\
\cmidrule(lr){3-5}
120 &  24 & 0.32 / 0.31 / 0.28 & 0.81 / 0.78 / 0.78 & 0.12 / 0.13 / 0.09 \\ 
  120 &  48 & 0.16 / 0.13 / 0.13 & 0.87 / 0.92 / 0.90 & 0.08 / 0.08 / 0.10 \\ 
  120 &  96 & 0.06 / 0.02 / 0.04 & 0.91 / 0.95 / 1.00 & 0.07 / 0.06 / 0.05 \\ 
  240 &  24 & 0.21 / 0.13 / 0.11 & 0.87 / 0.85 / 0.85 & 0.10 / 0.11 / 0.09 \\ 
  240 &  48 &  0.06 / -0.01 /  0.05 & 0.93 / 0.91 / 0.87 & 0.06 / 0.07 / 0.10 \\ 
  240 &  96 & 0.02 / 0.01 / 0.05 & 0.94 / 0.97 / 0.98 & 0.06 / 0.06 / 0.05 \\ 
   \midrule
&&\multicolumn{3}{c}{Heteroskedastic across Units}\\
\cmidrule(lr){3-5}
120 &  24 & 0.14 / 0.22 / 0.36 & 0.84 / 0.79 / 0.85 & 0.10 / 0.12 / 0.10 \\ 
  120 &  48 & 0.10 / 0.11 / 0.12 & 0.88 / 0.98 / 0.89 & 0.07 / 0.06 / 0.08 \\ 
  120 &  96 & 0.10 / 0.06 / 0.08 & 0.96 / 0.97 / 0.99 & 0.07 / 0.06 / 0.05 \\ 
  240 &  24 & 0.20 / 0.02 / 0.15 & 0.92 / 0.81 / 0.84 & 0.08 / 0.11 / 0.10 \\ 
  240 &  48 & 0.06 / 0.06 / 0.08 & 0.95 / 0.88 / 0.90 & 0.07 / 0.08 / 0.08 \\ 
  240 &  96 & 0.03 / 0.03 / 0.04 & 0.92 / 0.94 / 1.02 & 0.07 / 0.07 / 0.04 \\ 
   \midrule
&&\multicolumn{3}{c}{Heteroskedastic across Units and over Time}\\
\cmidrule(lr){3-5}
120 &  24 & 0.10 / 0.43 / 0.36 & 0.81 / 0.83 / 0.82 & 0.13 / 0.12 / 0.11 \\ 
  120 &  48 & 0.07 / 0.26 / 0.19 & 0.87 / 0.94 / 0.87 & 0.09 / 0.07 / 0.10 \\ 
  120 &  96 & 0.07 / 0.07 / 0.06 & 0.94 / 0.95 / 0.91 & 0.07 / 0.06 / 0.07 \\ 
  240 &  24 & 0.26 / 0.23 / 0.21 & 0.85 / 0.78 / 0.80 & 0.10 / 0.15 / 0.11 \\ 
  240 &  48 & 0.07 / 0.00 / 0.10 & 0.91 / 0.89 / 0.90 & 0.07 / 0.06 / 0.08 \\ 
  240 &  96 & 0.04 / 0.06 / 0.02 & 0.93 / 0.93 / 0.92 & 0.07 / 0.06 / 0.07 \\ 
   \bottomrule
		\end{tabular}
		\begin{tablenotes}
			\footnotesize
			\item \emph{Note:} $\psi$ denotes the share of missing observations; Bias refers to relative biases in percentage, Ratio denotes the average ratios of standard errors to standard deviations, and Size is the empirical size of $z$-tests with 5\% nominal size; results are based on $500$ replications.
		\end{tablenotes}
	\end{threeparttable}
\end{table}

\begin{table}[!htbp]
	\centering
	\begin{threeparttable}
		\caption{Properties of $\tilde{\beta}$ - Missing Data Pattern 3}
		\label{tab:prop_pattern3}
		\begin{tabular}{@{}*{2}{l}*{3}{c}@{}}
			\toprule
 $\overline{N}$&$\overline{T}$&\multicolumn{3}{c}{$\psi = 0.0 \; / \; \psi = 0.2 \; / \; \psi = 0.4$}\\
 \cmidrule(lr){3-5}
 &&Bias&Ratio&Size\\
 \midrule
&&\multicolumn{3}{c}{Homoskedastic}\\
\cmidrule(lr){3-5}
120 &  24 & 0.32 / 0.29 / 0.32 & 0.79 / 0.89 / 0.84 & 0.12 / 0.10 / 0.12 \\ 
  120 &  48 & 0.03 / 0.05 / 0.16 & 0.91 / 0.93 / 0.88 & 0.06 / 0.06 / 0.08 \\ 
  120 &  96 & -0.03 /  0.07 /  0.12 & 0.97 / 0.95 / 0.93 & 0.06 / 0.07 / 0.08 \\ 
  240 &  24 & 0.17 / 0.03 / 0.07 & 0.83 / 0.85 / 0.89 & 0.10 / 0.10 / 0.08 \\ 
  240 &  48 & 0.10 / 0.04 / 0.07 & 0.89 / 0.88 / 0.91 & 0.08 / 0.08 / 0.07 \\ 
  240 &  96 & 0.05 / 0.01 / 0.01 & 0.97 / 0.98 / 0.99 & 0.05 / 0.06 / 0.06 \\ 
   \midrule
&&\multicolumn{3}{c}{Homoskedastic with Fat Tails}\\
\cmidrule(lr){3-5}
120 &  24 & 0.32 / 0.30 / 0.31 & 0.81 / 0.83 / 0.83 & 0.12 / 0.10 / 0.11 \\ 
  120 &  48 & 0.16 / 0.20 / 0.18 & 0.87 / 0.91 / 0.91 & 0.08 / 0.08 / 0.08 \\ 
  120 &  96 & 0.06 / 0.02 / 0.01 & 0.91 / 0.95 / 0.94 & 0.07 / 0.04 / 0.07 \\ 
  240 &  24 & 0.21 / 0.04 / 0.16 & 0.87 / 0.77 / 0.78 & 0.10 / 0.13 / 0.12 \\ 
  240 &  48 & 0.06 / 0.08 / 0.05 & 0.93 / 0.98 / 0.90 & 0.06 / 0.05 / 0.07 \\ 
  240 &  96 & 0.02 / 0.07 / 0.06 & 0.94 / 0.94 / 0.93 & 0.06 / 0.06 / 0.07 \\ 
   \midrule
&&\multicolumn{3}{c}{Heteroskedastic across Units}\\
\cmidrule(lr){3-5}
120 &  24 & 0.14 / 0.25 / 0.22 & 0.84 / 0.82 / 0.84 & 0.10 / 0.11 / 0.11 \\ 
  120 &  48 & 0.10 / 0.07 / 0.13 & 0.88 / 0.92 / 0.91 & 0.07 / 0.08 / 0.08 \\ 
  120 &  96 & 0.10 / 0.04 / 0.09 & 0.96 / 0.92 / 0.95 & 0.07 / 0.06 / 0.06 \\ 
  240 &  24 & 0.20 / 0.19 / 0.12 & 0.92 / 0.79 / 0.74 & 0.08 / 0.12 / 0.15 \\ 
  240 &  48 & 0.06 / 0.17 / 0.03 & 0.95 / 0.90 / 0.90 & 0.07 / 0.09 / 0.08 \\ 
  240 &  96 & 0.03 / 0.00 / 0.01 & 0.92 / 0.96 / 0.93 & 0.07 / 0.06 / 0.07 \\ 
   \midrule
&&\multicolumn{3}{c}{Heteroskedastic across Units and over Time}\\
\cmidrule(lr){3-5}
120 &  24 & 0.10 / 0.32 / 0.24 & 0.81 / 0.80 / 0.83 & 0.13 / 0.12 / 0.10 \\ 
  120 &  48 & 0.07 / 0.06 / 0.07 & 0.87 / 0.88 / 0.89 & 0.09 / 0.08 / 0.07 \\ 
  120 &  96 & 0.07 / 0.06 / 0.03 & 0.94 / 0.97 / 0.96 & 0.07 / 0.07 / 0.06 \\ 
  240 &  24 & 0.26 / 0.11 / 0.13 & 0.85 / 0.86 / 0.86 & 0.10 / 0.10 / 0.09 \\ 
  240 &  48 & 0.07 / 0.11 / 0.04 & 0.91 / 0.92 / 0.94 & 0.07 / 0.07 / 0.06 \\ 
  240 &  96 & 0.04 / 0.03 / 0.01 & 0.93 / 0.97 / 0.95 & 0.07 / 0.07 / 0.07 \\ 
   \bottomrule
		\end{tabular}
		\begin{tablenotes}
			\footnotesize
			\item \emph{Note:} $\psi$ denotes the share of missing observations; Bias refers to relative biases in percentage, Ratio denotes the average ratios of standard errors to standard deviations, and Size is the empirical size of $z$-tests with 5\% nominal size; results are based on $500$ replications.
		\end{tablenotes}
	\end{threeparttable}
\end{table}

Second, we analyze the estimators for the number of factors proposed by \textcites{bn2002}{o2010}{ah2013}{do2019}. For $\psi > 0$, we apply the estimators to $\mathcal{P}_{\mathcal{D}}^{\phantom{\perp}}(\boldsymbol{\Gamma}(\hat{\beta}_{\bar{R}})) / (1 - \psi)$ as suggested by \textcite{jms2021}, where $\hat{\beta}_{\bar{R}}$ denotes the initial estimator with $R = \lceil 12 (\min(\overline{N}, \overline{T}) / 100)^{1 / 4} \rceil$.\footnote{This rule of thumb was suggested by \textcite{bn2002} in footnote 10 and originates in \textcite{s1989}.} For ER and GR, we use the mock eigenvalue of \textcite{ah2013} to accommodate the possibility of selecting zero factors. All results are summarized in Tables \ref{tab:nof_pattern1}--\ref{tab:nof_pattern3}. In the balanced case, all estimators exhibit little bias for sufficiently large $T$; moreover, $\text{IC}_{2}$, $\text{BIC}_{3}$, ED, and PA show low bias regardless of sample size, while ER and GR slightly underestimate the true number of factors. For unbalanced panels, the fraction and pattern of missing data affect the performance of all estimators, though in different ways. In general, ER and GR tend to underestimate the number of factors, whereas the remaining estimators tend to overestimate it. The accuracy of the different estimators in Pattern 1 remains close to that in the balanced case, but this holds only partially for Patterns 2 and 3. As in Section \ref{sec:simulation}, the results support the conjecture of \textcite{mw2015} that their main findings extend beyond the case of independent and identically normally distributed errors.

\begin{sidewaystable}[!htbp]
	\footnotesize
	\centering
	\begin{threeparttable}
		\caption{Average of $\widehat{R}$ - Missing Data Pattern 1}
		\label{tab:nof_pattern1}
		\begin{tabular}{@{}*{2}{l}*{6}{c}@{}}
			\toprule
 $\overline{N}$&$\overline{T}$&\multicolumn{6}{c}{$\psi = 0.0 \; / \; \psi = 0.2 \; / \; \psi = 0.4$}\\
 \cmidrule(lr){3-8}
 &&$\text{IC}_{2}$&$\text{BIC}_{3}$&ER&GR&ED&PA\\
 \midrule
&&\multicolumn{6}{c}{Homoskedastic}\\
\cmidrule(lr){3-8}
120 &  24 & 1.96 / 1.96 / 1.83 & 1.87 / 1.56 / 1.11 & 1.60 / 1.51 / 1.35 & 1.75 / 1.65 / 1.43 & 2.01 / 2.04 / 2.12 & 1.95 / 1.96 / 1.91 \\ 
  120 &  48 & 2.00 / 2.00 / 2.00 & 1.99 / 1.92 / 1.61 & 1.87 / 1.79 / 1.59 & 1.96 / 1.89 / 1.70 & 2.01 / 2.05 / 2.15 & 2.00 / 2.00 / 2.00 \\ 
  120 &  96 & 2.00 / 2.00 / 2.00 & 2.00 / 2.00 / 1.98 & 1.98 / 1.98 / 1.90 & 2.00 / 2.00 / 1.95 & 2.02 / 2.04 / 2.12 & 2.00 / 2.00 / 2.00 \\ 
  240 &  24 & 1.98 / 1.97 / 1.95 & 1.76 / 1.38 / 1.05 & 1.72 / 1.59 / 1.41 & 1.85 / 1.72 / 1.49 & 2.02 / 2.08 / 2.19 & 2.00 / 1.98 / 1.97 \\ 
  240 &  48 & 2.00 / 2.00 / 2.00 & 1.99 / 1.93 / 1.61 & 1.96 / 1.88 / 1.72 & 1.99 / 1.95 / 1.83 & 2.01 / 2.09 / 2.23 & 2.00 / 2.00 / 2.01 \\ 
  240 &  96 & 2.00 / 2.00 / 2.00 & 2.00 / 2.00 / 2.00 & 2.00 / 2.00 / 1.97 & 2.00 / 2.00 / 1.99 & 2.01 / 2.08 / 2.19 & 2.00 / 2.00 / 2.00 \\ 
   \midrule
&&\multicolumn{6}{c}{Homoskedastic with Fat Tails}\\
\cmidrule(lr){3-8}
120 &  24 & 2.00 / 1.95 / 1.89 & 1.91 / 1.61 / 1.11 & 1.58 / 1.54 / 1.34 & 1.73 / 1.67 / 1.45 & 2.06 / 2.08 / 2.15 & 1.96 / 1.94 / 1.90 \\ 
  120 &  48 & 2.01 / 2.01 / 2.00 & 2.00 / 1.93 / 1.59 & 1.83 / 1.74 / 1.59 & 1.94 / 1.86 / 1.70 & 2.08 / 2.08 / 2.13 & 2.00 / 2.00 / 2.00 \\ 
  120 &  96 & 2.00 / 2.01 / 2.00 & 2.00 / 2.00 / 1.99 & 1.97 / 1.95 / 1.88 & 2.00 / 1.99 / 1.95 & 2.13 / 2.10 / 2.15 & 2.01 / 2.01 / 2.01 \\ 
  240 &  24 & 2.00 / 1.99 / 1.94 & 1.81 / 1.42 / 1.06 & 1.69 / 1.59 / 1.38 & 1.82 / 1.72 / 1.49 & 2.05 / 2.09 / 2.23 & 2.00 / 1.99 / 1.96 \\ 
  240 &  48 & 2.00 / 2.00 / 2.00 & 1.99 / 1.93 / 1.65 & 1.92 / 1.88 / 1.71 & 1.98 / 1.95 / 1.81 & 2.06 / 2.11 / 2.21 & 2.00 / 2.00 / 2.02 \\ 
  240 &  96 & 2.00 / 2.00 / 2.00 & 2.00 / 2.00 / 2.00 & 2.00 / 2.00 / 1.96 & 2.00 / 2.00 / 1.98 & 2.09 / 2.11 / 2.18 & 2.00 / 2.01 / 2.02 \\ 
   \midrule
&&\multicolumn{6}{c}{Heteroskedastic across Units}\\
\cmidrule(lr){3-8}
120 &  24 & 1.98 / 1.94 / 1.86 & 1.90 / 1.59 / 1.11 & 1.58 / 1.52 / 1.34 & 1.74 / 1.63 / 1.42 & 2.01 / 2.04 / 2.10 & 1.96 / 1.96 / 1.94 \\ 
  120 &  48 & 2.00 / 2.00 / 2.00 & 2.00 / 1.94 / 1.57 & 1.88 / 1.73 / 1.56 & 1.95 / 1.83 / 1.67 & 2.02 / 2.02 / 2.12 & 2.00 / 2.00 / 2.00 \\ 
  120 &  96 & 2.00 / 2.00 / 2.00 & 2.00 / 2.00 / 1.99 & 1.97 / 1.94 / 1.88 & 2.00 / 1.98 / 1.95 & 2.01 / 2.02 / 2.12 & 2.00 / 2.00 / 2.00 \\ 
  240 &  24 & 1.99 / 1.97 / 1.94 & 1.82 / 1.40 / 1.03 & 1.71 / 1.59 / 1.42 & 1.84 / 1.73 / 1.49 & 2.01 / 2.08 / 2.20 & 1.99 / 1.98 / 1.98 \\ 
  240 &  48 & 2.00 / 2.00 / 2.00 & 2.00 / 1.95 / 1.62 & 1.93 / 1.88 / 1.71 & 1.99 / 1.95 / 1.83 & 2.01 / 2.08 / 2.24 & 2.00 / 2.00 / 2.01 \\ 
  240 &  96 & 2.00 / 2.00 / 2.00 & 2.00 / 2.00 / 2.00 & 2.00 / 1.99 / 1.97 & 2.00 / 1.99 / 1.99 & 2.01 / 2.04 / 2.16 & 2.00 / 2.00 / 2.00 \\ 
   \midrule
&&\multicolumn{6}{c}{Heteroskedastic across Units and over Time}\\
\cmidrule(lr){3-8}
120 &  24 & 1.98 / 1.96 / 1.89 & 1.92 / 1.60 / 1.13 & 1.56 / 1.46 / 1.38 & 1.68 / 1.60 / 1.46 & 2.01 / 2.04 / 2.15 & 1.96 / 1.95 / 1.93 \\ 
  120 &  48 & 2.00 / 2.00 / 2.00 & 2.00 / 1.94 / 1.57 & 1.84 / 1.75 / 1.59 & 1.94 / 1.86 / 1.69 & 2.02 / 2.03 / 2.10 & 2.00 / 2.00 / 2.00 \\ 
  120 &  96 & 2.00 / 2.00 / 2.00 & 2.00 / 2.00 / 1.98 & 1.98 / 1.96 / 1.88 & 2.00 / 2.00 / 1.94 & 2.02 / 2.02 / 2.07 & 2.00 / 2.00 / 2.00 \\ 
  240 &  24 & 1.99 / 1.97 / 1.94 & 1.84 / 1.46 / 1.05 & 1.61 / 1.56 / 1.37 & 1.77 / 1.67 / 1.46 & 2.01 / 2.04 / 2.17 & 1.98 / 1.97 / 1.96 \\ 
  240 &  48 & 2.00 / 2.00 / 2.00 & 2.00 / 1.95 / 1.62 & 1.93 / 1.85 / 1.68 & 1.98 / 1.93 / 1.78 & 2.01 / 2.08 / 2.19 & 2.00 / 2.00 / 2.01 \\ 
  240 &  96 & 2.00 / 2.00 / 2.00 & 2.00 / 2.00 / 2.00 & 2.00 / 1.99 / 1.97 & 2.00 / 2.00 / 1.99 & 2.01 / 2.04 / 2.17 & 2.00 / 2.00 / 2.00 \\ 
   \bottomrule
		\end{tabular}
		\begin{tablenotes}
			\scriptsize
			\item \emph{Note:} $\psi$ denotes the share of missing observations; $\text{IC}_2$ and $\text{BIC}_{3}$ denote the information criteria of \textcite{bn2002}, ER and GR are the estimators of \textcite{ah2013}, ED is the estimator of \textcite{o2010}, and PA is the parallel analysis described in \textcite{do2019}. The true number of factors is two. The initial estimator for $\beta$ uses $R = \lceil 12 (\min(\overline{N}, \overline{T}) / 100)^{1 / 4} \rceil$ factors. Results are based on $500$ replications.
		\end{tablenotes}
	\end{threeparttable}
\end{sidewaystable}

\begin{sidewaystable}[!htbp]
	\footnotesize
	\centering
	\begin{threeparttable}
		\caption{Average of $\widehat{R}$ - Missing Data Pattern 2}
		\label{tab:nof_pattern2}
		\begin{tabular}{@{}*{2}{l}*{6}{c}@{}}
			\toprule
 $\overline{N}$&$\overline{T}$&\multicolumn{6}{c}{$\psi = 0.0 \; / \; \psi = 0.2 \; / \; \psi = 0.4$}\\
 \cmidrule(lr){3-8}
 &&$\text{IC}_{2}$&$\text{BIC}_{3}$&ER&GR&ED&PA\\
 \midrule
&&\multicolumn{6}{c}{Homoskedastic}\\
\cmidrule(lr){3-8}
120 &  24 & 1.96 / 2.14 / 2.49 & 1.87 / 1.84 / 1.99 & 1.60 / 1.40 / 1.43 & 1.75 / 1.52 / 1.63 & 2.01 / 2.74 / 2.79 & 1.95 / 2.05 / 2.28 \\ 
  120 &  48 & 2.00 / 2.51 / 2.93 & 1.99 / 2.02 / 2.29 & 1.87 / 1.38 / 1.48 & 1.96 / 1.57 / 1.87 & 2.01 / 3.13 / 3.26 & 2.00 / 2.73 / 2.88 \\ 
  120 &  96 & 2.00 / 2.91 / 3.02 & 2.00 / 2.14 / 2.80 & 1.98 / 1.44 / 1.65 & 2.00 / 1.87 / 2.19 & 2.02 / 3.75 / 3.93 & 2.00 / 3.00 / 3.14 \\ 
  240 &  24 & 1.98 / 2.23 / 2.63 & 1.76 / 1.79 / 1.93 & 1.72 / 1.39 / 1.46 & 1.85 / 1.54 / 1.65 & 2.02 / 2.92 / 2.94 & 2.00 / 2.20 / 2.41 \\ 
  240 &  48 & 2.00 / 2.75 / 2.98 & 1.99 / 2.00 / 2.26 & 1.96 / 1.41 / 1.69 & 1.99 / 1.71 / 2.14 & 2.01 / 3.56 / 3.47 & 2.00 / 2.92 / 2.94 \\ 
  240 &  96 & 2.00 / 3.00 / 3.15 & 2.00 / 2.22 / 2.94 & 2.00 / 1.47 / 1.76 & 2.00 / 1.98 / 2.38 & 2.01 / 3.99 / 3.99 & 2.00 / 3.07 / 3.33 \\ 
   \midrule
&&\multicolumn{6}{c}{Homoskedastic with Fat Tails}\\
\cmidrule(lr){3-8}
120 &  24 & 2.00 / 2.19 / 2.56 & 1.91 / 1.86 / 2.00 & 1.58 / 1.36 / 1.37 & 1.73 / 1.48 / 1.51 & 2.06 / 2.64 / 2.75 & 1.96 / 2.08 / 2.29 \\ 
  120 &  48 & 2.01 / 2.48 / 2.93 & 2.00 / 2.01 / 2.30 & 1.83 / 1.41 / 1.47 & 1.94 / 1.61 / 1.80 & 2.08 / 3.11 / 3.19 & 2.00 / 2.71 / 2.87 \\ 
  120 &  96 & 2.00 / 2.87 / 3.03 & 2.00 / 2.15 / 2.81 & 1.97 / 1.49 / 1.62 & 2.00 / 1.84 / 2.24 & 2.13 / 3.59 / 3.89 & 2.01 / 3.01 / 3.13 \\ 
  240 &  24 & 2.00 / 2.23 / 2.61 & 1.81 / 1.76 / 1.95 & 1.69 / 1.36 / 1.41 & 1.82 / 1.50 / 1.62 & 2.05 / 2.88 / 2.91 & 2.00 / 2.19 / 2.43 \\ 
  240 &  48 & 2.00 / 2.77 / 2.99 & 1.99 / 2.01 / 2.29 & 1.92 / 1.38 / 1.66 & 1.98 / 1.70 / 2.01 & 2.06 / 3.43 / 3.48 & 2.00 / 2.92 / 2.96 \\ 
  240 &  96 & 2.00 / 3.00 / 3.16 & 2.00 / 2.25 / 2.94 & 2.00 / 1.46 / 1.73 & 2.00 / 2.02 / 2.39 & 2.09 / 4.02 / 4.03 & 2.00 / 3.08 / 3.32 \\ 
   \midrule
&&\multicolumn{6}{c}{Heteroskedastic across Units}\\
\cmidrule(lr){3-8}
120 &  24 & 1.98 / 2.15 / 2.55 & 1.90 / 1.85 / 2.02 & 1.58 / 1.36 / 1.41 & 1.74 / 1.48 / 1.59 & 2.01 / 2.58 / 2.74 & 1.96 / 2.08 / 2.32 \\ 
  120 &  48 & 2.00 / 2.57 / 2.92 & 2.00 / 2.01 / 2.34 & 1.88 / 1.38 / 1.50 & 1.95 / 1.57 / 1.85 & 2.02 / 3.01 / 3.11 & 2.00 / 2.76 / 2.90 \\ 
  120 &  96 & 2.00 / 2.90 / 3.04 & 2.00 / 2.20 / 2.88 & 1.97 / 1.43 / 1.59 & 2.00 / 1.72 / 2.16 & 2.01 / 3.37 / 3.77 & 2.00 / 3.00 / 3.18 \\ 
  240 &  24 & 1.99 / 2.26 / 2.61 & 1.82 / 1.78 / 1.93 & 1.71 / 1.35 / 1.43 & 1.84 / 1.46 / 1.64 & 2.01 / 2.89 / 2.90 & 1.99 / 2.26 / 2.42 \\ 
  240 &  48 & 2.00 / 2.77 / 2.98 & 2.00 / 2.00 / 2.29 & 1.93 / 1.43 / 1.59 & 1.99 / 1.71 / 1.99 & 2.01 / 3.33 / 3.33 & 2.00 / 2.91 / 2.96 \\ 
  240 &  96 & 2.00 / 3.00 / 3.18 & 2.00 / 2.29 / 2.94 & 2.00 / 1.51 / 1.75 & 2.00 / 2.02 / 2.37 & 2.01 / 3.94 / 3.99 & 2.00 / 3.08 / 3.35 \\ 
   \midrule
&&\multicolumn{6}{c}{Heteroskedastic across Units and over Time}\\
\cmidrule(lr){3-8}
120 &  24 & 1.98 / 2.16 / 2.50 & 1.92 / 1.87 / 2.01 & 1.56 / 1.35 / 1.36 & 1.68 / 1.48 / 1.54 & 2.01 / 2.48 / 2.60 & 1.96 / 2.05 / 2.23 \\ 
  120 &  48 & 2.00 / 2.54 / 2.92 & 2.00 / 2.01 / 2.40 & 1.84 / 1.39 / 1.49 & 1.94 / 1.57 / 1.87 & 2.02 / 3.01 / 3.10 & 2.00 / 2.72 / 2.88 \\ 
  120 &  96 & 2.00 / 2.89 / 3.03 & 2.00 / 2.18 / 2.87 & 1.98 / 1.45 / 1.62 & 2.00 / 1.82 / 2.15 & 2.02 / 3.48 / 3.79 & 2.00 / 3.01 / 3.14 \\ 
  240 &  24 & 1.99 / 2.24 / 2.66 & 1.84 / 1.82 / 1.95 & 1.61 / 1.39 / 1.40 & 1.77 / 1.51 / 1.55 & 2.01 / 2.79 / 2.82 & 1.98 / 2.21 / 2.36 \\ 
  240 &  48 & 2.00 / 2.74 / 2.98 & 2.00 / 2.01 / 2.34 & 1.93 / 1.39 / 1.61 & 1.98 / 1.65 / 2.03 & 2.01 / 3.21 / 3.22 & 2.00 / 2.88 / 2.93 \\ 
  240 &  96 & 2.00 / 3.00 / 3.15 & 2.00 / 2.28 / 2.93 & 2.00 / 1.44 / 1.77 & 2.00 / 2.07 / 2.31 & 2.01 / 3.92 / 3.98 & 2.00 / 3.06 / 3.30 \\ 
   \bottomrule
		\end{tabular}
		\begin{tablenotes}
			\scriptsize
			\item \emph{Note:} $\psi$ denotes the share of missing observations; $\text{IC}_2$ and $\text{BIC}_{3}$ denote the information criteria of \textcite{bn2002}, ER and GR are the estimators of \textcite{ah2013}, ED is the estimator of \textcite{o2010}, and PA is the parallel analysis described in \textcite{do2019}. The true number of factors is two. The initial estimator for $\beta$ uses $R = \lceil 12 (\min(\overline{N}, \overline{T}) / 100)^{1 / 4} \rceil$ factors. Results are based on $500$ replications.
		\end{tablenotes}
	\end{threeparttable}
\end{sidewaystable}

\begin{sidewaystable}[!htbp]
	\footnotesize
	\centering
	\begin{threeparttable}
		\caption{Average of $\widehat{R}$ - Missing Data Pattern 3}
		\label{tab:nof_pattern3}
		\begin{tabular}{@{}*{2}{l}*{6}{c}@{}}
			\toprule
 $\overline{N}$&$\overline{T}$&\multicolumn{6}{c}{$\psi = 0.0 \; / \; \psi = 0.2 \; / \; \psi = 0.4$}\\
 \cmidrule(lr){3-8}
 &&$\text{IC}_{2}$&$\text{BIC}_{3}$&ER&GR&ED&PA\\
 \midrule
&&\multicolumn{6}{c}{Homoskedastic}\\
\cmidrule(lr){3-8}
120 &  24 & 1.96 / 1.99 / 2.61 & 1.87 / 1.76 / 1.70 & 1.60 / 1.48 / 1.23 & 1.75 / 1.59 / 1.34 & 2.01 / 2.27 / 3.02 & 1.95 / 1.98 / 2.71 \\ 
  120 &  48 & 2.00 / 2.00 / 3.08 & 1.99 / 1.99 / 2.21 & 1.87 / 1.63 / 1.18 & 1.96 / 1.81 / 1.31 & 2.01 / 2.78 / 4.06 & 2.00 / 2.04 / 3.57 \\ 
  120 &  96 & 2.00 / 2.04 / 3.67 & 2.00 / 2.00 / 2.82 & 1.98 / 1.73 / 1.14 & 2.00 / 1.87 / 1.25 & 2.02 / 3.69 / 4.87 & 2.00 / 2.55 / 4.55 \\ 
  240 &  24 & 1.98 / 1.98 / 2.79 & 1.76 / 1.63 / 1.60 & 1.72 / 1.47 / 1.23 & 1.85 / 1.63 / 1.35 & 2.02 / 2.62 / 3.53 & 2.00 / 1.99 / 2.96 \\ 
  240 &  48 & 2.00 / 2.02 / 3.39 & 1.99 / 1.99 / 2.29 & 1.96 / 1.66 / 1.17 & 1.99 / 1.80 / 1.30 & 2.01 / 3.49 / 4.59 & 2.00 / 2.12 / 3.90 \\ 
  240 &  96 & 2.00 / 2.24 / 4.15 & 2.00 / 2.00 / 2.98 & 2.00 / 1.84 / 1.09 & 2.00 / 1.93 / 1.27 & 2.01 / 4.12 / 5.22 & 2.00 / 3.21 / 4.91 \\ 
   \midrule
&&\multicolumn{6}{c}{Homoskedastic with Fat Tails}\\
\cmidrule(lr){3-8}
120 &  24 & 2.00 / 2.00 / 2.57 & 1.91 / 1.77 / 1.74 & 1.58 / 1.48 / 1.22 & 1.73 / 1.60 / 1.35 & 2.06 / 2.23 / 2.91 & 1.96 / 1.95 / 2.71 \\ 
  120 &  48 & 2.01 / 2.01 / 3.10 & 2.00 / 1.99 / 2.25 & 1.83 / 1.63 / 1.22 & 1.94 / 1.78 / 1.35 & 2.08 / 2.69 / 3.86 & 2.00 / 2.06 / 3.57 \\ 
  120 &  96 & 2.00 / 2.03 / 3.68 & 2.00 / 2.00 / 2.84 & 1.97 / 1.77 / 1.12 & 2.00 / 1.89 / 1.27 & 2.13 / 3.42 / 4.78 & 2.01 / 2.49 / 4.49 \\ 
  240 &  24 & 2.00 / 1.99 / 2.81 & 1.81 / 1.69 / 1.64 & 1.69 / 1.53 / 1.19 & 1.82 / 1.66 / 1.28 & 2.05 / 2.45 / 3.43 & 2.00 / 1.99 / 2.95 \\ 
  240 &  48 & 2.00 / 2.03 / 3.41 & 1.99 / 2.00 / 2.28 & 1.92 / 1.67 / 1.16 & 1.98 / 1.80 / 1.31 & 2.06 / 3.21 / 4.56 & 2.00 / 2.12 / 3.95 \\ 
  240 &  96 & 2.00 / 2.18 / 4.19 & 2.00 / 2.00 / 2.97 & 2.00 / 1.84 / 1.11 & 2.00 / 1.94 / 1.24 & 2.09 / 3.99 / 5.23 & 2.00 / 3.17 / 4.90 \\ 
   \midrule
&&\multicolumn{6}{c}{Heteroskedastic across Units}\\
\cmidrule(lr){3-8}
120 &  24 & 1.98 / 1.99 / 2.66 & 1.90 / 1.79 / 1.75 & 1.58 / 1.48 / 1.18 & 1.74 / 1.60 / 1.29 & 2.01 / 2.13 / 3.02 & 1.96 / 1.97 / 2.79 \\ 
  120 &  48 & 2.00 / 2.01 / 3.15 & 2.00 / 1.99 / 2.24 & 1.88 / 1.66 / 1.17 & 1.95 / 1.81 / 1.32 & 2.02 / 2.45 / 3.76 & 2.00 / 2.04 / 3.63 \\ 
  120 &  96 & 2.00 / 2.04 / 3.68 & 2.00 / 2.00 / 2.87 & 1.97 / 1.76 / 1.15 & 2.00 / 1.88 / 1.25 & 2.01 / 3.27 / 4.60 & 2.00 / 2.61 / 4.62 \\ 
  240 &  24 & 1.99 / 2.00 / 2.73 & 1.82 / 1.68 / 1.62 & 1.71 / 1.46 / 1.23 & 1.84 / 1.62 / 1.33 & 2.01 / 2.46 / 3.40 & 1.99 / 1.98 / 2.97 \\ 
  240 &  48 & 2.00 / 2.02 / 3.40 & 2.00 / 2.00 / 2.28 & 1.93 / 1.73 / 1.16 & 1.99 / 1.84 / 1.28 & 2.01 / 3.14 / 4.54 & 2.00 / 2.13 / 3.95 \\ 
  240 &  96 & 2.00 / 2.21 / 4.17 & 2.00 / 2.00 / 2.99 & 2.00 / 1.81 / 1.11 & 2.00 / 1.91 / 1.23 & 2.01 / 3.89 / 5.09 & 2.00 / 3.25 / 4.92 \\ 
   \midrule
&&\multicolumn{6}{c}{Heteroskedastic across Units and over Time}\\
\cmidrule(lr){3-8}
120 &  24 & 1.98 / 1.98 / 2.61 & 1.92 / 1.81 / 1.73 & 1.56 / 1.46 / 1.23 & 1.68 / 1.60 / 1.32 & 2.01 / 2.14 / 2.76 & 1.96 / 1.96 / 2.67 \\ 
  120 &  48 & 2.00 / 2.01 / 3.17 & 2.00 / 1.98 / 2.26 & 1.84 / 1.60 / 1.16 & 1.94 / 1.74 / 1.29 & 2.02 / 2.47 / 3.80 & 2.00 / 2.05 / 3.62 \\ 
  120 &  96 & 2.00 / 2.03 / 3.68 & 2.00 / 2.00 / 2.83 & 1.98 / 1.74 / 1.14 & 2.00 / 1.86 / 1.26 & 2.02 / 3.35 / 4.69 & 2.00 / 2.50 / 4.50 \\ 
  240 &  24 & 1.99 / 2.00 / 2.81 & 1.84 / 1.71 / 1.65 & 1.61 / 1.48 / 1.22 & 1.77 / 1.63 / 1.29 & 2.01 / 2.34 / 3.14 & 1.98 / 1.99 / 2.97 \\ 
  240 &  48 & 2.00 / 2.01 / 3.38 & 2.00 / 2.00 / 2.28 & 1.93 / 1.68 / 1.14 & 1.98 / 1.84 / 1.26 & 2.01 / 3.10 / 4.33 & 2.00 / 2.11 / 3.87 \\ 
  240 &  96 & 2.00 / 2.15 / 4.17 & 2.00 / 2.00 / 2.98 & 2.00 / 1.81 / 1.11 & 2.00 / 1.93 / 1.25 & 2.01 / 3.87 / 5.09 & 2.00 / 3.07 / 4.88 \\ 
   \bottomrule
		\end{tabular}
		\begin{tablenotes}
			\scriptsize
			\item \emph{Note:} $\psi$ denotes the share of missing observations; $\text{IC}_2$ and $\text{BIC}_{3}$ denote the information criteria of \textcite{bn2002}, ER and GR are the estimators of \textcite{ah2013}, ED is the estimator of \textcite{o2010}, and PA is the parallel analysis described in \textcite{do2019}. The true number of factors is two. The initial estimator for $\beta$ uses $R = \lceil 12 (\min(\overline{N}, \overline{T}) / 100)^{1 / 4} \rceil$ factors. Results are based on $500$ replications.
		\end{tablenotes}
	\end{threeparttable}
\end{sidewaystable}

\end{document}